\documentclass[twocolumn]{aastex631}

\let\tablenum\relax
\usepackage{siunitx}
\usepackage{booktabs}
\usepackage{ragged2e}
\usepackage{blindtext}
\usepackage{tabularx}
\usepackage{amsmath}
\usepackage{mathtools}
\usepackage{xspace}
\usepackage{xcolor}
\usepackage{placeins}
\usepackage{upgreek}
\usepackage{rotating}  
\font\myfont=cmr12 at 11pt

\usepackage{graphicx}
\usepackage{hyperref}
\newcommand{\dmunit}{pc\,cm$^{-3}$\xspace}

\begin{document}

\title{{\myfont VLBI OBSERVATIONS OF SN 2012AU REVEAL A COMPACT RADIO SOURCE A DECADE POST EXPLOSION}}

\author[0000-0002-5857-4264]{Mattias Lazda}
\affiliation{Dunlap Institute for Astronomy \& Astrophysics, University of Toronto, 50 St. George Street, Toronto, ON M5S 3H4, Canada}
\affiliation{David A. Dunlap Department of Astronomy \& Astrophysics, University of Toronto, 50 St. George Street, Toronto, ON M5S 3H4, Canada}
\author[0000-0003-0510-0740]{Kenzie Nimmo}
\altaffiliation{NHFP Einstein Fellow}
\affiliation{MIT Kavli Institute for Astrophysics and Space Research, Massachusetts Institute of Technology, 77 Massachusetts Ave, Cambridge, MA 02139, USA}
\affiliation{Center for Interdisciplinary Exploration and Research in Astronomy (CIERA), Northwestern University, 1800 Sherman Avenue, Evanston, IL 60201, USA}
\author[0000-0001-7081-0082]{Maria R. Drout}
\affiliation{David A. Dunlap Department of Astronomy \& Astrophysics, University of Toronto, 50 St. George Street, Toronto, ON M5S 3H4, Canada}
\author[0000-0001-9814-2354]{Benito Marcote}
\affiliation{Anton Pannekoek Institute for Astronomy, University of Amsterdam, Science Park 904, 1098 XH Amsterdam, The Netherlands}
\affiliation{Joint Institute for VLBI ERIC, Oude Hoogeveensedijk 4, 7991~PD Dwingeloo, The Netherlands}
\author[0000-0003-2317-1446]{Jason W.T. Hessels}
\affiliation{Anton Pannekoek Institute for Astronomy, University of Amsterdam, Science Park 904, 1098 XH Amsterdam, The Netherlands}
\affiliation{ASTRON, Netherlands Institute for Radio Astronomy, Oude Hoogeveensedijk 4, 7991 PD Dwingeloo, The Netherlands}
\affiliation{Trottier Space Institute, McGill University, 3550 rue University, Montr\'eal, QC H3A 2A7, Canada}
\affiliation{Department of Physics, McGill University, 3600 rue University, Montr\'eal, QC H3A 2T8, Canada}
\author[0009-0002-4843-2913]{Eli Wiston}
\affiliation{Department of Astronomy, University of California, Berkeley, CA 94720-3411, USA}
\affiliation{Berkeley Center for Multi-messenger Research on AstroPhysical Transients and OutReach (Multi-RAPTOR), University of California, Berkeley, CA}
\author[0000-0003-4768-7586]{Raffaella Margutti}
\affiliation{Department of Astronomy, University of California, Berkeley, CA 94720-3411, USA}
\affiliation{Department of Physics, University of California, 366 Physics North MC 7300, Berkeley, CA 94720, USA}
\affiliation{Berkeley Center for Multi-messenger Research on AstroPhysical Transients and OutReach (Multi-RAPTOR), University of California, Berkeley, CA}
\author[0000-0001-9381-8466]{Omar Ould-Boukattine}
\affiliation{Anton Pannekoek Institute for Astronomy, University of Amsterdam, Science Park 904, 1098 XH Amsterdam, The Netherlands}
\affiliation{ASTRON, Netherlands Institute for Radio Astronomy, Oude Hoogeveensedijk 4, 7991 PD Dwingeloo, The Netherlands}
\author[0000-0003-1792-2338]{Tanmoy Laskar}
\affiliation{Department of Physics \& Astronomy, University of Utah, Salt Lake City, UT 84112, USA}
\author[0000-0002-4708-4219
]{Jacco Vink}
\affiliation{Anton Pannekoek Institute for Astronomy, University of Amsterdam, Science Park 904, 1098 XH Amsterdam, The Netherlands}
\author[0000-0002-7706-5668]{Ryan Chornock}
\affiliation{Department of Astronomy, University of California, Berkeley, CA 94720-3411, USA}
\affiliation{Berkeley Center for Multi-messenger Research on AstroPhysical Transients and OutReach (Multi-RAPTOR), University of California, Berkeley, CA}
\author[0000-0002-9415-3766]{James K.\ Leung}
\affiliation{Dunlap Institute for Astronomy \& Astrophysics, University of Toronto, 50 St. George Street, Toronto, ON M5S 3H4, Canada}
\affiliation{David A. Dunlap Department of Astronomy \& Astrophysics, University of Toronto, 50 St. George Street, Toronto, ON M5S 3H4, Canada}
\affiliation{Racah Institute of Physics, The Hebrew University of Jerusalem, Jerusalem 91904, Israel}
\author[0000-0001-5126-6237]{Deanne L. Coppejans}
\affiliation{Department of Physics, University of Warwick, Coventry CV4 7AL, UK}
\author[0000-0002-0763-3885]{Dan Milisavljevic}
\affiliation{Department of Physics and Astronomy, Purdue University, 525 Northwestern Ave, West Lafayette, IN 47907, USA}
\author[0000-0002-0772-9326]{Juan Mena-Parra}
\affiliation{Dunlap Institute for Astronomy \& Astrophysics, University of Toronto, 50 St. George Street, Toronto, ON M5S 3H4, Canada}
\affiliation{David A. Dunlap Department of Astronomy \& Astrophysics, University of Toronto, 50 St. George Street, Toronto, ON M5S 3H4, Canada}
\author[0000-0002-7507-8115]{Dan Patnaude}
\affiliation{Center for Astrophysics—Harvard \& Smithsonian, 60 Garden Street, Cambridge, MA 02138, USA}

\begin{abstract}

Three leading models have been put forth to justify the observed radio re-brightening associated with stripped-envelope supernovae (SESNe) years post-explosion: radiation from an emerging pulsar wind nebula (PWN), shock interaction with a dense circumstellar medium (CSM), or emission from off-axis, relativistic jets. SN~2012au is a particularly intriguing SESN in this regard as observations obtained $\gtrsim$ 6 years post-explosion have shown \emph{both} (i) optical emission features consistent with a young PWN and (ii) a radio re-brightening. We present the results of our Very-Long-Baseline-Interferometric (VLBI) observations of SN 2012au performed between 8 to 13 years post core-collapse. Our VLBI observations reveal a luminous, steadily fading radio source that remains compact ($\leq1.4\times10^{17}~\mathrm{cm}$) and stationary ($\leq0.36c$) over the course of our campaign.  Overall, we find that our VLBI measurements can be readily explained by a $\sim$decade-old PWN, potentially  explained by shock interaction with specific CSM geometries, and are unlikely to be explained by emission from an off-axis, relativistic jet. Assuming a PWN origin, our observations require that the initial spin-down luminosity of the central pulsar be between $10^{36}~\mathrm{erg~s^{-1}}\leq\dot{E}_0\leq {4\times10^{42}}~\mathrm{erg~s^{-1}}$ and radio efficiency factor be $\eta_\mathrm{R}\geq {3\times10^{-7}}$ (both quoted at the  $ {99.7\%}$ confidence interval). These results are consistent with independent inferences obtained using optical spectroscopy of SN~2012au, alongside inferences of known Galactic systems. If a PWN origin is confirmed, SN 2012au would represent the first extragalactic PWN emerging from a modern day SN, providing a novel opportunity to study the formation properties of a decade-old pulsar.
\end{abstract}

\newpage

  
2\section{Introduction}\label{s:introduction}

Core-collapse supernovae (CCSNe) are a diverse class of stellar explosions that mark the catastrophic end of massive stars. Broadly, CCSNe are divided into hydrogen-rich and hydrogen-poor, characterized by the amount of hydrogen present in their spectra. Those with a deficit of hydrogen are denoted stripped-envelope events and include the spectroscopic classes Type IIb, Ib, and Ic \citep{1997_filippenko}. The proposed mechanisms leading to H-envelope removal often involve a binary companion, stellar winds or eruptions, although the exact mechanism is an area of active research (see \citealt{smith_2014_review} for a review).

SN~2012au was an atypically energetic SN Ib ($E_k\cong10^{52}~\mathrm{erg}$, where $E_k$ is the kinetic energy coupled with the optically emitting ejecta) that garnered significant interest due to its spectral evolution \citep{Milisavljevic_2013,Takaki_2013,Milisavljevic_2018}. Localized to NGC 4790 at a luminosity distance of $D_L = 23.5\pm3.1~\mathrm{Mpc}$\footnote{\cite{Milisavljevic_2013} originally quote an uncertainty of $\pm0.5~\mathrm{Mpc}$ on the distance to NGC 4790. Upon revisiting available distance estimates for NGC 4790, we found that this uncertainty was likely underestimated. Our updated uncertainty reflects the standard deviation across independent measurements, retrieved from \cite{1985A&AS...59...43B,2007A&A...465...71T,2013MNRAS.429.2677K}. }, SN~2012au gradually evolved from a Type Ib to Ic classification, eventually exhibiting nebular emission lines resembling the engine-driven SN 1998bw \citep{1998Natur.395..670G}, SN 1997dq \citep{2001AJ....121.1648M} and H-poor superluminous SN 2007bi \citep{2009Natur.462..624G}. The similarities in the late-time ($\sim1~\mathrm{yr}$) spectral evolution of SN~2012au and these events prompted \cite{Milisavljevic_2013} to suggest that a common explosion mechanism may unify this subset of SNe over a range of luminosities. Additionally, subsequent re-analysis of the SN~2012au's optical evolution prompted \cite{Pandey_2021} and \cite{2024Natur.628..733R} to suggest that the slow optical fading could be due to energy injection from a central engine. A common model for such central engines invokes the rapid spin-down of a magnetar: a highly magnetized neutron star (NS) with a dipole magnetic field strength in excess of $B\gtrsim 10^{14}~\mathrm{G}$, whose rotational energy is converted into magnetic dipole radiation and deposited into the surrounding ejecta \citep{2010ApJ...717..245K,2010ApJ...719L.204W,eiden2025dynamicsobservationalsignaturescorecollapse}. \par

Targeted optical observations of SN 2012au six years post explosion revealed forbidden oxygen and sulfur emission lines \emph{unaccompanied} by narrow H Balmer lines  \citep{Milisavljevic_2018}. Narrow emission features detected many years post core-collapse are often explained by invoking delayed shock interaction with extended, dense, H-rich circumstellar material, arising from the mass loss of the progenitor star (see, e.g., SN 2014C and 2001em; \citealt{2006ApJ...641.1051C,2015ApJ...815..120M,Margutti2017,Chandra_2020}). For SN 2012au, however, the lack of accompanying narrow H Balmer lines ruled out the presence of such material. \cite{Milisavljevic_2018} instead argued that a plausible scenario involved shock interaction driven by a newly-formed pulsar wind nebula (PWN), an astrophysical object formed when relativistic winds, driven by the dissipation of the rotational energy of a pulsar, are confined to a dense region surrounding a young NS (\citealt{1984ApJ...278..630R,Gaensler_2006}).\par

The possibility of a candidate, extragalactic PWN emerging from SN 2012au is notable for two key reasons. First, while more than $200$ PWNe have been discovered to date, they all reside within the Milky Way and Small/Large Magellanic Clouds (e.g., \citealt{1997ApJ...480..364F,Haberl_2012,Carli_2024,Fransson_2024}). Consequently, confirmation of a PWN in NGC 4790 would allow us to explore the similarities and differences between an extragalactic PWN and those within our Local Group, broadening the sample of PWNe available for study. Second, all confirmed PWNe are $\gtrsim900$ years old \citep{Gaensler_2006}, and since no PWN has been confirmed to emerge from a modern-day SN\footnote{SN 1987A in the LMC remains the one potential exception, with recent evidence using the \textit{JWST} favoring the emergence of a cooling NS or potential PWN \citep{Fransson_2024}. However, its exact nature remains to be confirmed and, as yet, it has no detectable radio counterpart below a few hundred GHz \citep{2018ApJ...867...65C,2025arXiv251013629B}.}, our understanding of these systems at formation has relied solely on theoretical models (e.g., \citealt{1984ApJ...278..630R,2009ApJ...703.2051G}). Consequently, the confirmation of a $\sim$ decade-old PWN would provide both immediate constraints on the formation properties of this system and insight into the early evolution of a NS's rotational properties for the first time. \par

Alongside nebular emission features, theoretical models predict that radio emission from a young PWN could emerge anywhere between years to centuries post explosion, depending on the properties of the SN and formation properties of the NS (e.g., \citealt{1984ApJ...278..630R,Gaensler_2006}). As such, radio emission offers another means to  probe the possible emergence of a young PWN. SN 2012au was initially observed in the radio by \cite{Kamble_2014} over 4 months after the explosion with the Jansky Very-Large-Array (VLA), inferring a circumstellar medium (CSM) with a wind-like density profile.  Subsequently, approximately 7 years after the initial collapse, the radio flux was found to have increased by a factor of $\sim4$ at GHz frequencies relative to what it was at 4 months post-explosion \citep{Stroh_2021}. However, the detection of a radio re-brightening is not \textit{necessarily} confirmation of an emerging PWN. Indeed, SN~2012au is one of a growing population of SNe (many of which are stripped-envelope events) in which late-time radio re-brightening has been detected in recent years, and for which numerous models have been put forth to explain the observed radio evolution for these SNe \citep{2014c,Chandra_2020,Stroh_2021,rose2024latetimesupernovaeradiorebrightening}. \par

While emission from an emerging PWN remains one plausible model to explain the late-time radio re-brightening, alternative models such as delayed shock interaction with a dense CSM or emission from off-axis jets often provide equally plausible explanations (see, e.g., \citealt{Soderberg_2006,Bietenholz_2017,Margutti2017, Stroh_2021}). In particular, whether it be emission arising from shocked material driven by the interaction between (i) the highly magnetized wind powered by the central NS and the inner SN ejecta, (ii) the expanding SN ejecta and an extended CSM component, or (iii) the expanding jet and surrounding stellar environment, each model successfully predicts a late-time radio re-brightening.\par

One approach to disentangle these competing models involves performing  sufficiently high angular resolution radio imaging to successfully resolve the emitting region. Monitoring over the course of years-to-decades can enable direct constraints on the geometry and proper motion of the emitting region, allowing for direct comparisons with theoretical predictions. For example, observations of the Type II SN 1987A with the Australian Telescope Compact Array (ATCA) revealed a radio re-brightening $\sim3$ years post explosion \citep{1992Natur.355..147S}. Continuous monitoring later revealed a resolved, extended ring and the radio emission subsequently interpreted as shock interaction with a dense CSM component \citep{1997ApJ...479..845G}.  While traditional connected-element interferometers, like ATCA and the VLA, are sufficient for studying CCSNe within the Local Group like SN 1987A, the vast majority of CCSNe reside beyond the resolvable limit of traditional interferometers. As such, one must employ the technique of Very-Long-Baseline Interferometry (VLBI) to resolve the emitting region. VLBI observations of the Type IIn SN 1986J over the span of 30 years revealed the late-time emergence of a marginally resolved central component embedded in an extended shell \citep{Bietenholz_2017}. With a measurement of the size of the emitting region and an upper limit on the proper motion, \cite{Bietenholz_2017} used this information to favor an interpretation of the compact central component as a dense CSM surrounding the progenitor, resulting from a period of common-envelope evolution of the progenitor \citep{2012ApJ...752L...2C}, although a NS origin has also been considered \citep{Bietenholz_2002,2019arXiv190506690B}. Both examples showcase how resolved radio imaging can be leveraged to better understand the nature of the radio emission detected decades post explosion. \par 

Unfortunately, the majority of optically detected SNe have either no detectable radio counterpart, or one that fades too rapidly to adequately monitor its evolution. As a direct consequence, VLBI observations have, to date, only been performed on a small subset of core-collapse events (e.g., \citealt{1996cr_vlbi,bietenholz2011radioimagingsn1993j,Bietenholz_2017,Ng_2011,1979c_vlbi,2001em_vlbi,2012ApJ...751..125B}). However, as the sample of CCSNe with delayed radio re-brightening grows \citep{Stroh_2021,rose2024latetimesupernovaeradiorebrightening}, establishing a framework on how to effectively use VLBI observations to disentangle the competing models is necessary to improve our understanding of these complex systems. \par

With optical analysis favoring the emergence of a young PWN and detected radio re-brightening of SN~2012au six years post-collapse, we carried out a VLBI campaign to monitor SN~2012au on milliarcsecond angular scales. Additionally, with some models favoring a young magnetar origin for the SN, we performed a dedicated search for fast radio bursts (FRBs; see \citealt{Petroff22} for a review). We performed our VLBI observations of SN~2012au using the European VLBI Network (EVN) and the Very Long Baseline Array (VLBA). Here, we present our results from a multi-year campaign of observations that were performed at $5$, $22$ and $24$ GHz from $2020$--$2025$ (corresponding to 8 to 13 years post explosion). We confirm that SN~2012au has re-brightened relative to radio observations performed months post explosion \citep{Kamble_2014,Stroh_2021} and find that the source remains unresolved on VLBI scales, stationary within uncertainties and is now gradually fading. We investigate in our companion paper (Wiston et al. in prep., Paper II hereafter) the origin of this late-time radio re-brightening using multi-epoch broad-band radio spectral energy distributions (SEDs), alongside updated X-ray observations. In this analysis, we focus on implications of the VLBI observations. In Section \ref{s:observations}, we provide details of our observations,  data reduction steps and results of our FRB search. In Section \ref{s:results}, we present VLBI images and flux measurements, along with constraints on the size and proper motion of the emitting region. In Sections \ref{s:pwn}, \ref{s:csm} and \ref{s:jet}, we explore whether the radio emission is consistent with a PWN, CSM interaction, and off-axis jet, respectively. Finally, we summarize and conclude our analysis in Section \ref{s:conclusion}.\par
Unless otherwise stated, we adopt an explosion date of February 29.5 2012 (MJD 55986.5; \citealt{2023ApJ...955...71R,2012CBET.3052....1H,Milisavljevic_2013}). All timestamps quoted throughout this analysis are relative to the aforementioned explosion date. At a spectroscopic redshift of $z = 0.004540$ \citep{Kourkchi_2020}, we translate the Tully-Fisher distance to NGC 4790 to an angular diameter distance of $D_A=23.3\pm3.1~\mathrm{Mpc}$. Finally, all uncertainties are quoted at the $68\%$ confidence interval unless stated otherwise.

\section{Observations \& Data Reduction} \label{s:observations}
We observed SN~2012au using the EVN and the VLBA at 8 epochs between Feb 2020 and Jan 2025 (roughly 8--13 years post-explosion; see Table \ref{tab:radio-obs}). The relevant project codes are EN006[A-E] (PI: Nimmo), EL071[A-B] (PI: Lazda) and VLBA/24B-252 (PI: Margutti \& Drout). The 7 epochs of EVN observations were conducted in either the C- or K-band (central frequency of $4.93$ and $22.24~\mathrm{GHz}$, respectively). These data were recorded at a rate of 2048 Mbps (full polarization, eight subbands per polarization, $32~\mathrm{MHz}$ per subband). In addition, a single VLBA observation was performed at a central frequency of $23.57~\mathrm{GHz}$ with data recorded at 4096 Mbps (full polarization, four subbands, 128 MHz per subband). The integrated time resolution of all VLBI observations was $2$\,s per sample. Details regarding specific configurations, along with participating antennas in each observation, are provided in Table~\ref{tab:radio-obs}. The EVN data were correlated at the Joint Institute for VLBI ERIC (JIVE) while the VLBA data were correlated at the National Radio Astronomy Observatory (NRAO). Visibilities were correlated using the Super FX Correlator (EVN; \citealt{2015ExA....39..259K}) and DiFX (VLBA; \citealt{Deller_2007}). In the subsections below we further describe the calibration and imaging of this VLBI data, as well as our search for short milli-second timescale transients. 

\begin{deluxetable*}{clcccc}
\tabletypesize{\footnotesize}
\tablecaption{Summary of VLBI observations of SN~2012au using the EVN and VLBA.\label{tab:radio-obs}}
\tablehead{\colhead{Epoch} & \colhead{Date} & \colhead{Age$^d$} & \colhead{Code} & \colhead{$\nu^b$} & \colhead{Participating Antennas$^a$} \\
\colhead{} & \colhead{} & \colhead{(days)} & \colhead{} & \colhead{(GHz)} & \colhead{}}
\startdata  
     {$1$} & {2020 Feb 28} & {2920} &{EN006A} & $4.93$ & {Jb,O8,Wb,T6,Ys,Hh,Sv,Zc,Bd,Ir,De,Kn,Da} \\
     {$2$} & {2020 Mar 04} & {2925} & {EN006B} & $22.24$ & {Jb,Ef,Wb,O8,T6,Ur,Hh,Sv,Zc,Bd,Da,De,Kn,Pi} \\
     {$3$} & {2020 Jun 08} & {3021} &{EN006C} & $22.24$ & {Ef,Zc,Ur,Ys,Hh,Sv,Bd,Mh} \\
     {$4$} & {2020 Jun 17} & {3030} &{EN006D} & $4.93$ & {Jb,Ef,Wb,O8,T6,Ur,Hh,Sv,Zc,Bd,Da,De,Kn,Pi} \\
     {$5$} & {2021 Mar 03} & {3289} &{EN006E} & $4.93$ & {Jb, Wb,Ef,Mc,O8,Ur,Hh,Sv,Zc,Ir} \\
     {$6$} & {2024 Feb 29} & {4382} &{EL071A} & $4.93$ & {Wb,Ef,Mc,O8,Jb,T6,Ur,Tr,Ys,Hh,Ir,Sr} \\
     {$~~\!7^{c}$} & {2024 Mar 11}  & {4393} & {EL071B} & $22.24$ & {T6,Ky,Ku,Hh,Ys,Mc,Sr,Ur,Ef,Jb} \\
     {$8$} & {2025 Jan 23} & {4711} & {24B-252} & $23.57$ & {Sc,Mk,Br,Ov,Kp,Pt,Fd,Nl,Hn}\\
\enddata
\tablenotetext{a}{Ef=Effelsberg (100-m), Jb=Jodrell Bank (38-m $\times$ 25-m), Wb=Westerbork (25-m), Mc=Medicina (32-m), Sr=Sardinia (65-m), O8=Onsala (25-m),T6=Tianma (65-m), Ur=Urumqi (25-m), Ys=Yebes (40-m), Ir=Irbene (32-m), Mh=Metsahovi (14-m), Hh=Hartebeesthoek (26-m), Zc=Zelenchukskaya (32-m), Sv=Svetloe (32-m), Bd=Badary (32-m),Ky=KVN Yonsei (21-m), Ku=KVN Ulsan (21-m), Da=Darnhall (25-m), De=Defford (25-m), Kn=Knockin (25-m), Pi=Pickmere (25-m), Sc=Saint-Croix (25-m), Mk=Mauna Kea (25-m), Br=Brewster (25-m), Ov=Owens Valley (25-m), Kp=Kitt Peak (25-m), Pt=Pie Town (25-m), Fd=Fort Davis (25-m), Nl=North Liberty (25-m), Hn=Hancock (25-m).}
\tablenotetext{b}{Central observing frequency.}
\tablenotetext{c}{Observation omitted from analysis because of failure to obtain robust calibration solutions resulting from faint (\(\sim\mathrm{mJy}\)) phase-calibrator.}
\tablenotetext{d}{Days since date of explosion on MJD 55986.5 \citep{2023ApJ...955...71R,2012CBET.3052....1H,Milisavljevic_2013}}.

\end{deluxetable*}

\subsection{Calibration of Interferometric Data}\label{ss:calibration}
\subsubsection{Observing Strategy}\label{ss:obs_strategy}

We calibrated our interferometric data using phase-referencing, a process which derives calibration solutions from a nearby calibrator source and applies them to the target. We observed a minimum of four sources per observation: RFC J1159$+$2914, which served as the fringe finder; RFC J1305$-$1033, the phase calibrator; ICRF J130313.8$-$105117, the check source  (J1303$-$1051 hereafter), and the target, SN~2012au  \citep{2015AJ....150...58F,2022ApJS..263...24A}. In general, the structure of the scans was organized as follows. First, each epoch began with $\sim 4$~minute scan of the fringe finder. Fringe finder scans were interspersed throughout the observation on $\sim 2~\mathrm{hr}$ cycles. Then, we scheduled a phase-referencing cycle of $1.5$~minutes on J1305$-$1033 and $3.5$~minutes on SN~2012au at C-band. For EVN K-band observations, we shortened the cycle to $0.75$~minutes on J1305$-$1033 and $1.5$~minutes on SN~2012au to account for more rapidly varying atmospheric conditions at higher frequencies. For the single VLBA observation, we further reduced this to $40~\mathrm{sec}$ on J1305$-$1033 and $40~\mathrm{sec}$ on SN 2012au\footnote{Slew times of individual EVN dishes increased the overhead of EVN observations, resulting in longer switching intervals at K-band. The more rapid slew-times with the VLBA allowed us to reduce this interval to achieve better phase coherence in this observation.}. Finally, every 4 to 5 cycles, we interleaved observations of the check source J1303-1051.\par
Throughout this paper we assume a position of J1305$-$1033 of $\alpha(\mathrm{J}2000) = 13^\mathrm{hr}05^\mathrm{m}33^\mathrm{s}.015036(4)$, $\delta(\mathrm{J}2000) =-10^\circ 33'19''.4282(1)$ and a position of J1303$-$1051 of $\alpha(\mathrm{J}2000) = 13^\mathrm{hr}03^\mathrm{m}13^\mathrm{s}.867940(53)$, $\delta(\mathrm{J}2000) =-10^\circ 51'17''.12905(18)$\footnote{Positions and uncertainties at time of schedule are retrieved from \href{https://obs.vlba.nrao.edu/cst/}{https://obs.vlba.nrao.edu/cst/}, which query results from the third realization of the International Celestial Reference Frame and Radio Fundamental Catalog \citep{2020A&A...644A.159C,Petrov_2025}.}. This leads to a sky separation of $d = \sqrt{(\Delta\alpha\mathrm{cos}\delta)^2 + (\Delta\delta)^2} = 0.64^\circ$ between the phase calibrator and check source. Relative to the phase calibrator, SN~2012au has a sky separation of $d = 2.6^\circ$\footnote{Prior to our VLBI observations, the best known position of SN~2012au was $\alpha(\mathrm{J}2000) = 12^\mathrm{hr}54^\mathrm{m}52^\mathrm{s}.257(5)$, $\delta(\mathrm{J}2000) =-10^\circ 14'50''.5(3)$ \citep{Stroh_2021}. The updated position is provided in Section \ref{ss:astrometry}.}. In Epoch $7$, we attempted to use RFC J1254$-$1053 as the phase calibrator (in place of J1305$-$1033), because it is closer on the sky to SN~2012au ($d \sim 1^\circ$). However, the source was insufficiently bright to derive robust phase calibration solutions and thus this observation was omitted entirely from our analysis. 

\subsubsection{Reduction and Flagging of Visibilities}
The visibilities were reduced using standard procedures in the Astronomical Image Processing System (AIPS; \citealt{greisen_2003}). For EVN observations, a-priori gain calibration and system temperature measurements were applied using products obtained from the EVN pipeline\footnote{\href{https://evlbi.org/handling-evn-data}{https://evlbi.org/handling-evn-data}}. We manually inspected the system temperature measurements and flagged time ranges in which system temperatures varied widely over short time intervals, manually flagged visibilities contaminated by radio frequency interference (RFI; both as a function of time and frequency) and flagged data with extremely low amplitudes due to delays in arriving on source.  Finally, we flagged any data for which the elevation of the target fell below 15 degrees. 

For observations performed at C-band, we obtained and applied ionospheric corrections using the \texttt{vlbatecr} function after the initial round of flagging. 
In doing so, we observed an increase in the number of successful solution intervals by $\sim 1-5\%$ when deriving multiband delay calibration solutions. Thus, we opted to apply ionospheric corrections across all C-band observations for consistency. No ionospheric corrections were applied to the K-band observations given that delays introduced by the ionosphere are expected to be negligible at this observing band when phase referencing to a nearby calibrator. 

We derived instrumental delay calibration solutions using the fringe-finder scan(s), with the specific scans used varying on a per-epoch basis. In instances where not all antennas were participating during an individual fringe-finder scan, we derived solutions using a primary scan, referencing the solutions to a secondary scan of the fringe-finder in order to recover solutions for all antennas. For EVN observations, we opted to use the bandpass solutions calculated in the automatic EVN pipeline\footnote{These solutions are available for download on the EVN data archive: \href{https://archive.jive.nl/scripts/portal.php}{https://archive.jive.nl/scripts/portal.php}} after verifying that we could recover comparable solutions. For the VLBA observation, bandpass solutions were derived using scans of J1159$+$2914. Finally, we derived multiband delay calibration solutions independently per subband, with a chosen time interval to achieve, on average, two solutions per scan. The resulting solutions were inspected for any rapid variations in delays and subsequently flagged. \par

We applied our calibration solutions for each epoch to our target, SN~2012au, in addition to our phase calibrator and check source. Using \texttt{DIFMAP} \citep{1994BAAS...26..987S}, we imaged the calibrated visibilities for our phase calibrator J1305$-$1033. We derived phase and amplitude self-calibration solutions using the phase calibrator to improve our calibration solutions. These corrections were subsequently applied to the visibilities of SN~2012au and J1303$-$1051. \par

\subsubsection{Low Elevation Considerations}
The low declination of SN~2012au of $\delta < -10^\circ$ implies that systematic calibration errors induced by low elevations would be more prevalent than in a typical EVN observation. At low elevations, increased ground spill leads to an increase in system temperatures which could lead to degrading noise performance, exaggerating amplitude and phase errors. Additionally, the relatively large sky separation between SN~2012au and the phase calibrator of $d = 2.64^\circ$ (primarily dominant along the azimuthal axis) introduces longer slew times, increasing overhead and making it challenging to robustly track variations in the atmosphere which become more significant at higher frequencies. In fact, a systematic characterisation of astrometric errors performed using the EVN and VLBA has shown that systematic phase uncertainties begin to dominate below $\delta<0^\circ$, in some instances introducing uncertainties as large as $\sim0.5~\mathrm{mas}$ at $8.4~\mathrm{GHz}$ \citep{pradel2006astrometric-774}. For these reasons, we put a strong emphasis on regularly observing J1303$-$1051, which at a similar declination to SN~2012au, would be susceptible to similar elevation issues and could subsequently be used to identify problematic baselines, antennas, subbands, and time ranges. \par
Comparing our images of J1303$-$1051 to available historical images\footnote{Available historical images of J1303$-$1051 are limited: however, a subset are available for viewing using the astrogeo VLBI catalog, available here: \href{https://astrogeo.org/vlbi\_images/}{https://astrogeo.org/vlbi\_images/}.}, we manually inspected and identified baselines, antennas, subbands, and time ranges that caused its morphology or astrometry to visibly deviate from what is expected, flagging both J1303$-$1051's and SN 2012au's visibilities over these variables. The main caveat to this approach is that SN~2012au was observed for much longer than J1303$-$1051, allowing more time for subtle calibration errors to accumulate and for time-variable RFI to affect the data. Therefore, additional flagging was performed directly on the target to improve the fidelity of the target image.
\subsection{Imaging SN 2012au}
For all VLBI images of SN 2012au presented in this paper, we used natural weighting to improve sensitivity and mitigate residual calibration errors. We used the CLEAN algorithm \citep{CLEANalgo} to deconvolve the array response, adopting a cell size of roughly 10\% of the resulting synthesized beam. We identified the peak in the dirty map and cleaned until the residuals were below $\sim10$\% of the dirty image peak flux (with residual calibration errors limiting our ability to hit the thermal sensitivity limit). Note that we opted not to utilize the \texttt{modelfit} function provided by DIFMAP, which has the option of fitting either a circular or elliptical Gaussian directly to the visibilities. Forward modeling has been shown to, in some cases (e.g., \citealt{bietenholz2011radioimagingsn1993j}), provide more accurate results of measured parameters (e.g., flux and size) in comparison to fitting models to the image plane generated using the CLEAN algorithm. However, we observed that forward modeling introduced morphological structure resulting from residual calibration errors that we could identify in the check source. For this reason, we opted to use the CLEAN algorithm which allows the user to specify regions to subtract dominant components from the residuals while excluding regions containing systematically introduced sources of flux.

\subsection{Fast Radio Burst Search}\label{ss:frb_search}

In addition to our VLBI observations, we independently observed SN~2012au with the $25$-m Westerbork RT-1 telescope to search for FRBs. This effort was part of the HyperFlash program (PI: Ould-Boukattine), which focuses on high-cadence monitoring and follow-up of possible FRB sources. SN~2012au was included in this effort due to the possibility that the early light curve was powered by a young magnetar (see Section 1) as well as the possible connection between the late-time radio rebrightening observed and the ``persistent radio sources" identified at the location of some FRBs (see Section \ref{sss:prs-frb} for further discussion). Our campaign spanned from MJD $60410$ (2024 April 10) to MJD $60448$ (2024 May 18), corresponding to $4420$ to $4458$ days post explosion, totaling $41$~hours. 

The data recording and reduction setup is described in \cite{ouldboukattine_2024_arxiv} and \cite{kirsten_2024_natas}. In summary, we recorded and stored raw voltage (or baseband) data in \texttt{VDIF} format with dual circular polarization and 2-bit sampling \citep{whitney_2009_evlb}. We conducted $10$ observations which lasted $\sim4$~hours each using a $128$~MHz bandwidth from $1.207$ to $1.335$~GHz. Each observation was split into $15$~minute scans. Once a scan was recorded, we converted the scan to an 8-bit Stokes I (full intensity) channelized \texttt{filterbank} file using \texttt{digifil} from the \texttt{DSPSR} software package \cite{vanstraten_2011_pasa}. We created the filterbank data with a time and frequency resolution of $64~\upmu$s and $15.625$~kHz, respectively. We mitigated RFI by applying a static mask to compromised frequency channels. With this data product, we searched for dispersed transient signals using the burst searching algorithm \texttt{HEIMDALL}, where we set a signal-to-noise (S/N) cutoff of $7$. Since the dispersion measure (DM) was not known \textit{a priori} towards SN 2012au, we searched between $10-1500$~\dmunit. Identified candidates were then classified using the burst classifier \texttt{FETCH} where we made use of models A \& H  and set a probability threshold of $50\%$ \citep{agarwal_2020_mnras}. We then manually inspected all classified burst candidates in order to verify astrophysical origin. 

We did not find any bursts during our observational campaign. Using the radiometer equation we place a fluence upper limit of $7$~Jy~ms. Here, we assume an FRB width of $1$~ms spanning the full bandwidth of $128$~MHz, and use Westerbork's System Equivalent Flux Density (SEFD) value\footnote{Acquired from \url{https://www.evlbi.org/sites/default/files/shared/EVNstatus.txt}} of $420$\,Jy.

\begin{figure*}
    \centering
    \includegraphics[width=0.8\linewidth]{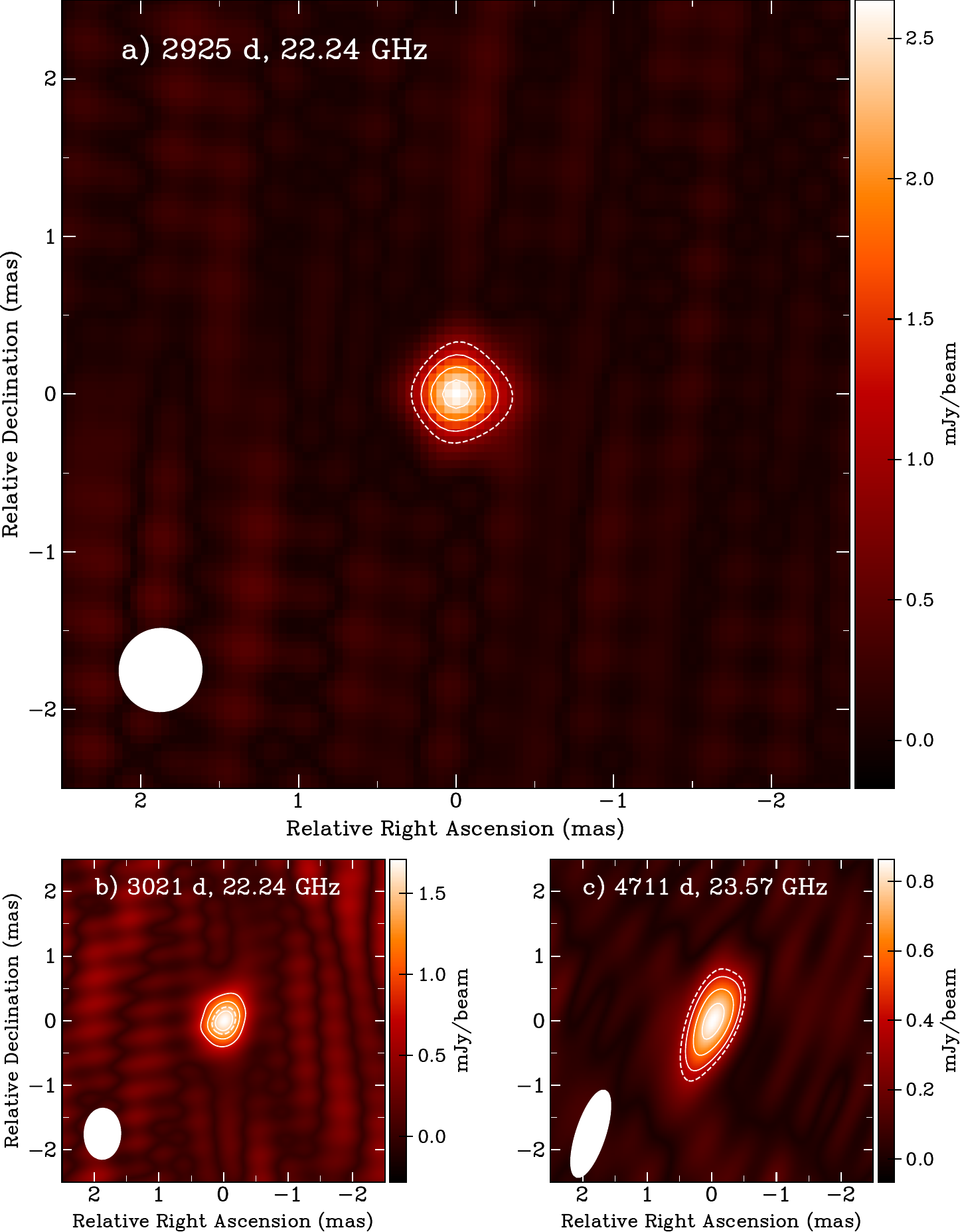}
    \caption{Cleaned K-band images of SN~2012au obtained with the EVN (panels a and b), and VLBA (panel c). 5$\sigma$ contours are indicated by the dashed line, where $\sigma$ is the root-mean-square of the image residuals. Solid contours are drawn at $50\%, 70\%$ and $90\%$ of the peak flux. The color scale is indicated in units of mJy/beam and the synthesized beam is shown as a white ellipse in the bottom left of each image. The label on each panel shows the time in days since the date of the original explosion \citep{Milisavljevic_2013} and the central observing frequency of these data. All images are centered on the peak flux of each image for visual clarity. We discuss the astrometry in Section \ref{ss:astrometry}. Panel ``a" provides the most constraining upper limit on the diameter of the emitting region of $d \lesssim(1.6\pm0.4)\times10^{17}~\mathrm{cm}$. }
    \label{fig:clean_2012au_kband}
\end{figure*}

\begin{figure*}
    \centering
    \includegraphics[width=1.0\linewidth]{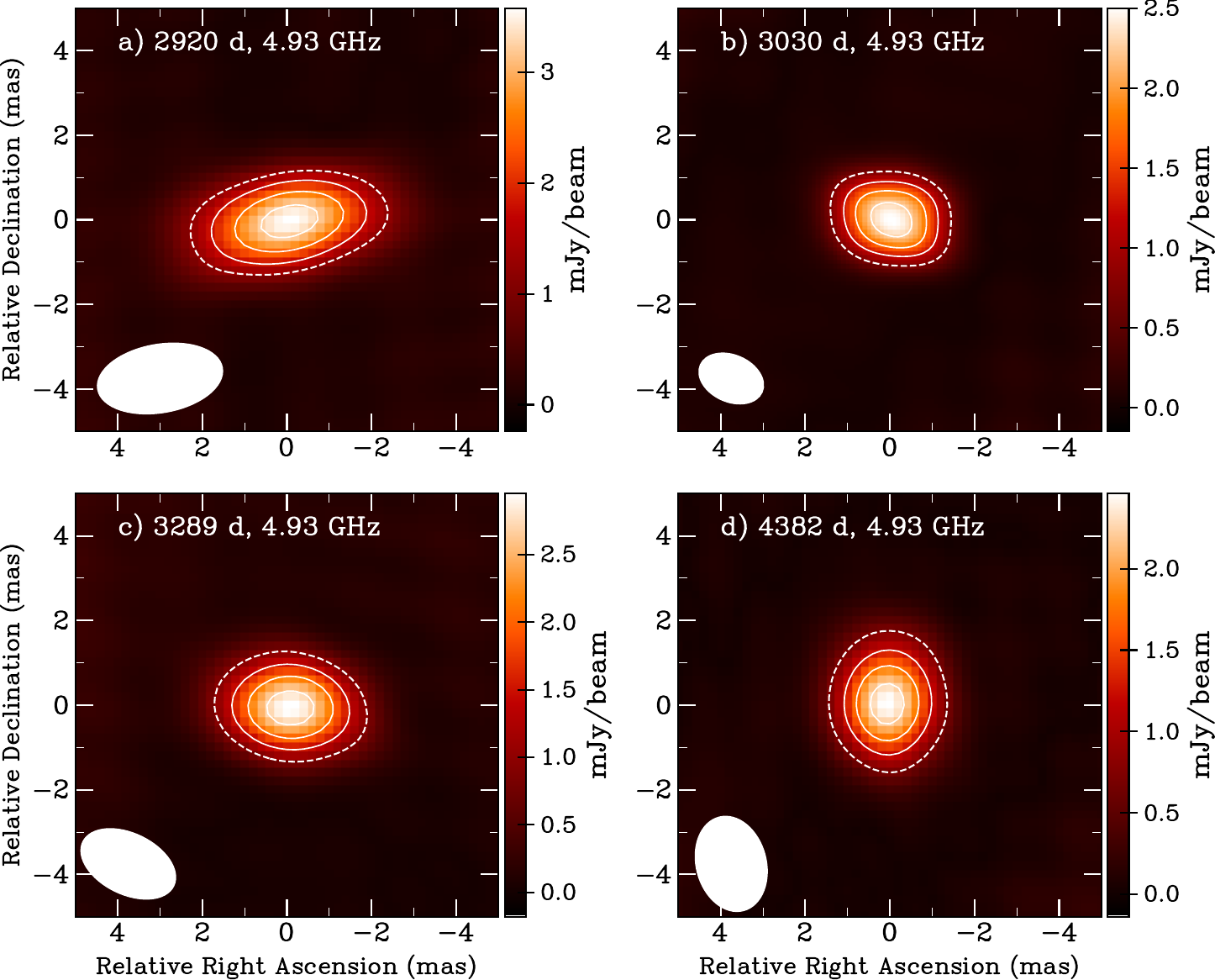}
    \caption{Same as Figure \ref{fig:clean_2012au_kband} but for C-band observations. All images were obtained using the EVN.}
    \label{fig:clean_2012au_cband}
\end{figure*}

\begin{deluxetable*}{ccccccccc}
\tabletypesize{\small}
\tablecaption{SN~2012au VLBI image properties and fit parameters.\label{tab:fitparams}}
\tablehead{
  \colhead{Epoch} &
  \colhead{Age$^a$} &
  \colhead{$\nu$$^b$} &
  \colhead{$\sigma$$^c$} &
  \colhead{$b_{\min}$$^d$} &
  \colhead{$b_{\mathrm{maj}}$$^d$} &
  \colhead{$d_{\min}$$^e$} &
  \colhead{$d_{\mathrm{maj}}$$^e$} &
  \colhead{$R$$^{f}$}
\\
  \colhead{} &
  \colhead{(days)} &
  \colhead{(GHz)} &
  \colhead{($\upmu$Jy\,beam$^{-1}$)} &
  \colhead{(mas)} &
  \colhead{(mas)} &
  \colhead{($\times10^{17}$\,cm)} &
  \colhead{($\times10^{17}$\,cm)} & 
  \colhead{($\times10^{17}$\,cm)}
}
\startdata  
  1 & 2920 &  4.93 & 240 & $\le1.9\pm0.4$ & $\le3.8\pm0.8$   &  $\le6.6\pm1.6$ & $\le13.0\pm3.2$ & $\leq5.7$ \\
  2 & 2925 & 22.24 & 170 & $\le0.5\pm0.1$  & $\le0.5\pm0.1$  &  $\le1.6\pm0.4$& $\le1.6\pm0.4$ & $\leq1.4$\\
  3 & 3021 & 22.24 & 280 & $\le0.7\pm0.1$  & $\le0.9\pm0.2$  &  $\le2.4\pm0.6$ & $\le3.0\pm0.8$ & $\leq2.1$ \\
  4 & 3030 &  4.93 & 150 & $\le2.3\pm0.5$  & $\le1.8\pm0.3$  &  $\le6.2\pm1.4$ & $\le8.0\pm2.0$ & $\leq5.2$ \\
  5 & 3289 &  4.93 & 180 & $\le2.8\pm0.6$  & $\le2.0\pm0.4$  &  $\le7.0\pm1.6$ & $\le9.6\pm2.4$ & $\leq5.9$\\
  6 & 4382 &  4.93 & 140 & $\le2.1\pm0.4$  & $\le2.5\pm0.5$  &  $\le7.2\pm1.8$ & $\le8.6\pm2.0$ & $\leq6.3$\\
  8 & 4711 & 23.57 &  68 & $\le0.7\pm0.1$  & $\le1.6\pm0.3$  &  $\le2.4\pm0.6$ & $\le5.4\pm1.4$ & $\leq2.1$\\
\enddata
\tablenotetext{a}{Days since date of explosion on MJD 55986.5 \citep{2023ApJ...955...71R,2012CBET.3052....1H,Milisavljevic_2013}.}
\tablenotetext{b}{Observing frequency.}
\tablenotetext{c}{Root‐mean‐square of residuals.}
\tablenotetext{d}{Minor and major FWHM of best-fit Gaussian fit to naturally-weighted images. A conservative $20\%$ systematic uncertainty is included to account for untracked systematic phase and amplitude errors, leading to uncertainty in the estimate of the angular size. This uncertainty is added in quadrature with the statistical uncertainty of the fit. }
\tablenotetext{e}{Minor and major FWHM at an angular diameter distance of $D_A = 23.3\pm3.1$ Mpc, accounting for both the systematic uncertainties in the angular diameters and distance to NGC 4790.}
\tablenotetext{f}{$99.7\%$ confidence interval radial upper limit assuming a spherical geometry of the radio source, calculated following $R=d_\mathrm{min}/2$. }
\end{deluxetable*}

\section{Results} \label{s:results}
We detect SN~2012au across all 7 epochs with usable data. 
We plot the cleaned K- and C-band images of SN~2012au in Figures \ref{fig:clean_2012au_kband} and \ref{fig:clean_2012au_cband}, respectively. We find that the source remains unresolved in both $5~\mathrm{GHz}$ and $22/24~\mathrm{GHz}$ images over the time span of $8$--$13$ years post explosion. 

\subsection{VLBI Size Measurements}\label{ss:vlbi_images}

To constrain the size of the emitting region, to each VLBI image we fit an elliptical Gaussian of the form: 
\begin{align}
\begin{split}
    S(x,y) &= A~\mathrm{exp}\left[-\left(\frac{X^2}{2\sigma_x^2} + \frac{Y^2}{2\sigma_y^2}\right) \right],\\
    X &\equiv (x-x_0)\mathrm{cos}~\theta + (y - y_0)\mathrm{sin}~\theta,\\
    Y &\equiv -(x-x_0)\mathrm{sin}~\theta + (y - y_0)\mathrm{cos}~\theta,
\end{split}
\end{align}
where $S$ is the flux density at the $x,y$ pixel, $x_0$ and $y_0$ are the center of the ellipse with standard deviation $\sigma_x$ and $\sigma_y$ and peak flux density $A$, rotated at an angle $\theta$.  The parameters are obtained by performing a linear least squares optimization routine using \texttt{curve\_fit} from the \texttt{scipy} Python package \citep{2020SciPy-NMeth}, signal-to-noise weighting the uncertainty of each pixel prior to performing the fit. While statistical correlations over the synthesized beam likely lead to some degree of bias in the fit parameters, we expect this contribution to be negligible in comparison to systematic contributions to the fit parameters of interest (namely, $\sigma_x,\sigma_y,x_0$ and $y_0$). For example, in some instances, we observed that removing individual antennas and re-fitting the model led to changes of up to $\sim20\%$ in $\sigma_x$ and $\sigma_y$. Thus for all final quoted fit parameters, we include a conservative $20\%$ systematic uncertainty, added in quadrature with the statistical uncertainty.  \par

A summary of all of the fit parameters and image properties is provided in Table \ref{tab:fitparams}. At a  distance of $D_A = 23.3\pm3.1~\mathrm{Mpc}$, we estimate the projected linear diameter by computing Gaussian full-width-half-maxima following $d_{\mathrm{min},\mathrm{maj}} \leq 2\sqrt{2\mathrm{ln}~2}\times\sigma_{x,y} \times D_A$, propagating both the uncertainty in the width and distance to NGC 4790. Finally, assuming a spherical geometry of the radio source, we quote the $99.7\%$ upper limit on the projected linear radius of the emitting region, assuming $R = d_\mathrm{min}/2$.  We find that SN 2012au is consistent with being unresolved across all epochs and find no evidence for extended emission at larger radii (see Appendix \ref{appendix:resolved?} for additional details).

\subsection{Astrometry \& Relative Motion}\label{ss:astrometry}
We investigate whether there is evidence for motion of the radio centroid over the course of four years of observations relative to our phase calibrator, J1305$-$1033. We estimate the total uncertainty on the relative position of the check source J1303$-$1051 and SN~2012au as
\begin{equation}
    \Delta_{\mathrm{tot},j} = \sqrt{\Delta^2_{\mathrm{stat},j} + \Delta^2_{\mathrm{pc},j} + \Delta^2_{\mathrm{pr},j}},
\end{equation}
where $\Delta_\mathrm{tot}$ is the total uncertainty, $j$ indexes R.A. and declination axes,  $\Delta_{\mathrm{stat}}$ is the statistical uncertainty of the gaussian fit position in Section \ref{ss:vlbi_images}, $\Delta_{\mathrm{pc}}$ is the statistical uncertainty of the phase calibrator's position quoted in Section \ref{ss:obs_strategy}, and $\Delta_{\mathrm{pr}}$ is the uncertainty introduced by phase referencing. We estimate $\Delta_{\mathrm{pr}}$ following \cite{pradel2006astrometric-774}: 
\begin{equation}\label{eq:phase_ref_err}
    \Delta_{\mathrm{pr},j} = \left(\Delta_j^{1^\circ} - 14\right)\times d + 14~(\upmu\mathrm{as}),
\end{equation}
where $d = \sqrt{((\alpha - \alpha_0)\mathrm{cos}\delta_0)^2 + (\delta - \delta_0)^2}$ is the separation between the target and phase calibrator in degrees, and $\Delta_j^{1^\circ}$ is an empirically determined constant which varies as a function of declination obtained from a sample of phase-referenced observations performed at X-band \citep{pradel2006astrometric-774}. At $\delta\lesssim-10^\circ$, no values for $\Delta^{1^\circ}_j$ are provided for the EVN alone. As such, for our C-band observations, we conservatively adopt the values quoted in \cite{pradel2006astrometric-774} for $\Delta^{1^\circ}_j$ at $\delta = -25^\circ$ (significantly lower than the $-10^\circ$ declination of SN~2012au) for the global VLBI array, which includes the EVN. For our K-band observations, given the increase in frequency from $\mathrm{X}$- to K-band where we expect tropospheric errors to dominate, we conservatively adopt the largest value reported in \cite{pradel2006astrometric-774} at $\delta = -25^\circ$: $\Delta^{1^\circ}_j = 481~\upmu\mathrm{as}$. \par

\begin{figure}
    \centering
    \includegraphics[width=1\linewidth]{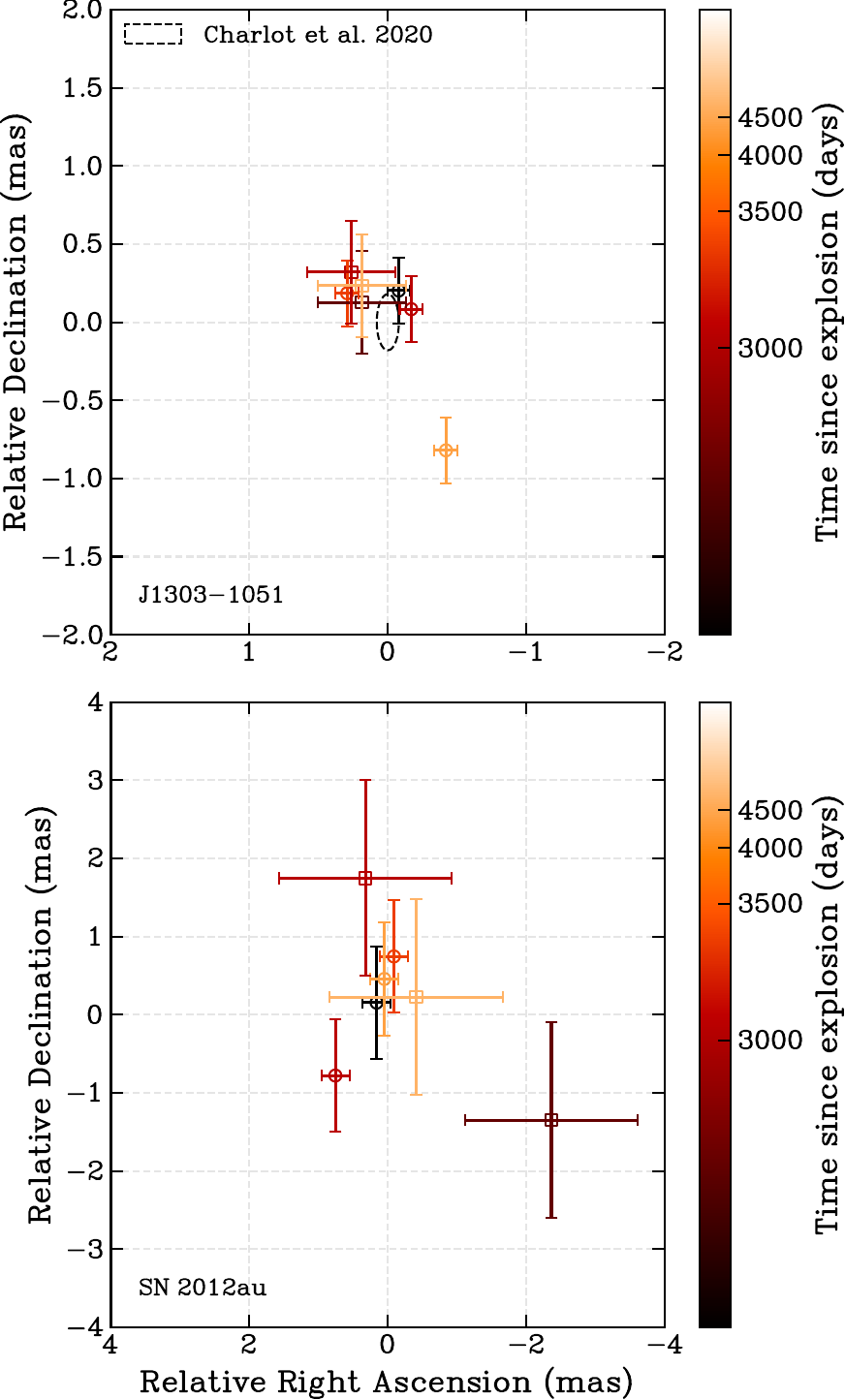}
    \caption{Offsets along the R.A. and declination axis relative to J2000 $(\alpha,\delta) = (13^\mathrm{hr}03^\mathrm{m}13^\mathrm{s}.867940(53),-10^\circ 51'17''.12905(18))$ for J1303$-$1051 (top) and $(\alpha,\delta) = (12^\mathrm{hr}54^\mathrm{m}52^\mathrm{s}.25525, -10^\circ 14'50''.634)$ for SN~2012au (bottom). We plot the total uncertainties which include both statistical and systematic contributions induced by phase referencing to the phase calibrator, J1305$-$1033 \citep{Petrov_2025}. In the top panel, we overplot the $1\sigma$-localization contour of the check source J1303$-$1051 from \cite{2020A&A...644A.159C}, highlighting consistency in our repeated phase referenced measurements. The one observation inconsistent with the remaining observations of J1303-1051 was omitted when measuring the proper motion of SN 2012au. Measurements performed at $\mathrm{K}$-band (22 \& 24\,GHz) are indicated as squares, C-band (5\,GHz) as circles. The colorscale of each data point is scaled relative to the original explosion date.}
    \label{fig:astrometry}
\end{figure}

In Figure \ref{fig:astrometry}, we plot the referenced positions of both the check-source, J1303$-$1051, and SN~2012au, along with their total uncertainties. The horizontal and vertical axes are scaled relative to the known position of J1303$-$1051 quoted in Section \ref{ss:obs_strategy} and $(\alpha,\delta) = (12^\mathrm{hr}54^\mathrm{m}52^\mathrm{s}.25525$ $, -10^\circ 14'50''.634)$ for SN~2012au. The latter is a fiducial position within the range of values found for SN~2012au that was chosen arbitrarily for visual clarity. For the check-source, J1303$-$1051, we plot archival $68\%$ confidence interval uncertainties quoted by \cite{2020A&A...644A.159C}. This comparison highlights that our measurements are nearly all in agreement with literature values, validating our phase-referencing techniques. We identify a single outlier event, corresponding to our observation on 2024 Feb 29 (Epoch 6), that lies significantly beyond the $1-\sigma$ contours of J1303$-$1051 (with a post-trials probability of  $\sim 4.3\times10^{-9}$, or $5.9\sigma$ Gaussian equivalent, of occurring by chance alone). Having no reasonable explanation to justify such a large offset, we decide to omit this epoch for any calculation relying on calibrated astrometric offsets. \par

Despite the increased angular resolution, the \textit{combined} positional uncertainties at $\gtrsim22~\mathrm{GHz}$ are dominated by systematic contributions introduced by tropospheric variations at low elevations, which is the reason why their measurements exhibit larger variance and larger uncertainties in comparison to the $5~\mathrm{GHz}$ measurements. We also note that the larger sky-separation between SN~2012au to the phase calibrator ($d = 2.6^\circ$) yields a $\sim4.3\times$ increase in uncertainties in comparison to the check source ($d = 0.6^\circ$) at similar frequencies, as defined by Equation \ref{eq:phase_ref_err}. We summarize the astrometric results in Table \ref{tab:astrometry}. Across all observations, $\Delta_\mathrm{pr}$ dominates the astrometric uncertainties for SN 2012au by an order of magnitude (or more) over $\Delta_\mathrm{stat}$ and $\Delta_\mathrm{pc}$.

\begin{table}[htbp]
  \centering
  \caption{Astrometry of SN\,2012au.  Offsets are quoted relative to J2000 $(\alpha,\delta) = (12^\mathrm{hr}54^\mathrm{m}52^\mathrm{s}.25525, -10^\circ 14'50''.634)$. The uncertainties are dominated by errors induced through phase referencing at low elevations. }
  \label{tab:astrometry}
    \begin{tabular}{l c c c }
      \toprule
      Date & Age (d)
           & \multicolumn{2}{c}{Astrometric Offset} \\
      \cmidrule(l){3-4}
           & 
           & $\Delta\alpha~\mathrm{cos\delta}$ (mas)    & $\Delta\delta$ (mas)    \\
      \midrule
      2020 Feb 28 &  2920   & $0.2\pm0.2$ & $0.2\pm0.7$         \\
      2020 Mar 04 &  2925  & $-2.4\pm1.3$ & $-1.3\pm1.3$         \\
      2020 Jun 08  & 3021  & $0.3\pm1.3$ & $1.7\pm1.3$  \\
      2020 Jun 17 &  3030  & $0.8\pm0.2$ & $-0.8\pm0.7$         \\
      2021 Mar 03 &  3289   & $-0.1\pm0.2$ & $0.7\pm0.7$         \\
      2024 Feb 29$^a$ &  4382  & $0.1\pm0.2$ & $0.5\pm0.7$         \\
      2025 Jan 23 & 4711   & $-0.4\pm1.3$  & $0.2\pm1.3$ \\
      \bottomrule
    \end{tabular}
    \tablenotetext{a}{Omitted when computing the proper motion, see Section \ref{ss:astrometry} for details. }

\end{table}

In order to investigate any possible proper motion associated with the radio counterpart to SN~2012au, we fit a linear model to the data in Table \ref{tab:astrometry}. We adopt a form of $\Delta\alpha~\mathrm{cos}\delta= v_\mathrm{RA}t +m$ and $\Delta\delta= v_\mathrm{dec}t +n$ to fit both the RA and declination offset as a function of time. Here, $t$ is the days post explosion, $v_\mathrm{RA}$ and $v_\mathrm{dec}$ are the proper motions along each axis, and $m$ and $n$ are constants. We exclude the $\delta t=4382~\mathrm{day}$ data point from the fit due to the observed astrometric offset in the check-source in the same epoch (see Figure \ref{fig:astrometry}). We infer a proper motion of $-300\pm200~\upmu\mathrm{as~yr}^{-1}$ along the R.A. axis and $100\pm 300 ~\upmu\mathrm{as~yr}^{-1}$ along the declination axis, corresponding to a net proper motion of $v_\mathrm{pm} = 320\pm210~\upmu\mathrm{as~yr^{-1}}$. At a distance of $D_A=23.3\pm3.1~\mathrm{Mpc}$, this translates to a projected sky proper motion of $v_\mathrm{pm}=0.12\pm0.08~c$, where $c$ is the speed of light. These results suggest that the source is consistent with being stationary at the $1.5\sigma$ level. We rule out proper motions exceeding $v_\mathrm{pm} >0.36c$ at the $99.7\%$ confidence interval.

\subsection{VLBI Flux Measurements}\label{ss:fluxes}

\begin{deluxetable*}{lcccc}
\tabletypesize{\small}
\tablecaption{VLBI flux densities and spectral luminosities of SN 2012au. \label{tab:flux-snc}}
\tablehead{
  \colhead{Date} & 
  \colhead{Age} & 
  \colhead{Observing Frequency} & 
  \colhead{$S_\nu$$^a$} & 
  \colhead{$\nu L_\nu$$^b$} \\
  \colhead{(UTC)} & 
  \colhead{(days)} & 
  \colhead{(GHz)} & 
  \colhead{(mJy)} & 
  \colhead{($\times10^{37}$ erg/s)}
}
\startdata  
2020 Feb 28 &  $2920$ & $4.93$ & $5.2 \pm 0.6$ & $1.7 \pm 0.5$\\
2020 Mar 04 &  $ {2925}$ &  $22.24$ & $2.2\pm0.5$ & $3.2 \pm 1.1$ \\
2020 Jun 08  & $ {3021}$ & $22.24$  & $2.2\pm0.5$ & $3.3 \pm 1.2$ \\
2020 Jun 17 &  $ {3030}$ & $4.93$ & $5.3\pm 0.6$ & $1.7 \pm 0.5$ \\
2021 Mar 03 &  $ {3289}$ & $4.93$ & $4.7\pm0.5$   & $1.5 \pm 0.4$ \\
2024 Feb 29 &  $ {4382}$ & $4.93$ & $3.3\pm0.4$   & $1.1 \pm 0.3$ \\
2025 Jan 23 & $ {4711}$ & $23.57$   & $1.4\pm0.3$ & $2.1 \pm 0.7$ \\
\enddata
\tablenotetext{a}{Uncertainties include statistical and systematic contributions, added in quadrature and quoted at the $68\%$ confidence interval. A flat $10\%$ systematic uncertainty is added to the $5~\mathrm{GHz}$ flux measurements, increased to $20\%$ for the $22/24~\mathrm{GHz}$ measurements. We caution relying on the quoted absolute K-band flux measurements due to significant differences when compared to VLA measurements obtained at similar epochs. Detailed modeling of the radio SEDs will be presented in Paper II.}
\tablenotetext{b}{Spectral luminosity at $D_L = 23.5\pm3.1~\mathrm{Mpc}$, accounting for both systematic uncertainties in the flux measurements and distance to NGC 4790.}
\end{deluxetable*}
Flux measurements are obtained by integrating CLEAN components within $5\sigma$ contours. To compensate for systematic uncertainties in our flux measurements driven by the low elevations that could, in turn, increase both phase and amplitude errors, we adopt the following scheme. At C-band, we add an additional $10\%$ uncertainty of the total flux in quadrature with the statistical uncertainty. At K-band, we increase the percentage to $20\%$ to attempt to reflect the additional increase in loss of flux due to untracked tropospheric phase variations. The adopted scheme reasonably compensates the C-band flux measurements in comparison to VLA measurements at similar epochs (Paper II). However, it is incapable of compensating at K-band, with observed differences between our flux measurements and the VLA (Paper II) by factors of up to $3$ to $5$, depending on the observation. 

We argue that it is more likely that the loss of flux is due to amplitude and phase errors driven by tropospheric variations, rather than our measurements resolving out extended structure (see Appendix \ref{appendix:resolved?} for additional details). Thus, we strongly caution against using the data presented here to make any inferences that would rely on accurate K-band fluxes. For this reason, in the following sections, we consider only the evolution of the $5$~GHz flux density measurements. We instead defer to Paper II for detailed broadband radio SED modeling of the late-time emission from SN~2012au.

To estimate the decay rate of the C-band fluxes, we perform a linear least squares optimization routine assuming a power-law model of the form $S_\nu\,= C(t/t_0)^\gamma$, where $S_\nu$ is the flux density in mJy at observing frequency $4.93~\mathrm{GHz}$, $t$ corresponds to the time in days post collapse, $t_0$ is the reference time which we fix to $3500~\mathrm{days}$ and $\alpha$ and $C$ are the fit parameters. We infer a power law index of $\gamma = -1.2^{+0.4}_{-0.4}$ and scaling constant of $C = 4.4\pm0.3~\mathrm{mJy}$, ruling out a flat/positive power-law index at the $99.7\%$ CI. Our measurements suggest that SN 2012au is, at present, steadily fading at a rate of $S_\nu\propto t^{-1.2\pm0.4}$ at $4.93~\mathrm{GHz}$.

\section{Implications of VLBI measurements for the Origin of the Late-time Radio Re-brightening}
In the following section, we consider multiple scenarios to interpret the \textit{compact} ($\leq 1.4\times10^{17}~\mathrm{cm}$) and \textit{stationary} ($\leq0.36c$) radio source associated with SN~2012au between 2920 to 4711 days post explosion, as revealed by our VLBI imaging campaign. Specifically, we consider three main scenarios that have been put forth to explain late-time radio rebrightening in core-collapse SNe: (i) a young, extragalactic PWN, (ii) shock interaction with a dense CSM and (iii) an off-axis jet. A summary of our main conclusions is provided in Section \ref{s:conclusion}. 
\subsection{An Extragalactic, Decade-old, Pulsar Wind Nebula}\label{s:pwn}
In the PWN model, a young, newly-formed pulsar converts its rotational energy into magnetic dipole radiation and deposits the energy into the surrounding environment (e.g., \citealt{Gaensler_2006}). This energy injection, which is confined to a dense environment surrounding the parent NS, produces a highly magnetized wind that generates synchrotron emission detectable at radio frequencies. Additionally, the radiative pressure introduced by the magnetized wind promotes an outward moving ``bubble" which expands into the cool, unshocked ejecta remaining from the original supernova explosion. The PWN is thus defined by the full dynamical system comprising both the central wind and wind-shocked SN ejecta as the bubble expands into the surrounding ejecta (see Figure 2 in \citealt{Gaensler_2006} for a cartoon depiction). 

While the emission from the PWN is originally obscured at early times due to free-free absorption (FFA) by the surrounding ejecta, a radio source contained within the expanding SN remnant (SNR) may eventually become detectable once the ejecta becomes optically thin. Thus, PWNe have been proposed as one of the main mechanisms that can lead to late-time radio rebrightening in CCSNe. As such, not only was a PWN already proposed for SN~2012au based on late-time optical emission (\citealt{Milisavljevic_2018}; See Section 1) but it also provides a natural way to produce a \emph{compact} late-time radio source as revealed by our VLBI observations. We further highlight that our current proper motion limits (Section \ref{ss:astrometry}) are consistent with theoretical expectations for a young PWN, whose centroid is expected to remain stationary at these distances (with typical kick velocities $\leq 500~\mathrm{km~s^{-1}}$; \citealt{Gaensler_2006}).\par
In the remainder of this section, we will explore the implications of our VLBI observations on the properties of the central NS, under the assumption that the late-time radio emission is from a young PWN. 

\subsubsection{Constraints on the Spin-down Luminosity and Radio Efficiency Factor}\label{sss:spindown}
Numerous theoretical models have been developed to predict the dynamical evolution of PWNe (e.g., \citealt{Chevalier1977,1984ApJ...278..630R,2005ApJ...619..839C,2009ApJ...703.2051G}). These models typically rely on a few key parameters: the initial spin-down luminosity of the pulsar $\dot{E}_0$ and the properties of the SN explosion (specifically, the kinetic energy $E_k$ and ejecta mass $M_\mathrm{ej}$ released during the explosion). Thus, if the late-time emission from SN~2012au is a PWN, the upper limits on the radius of the emitting region from our VLBI observations, coupled with inferences the SN explosion properties from early optical observations, directly constrain the pulsar spin-down luminosity.  Further, since $\dot{E}_0$ relates to the magnetic field strength $B$, initial spin period $P_0$ and initial spin period derivative $\dot{P}_0$ of the central NS, we can directly constrain its formation properties \citep{Gaensler_2006}.\par
The observed radio luminosity from a PWN is related to the spin-down luminosity of the pulsar via $\dot{E} = \frac{L_\mathrm{R}}{\eta_\mathrm{R}}$, where $L_\mathrm{R}$ is the radio luminosity of the PWN and $\eta_\mathrm{R}$ characterizes the efficiency at which the PWN shock converts the pulsar's rotational energy into radio synchrotron emission. Thus, by coupling constraints on $\dot{E_0}$ from our VLBI radius upper limit with the observed radio luminosity, we can also directly constrain the efficiency factor, $\eta_\mathrm{R}$. $\eta_\mathrm{R}$ has been directly measured for a subset of Galactic pulsars that are $\gtrsim900$ years-old, with values spanning many orders of magnitude: $10^{-6}\lesssim\eta_\mathrm{R}\lesssim10^{-3}$ \citep{1997ApJ...480..364F,Gaensler_2006,2000MNRAS.318...58G}. However, it has been suggested that $10^{-3}\lesssim\eta_\mathrm{R}\lesssim0.1$ may be a more representative range for decade-old PWNe since $\eta_\mathrm{R}$ represents the integrated history of the PWN's spin-down, rather than an instantaneous measurement \citep{Dong_2023,Gaensler_2006}. To date, however, this has not been directly tested with observations.

We aim to constrain the initial spin-down luminosity and efficiency factor of SN~2012au, assuming that the late-time compact radio emission is due to a young pulsar powering a surrounding nebula. Assuming that the radio luminosity is being powered by spin-down via magnetic dipole radiation, the pulsar outputs nearly constant energy up until a typical spin-down timescale of $\tau_0 = P_0/\left[(n-1)\dot{P_0}\right]$ where $n=3$ is the assumed ``braking index", a parameter which characterizes the rate at which the angular velocity of the central pulsar changes with time \citep{1973ApJ...186..249P,Gaensler_2006}. Throughout the remainder of this analysis, we further assume that the pulsar spins down from an initial period of $P_0$ such that $\dot\Omega = -k\Omega^n$, where $\Omega = 2\pi/P$ and $k = 2 M_\perp^2/(3Ic^3)$. Here, $I = 10^{45}~\mathrm{g~cm^{2}}$ is the NS moment of inertia for a constant density $1.4~M_\odot$ sphere of radius $R_\mathrm{NS}=10~\mathrm{km}$, $M_\perp$ is the component of the magnetic dipole moment orthogonal to the rotation axis, and $c$ is the speed of light \citep{Gaensler_2006}.  \par

\begin{figure*}
    \centering
    \includegraphics[width=0.8\linewidth]{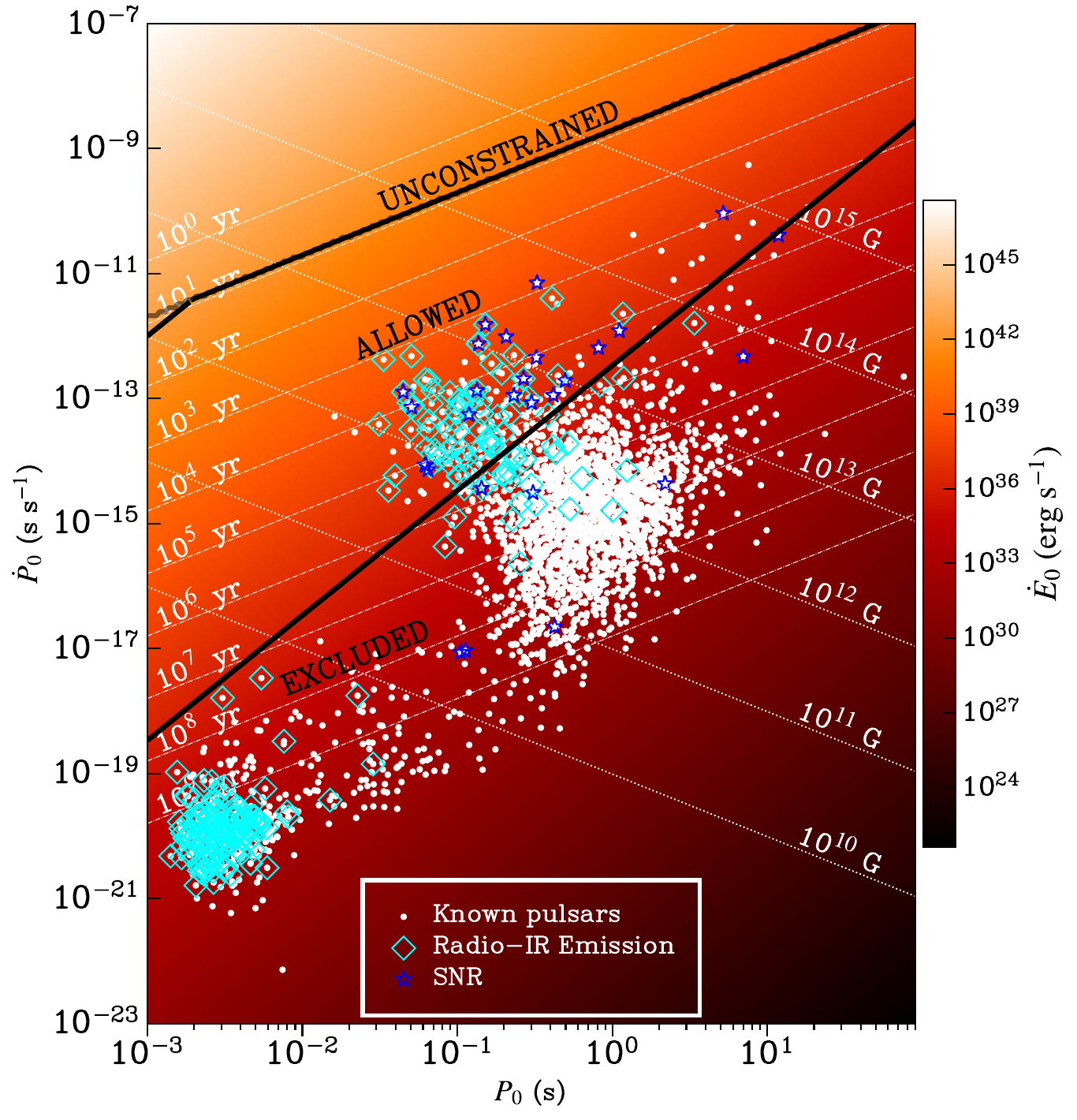}
    \caption{Constraints on the formation properties of the candidate NS at the center of SN 2012au, assuming the late-time radio emission is a PWN powered by spin-down via magnetic dipole radiation. The initial period (horizontal axis) and period derivative (vertical axis) are used to construct the initial spin-down luminosity  scales (heatmap). We overplot lines of constant dipole magnetic field strength and initial spin-down timescale as dotted and dash-dotted white lines, respectively. Allowable combinations of $P_0,\dot{P}_0$ and $B$ for SN 2012au are bounded by the central polygon. The region is bounded above by the requirement that the nebula must be contained to $R_\mathrm{PWN}\leq1.4\times10^{17}~\mathrm{cm}$ at $t=2925~\mathrm{days}$ post collapse (as constrained by our VLBI images). The region is bounded below by the requirement that the radio luminosity cannot be higher than the pulsar spin down luminosity. The region where no analytic constraint on the NS's formation properties can be established (e.g. where $\tau_0\lesssim8~\mathrm{yr}$, where $\tau_0$ is the typical spin-down timescale) is indicated by the ``unconstrained" region. We overplot present-day combinations of a large sample of known pulsars, including magnetars and recycled millisecond pulsars, for comparison \citep{2005AJ....129.1993M,2018JOSS....3..538P}. The allowable parameter space includes a number of present-day systems associated with a SNR or Radio-IR continuum emission, indicated by blue stars/cyan diamonds, respectively.  }
    \label{fig:ppdot1}
\end{figure*}

We consider the implications of our observations in the case where the spin-down timescale of the NS is greater than the age of the system (i.e. $\tau_0 \geq t$). In this case, the pulsar deposits roughly constant energy as a function of time (e.g., $\dot{E}(t) = \dot{E}_0$) up until $t \approx \tau_0$. We note that this assumption implicitly excludes all scenarios which invoke the rapid spin-down of a magnetar with $P_0 \leq0.1~\mathrm{s}$ (i.e. where $\tau_0\ll t$, see Figure \ref{fig:ppdot1}). The alternative case where $\tau_0 < t$ will be briefly discussed at the end of this section.

\paragraph{Upper Limit on Pulsar Spin-Down Luminosity from VLBI Radius} For the case where $\tau_0 \geq t$, and assuming a spherically symmetric nebula, the radial expansion of the PWN is expected to follow equation 8 in \cite{Gaensler_2006}. Re-arranging to isolate $\dot{E}_0$ and scaling to representative values, we find
\begin{align}
\begin{split}\label{eq:dotE}
    \dot{E}_0 = &1.1\times10^{39}~\mathrm{erg~s^{-1}} \left(\frac{R_\mathrm{PWN}}{0.01~\mathrm{pc}}\right)^5\left(\frac{E_k}{10^{51}~\mathrm{erg}}\right)^{-3/2}\\
    & \times\left(\frac{M_\mathrm{ej}}{5~M_\odot}\right)^{5/2}\left(\frac{t}{10~\mathrm{yr}}\right)^{-6},
\end{split}
\end{align}
where $R_\mathrm{PWN}$ is the radius of the PWN and $t$ is the time post explosion. 
With the most constraining upper limit on the radius imposed by our VLBI observations being $R_\mathrm{PWN}(t=2925~\mathrm{days) \leq 1.4\times10^{17}~\mathrm{cm}}$ ($99.7\%$ confidence interval; Table \ref{tab:fitparams}), we place constraints on the maximum allowable values of $\dot{E}_0$. Adopting the explosion parameters from \cite{Milisavljevic_2013} (i.e., $E_k \cong10^{52}~\mathrm{erg}$ and $M_\mathrm{ej} = 4\pm1~M_\odot$), we infer $\dot{E}_0\leq2\times10^{41}~\mathrm{erg~s^{-1}}$. If we instead assume the explosion parameters from \cite{Pandey_2021} (i.e., $E_k =4.8\pm0.6\times10^{51}~\mathrm{erg}$ and $M_\mathrm{ej} \simeq 5.1\pm0.7~M_\odot$), we infer $\dot{E}_0\leq {4\times10^{42}}~\mathrm{erg~s^{-1}}$. In both cases, the upper limit is quoted at the $99.7\%$ confidence interval accounting for the uncertainty in the radius and explosion parameters. From this point onward, we consider only $\dot{E}_0 \leq  {4\times10^{42}}~\mathrm{erg~s}^{-1}$, corresponding to the more conservative $99.7\%$ confidence interval upper limit on the spin-down luminosity. We note that at the time of writing, this upper limit corresponds to the most conservative upper limit on $\dot{E}_0$ based on available light-curve modeling of SN 2012au \citep{Milisavljevic_2013,Takaki_2013,Pandey_2021,2023ApJ...955...71R}. However, this upper limit is highly sensitive to the assumptions about $E_k$ and $M_\mathrm{ej}$. Consequently, any variations in the assumed explosion parameters could lead to significant variations in our quoted upper limit.

\paragraph{Constraints on Radio Efficiency Factor from Spin-Down Luminosity} Next, we can use the limit on the pulsar spin-down luminosity obtained from our VLBI size constraints, along with our measured radio luminosity, to place constraints on the radio efficiency factor. Specifically, we assume that $\dot{E}_0 \approx \nu L_\nu/\eta_\mathrm{R}$.  We can then explore constraints on $\eta_\mathrm{R}$ as:

\begin{align}
\begin{split}\label{eq:etar}
    \eta_\mathrm{R} \approx &~0.027\left(\frac{\nu L _{\nu\mathrm{,peak}}}{2\times10^{37}~\mathrm{erg~s^{-1}}}\right)\left(\frac{\dot{E}_0}{10^{39}~\mathrm{erg~s^{-1}}}\right)^{-1}.
\end{split}
\end{align}
Our measured peak 5 GHz luminosity is $\nu L_\nu = 1.7\pm0.5 \times10^{37}~\mathrm{erg~s^{-1}}$ (Table \ref{tab:flux-snc}). To be conservative, here we adopt $\nu L\nu >0.1\times10^{37}~\mathrm{erg~s^{-1}}$, which corresponds to the 99.7\% confidence interval lower limit. When combined with $\dot{E}_0\leq {4\times10^{42}}~\mathrm{erg~s^{-1}}$ (as determined by our size constraints above), we obtain $\eta_\mathrm{R}\geq {3\times10^{-7}}$ ($99.7\%$ confidence interval). This lower limit remains broadly consistent with the known population of Galactic pulsars with typical values extending between $10^{-6}\lesssim\eta_\mathrm{R} \lesssim10^{-3}$ \citep{Gaensler_2006}.\par

\paragraph{Strict Lower Limit on Pulsar Spin Down Luminosity from Measured Radio Flux} A conservative \textit{lower} limit on $\dot{E}_0$ can be established by asserting that $\eta_\mathrm{R}\leq1$ (i.e., that the observed radio luminosity can not exceed the total spin-down luminosity). Again adopting $\nu L_\nu \geq 0.1\times10^{37}~\mathrm{erg~s^{-1}}$, this yields a lower limit on the pulsar spin down luminosity of $\dot{E}_0\geq10^{36}~\mathrm{erg~s^{-1}}$ ($99.7\%$ confidence interval)\footnote{A more robust constraint relies on inferring a minimum radius by modeling the evolution of the radio SED, which will be presented in Paper II.}. Our observations thus bound the allowable initial spin-down luminosity of the putative NS (if the late-time compact radio emission from SN~2012au is from a PWN) to $10^{36}\leq \dot{E}_0 \leq  {4\times10^{42}}~\mathrm{erg~s^{-1}}$ ($99.7\%$ confidence interval). This interval is consistent with inferences for a number of Galactic PWNe with inferred initial spin-down luminosities ranging from $\dot{E}_0\sim10^{37}-10^{39}~\mathrm{erg~s^{-1}}$ (e.g., \citealt{10.1111/j.1365-2966.2012.22014.x,2021A&A...655A..41Z,2013ApJ...773..139V}). For example, inferences from modeling of $\sim900$ year old systems such as the Crab\footnote{It should be noted that while we draw on comparisons with the Crab PWN (a nebula powered by a standard pulsar and not a magnetar) due to it being the youngest and most well-studied PWN, we emphasize that the comparison should only be treated qualitatively. Modeling favors that the Crab PWN emerged from a low energy Type II SN \citep{2008ARA&A..46..127H}, in contrast to the atypically energetic Type Ib SN 2012au. As such, comparisons should not be treated as a direct statement regarding whether SN 2012au is a Crab-like PWN, but rather as a broad overall comparison between the two sources.} and G21.5$-$0.9 yield $\dot{E}_0=3.1-6.7\times10^{39}~\mathrm{erg~s^{-1}}$ and $0.58\times10^{38}~\mathrm{erg~s^{-1}}$, respectively \citep{10.1111/j.1365-2966.2012.22014.x,2021A&A...655A..41Z,2013ApJ...773..139V}. Inferences of $\gtrsim10~\mathrm{kyr}$-old systems, such as J1427$-$608 and  J1507$-$622, yield  $\dot{E}_0=1.2-5.5\times10^{38}~\mathrm{erg~s^{-1}}$, respectively \citep{2013ApJ...773..139V}. This range is also consistent with the independently inferred value for SN~2012au of $\dot{E}_0 \approx10^{40}~\mathrm{erg~s}^{-1}$ based on late-time optical spectroscopy \citep{Milisavljevic_2018}. 

\paragraph{Constraints on Allowed Magnetic Field Strength,  Initial Spin Period, and Period Derivative} Under the assumptions about the central NS defined at the beginning of this section, it can be shown that \citep{Dong_2023}
\begin{equation}
    \dot{E}_0 = 4\times10^{41}~\mathrm{erg~s^{-1}}\left(\frac{B}{10^{13}~\mathrm{G}}\right)^2\left(\frac{P_0}{10~\mathrm{ms}}\right)^{-4},
\end{equation}
where $P_0$ is the initial period and $B$ is the dipole magnetic field strength of the NS. With our constraints on $\dot{E}_0$ obtained from our VLBI measurements, we can further constrain $B$, $P_0$ and $\dot{P}_0$, which relate to one another via $\dot{E}_0 \propto\dot{P_0}/P_0^3$ and $B\propto(P_0\dot{P_0})^{1/2}$ \citep{Gaensler_2006}. \par
We plot the allowable region bounded by our observations and assumptions (i.e., $10^{36}\leq \dot{E}_0 \leq  {4\times10^{42}}~\mathrm{erg~s^{-1}}$ and $\tau_0 \gtrsim 8~\mathrm{yr}$) compared to present day values of the known NS population in Figure \ref{fig:ppdot1}. This was created using a modified version of \texttt{psrqpy.ppdot} \citep{2018JOSS....3..538P}. In the case that the compact radio emission we detect at the location of SN~2012au is a young PWN, we rule out a number of combinations of $P_0$ and $\dot{P}_0$.  In particular, we exclude the region that contains old, evolved NS populations (e.g. known millisecond pulsars and pulsars with $0.1~\mathrm{s}-1~\mathrm{s}$ periods).  Notably, however, the region consistent with our observations includes a number of the youngest set of known pulsars, many of which are observed to evolve within SNRs or have associated radio-IR emission and characteristic ages of $\tau \sim 10^{3}~\mathrm{to}~10^{4}~\mathrm{yr}$. Further, across all allowable values of $P_0$, our observations currently allow for larger values of $\dot{P_0}$ in comparison to present-day systems with similar spin periods. This potentially suggests that the central NS could be experiencing more extreme spin-down conditions early on its evolution in comparison to more evolved systems. 

\paragraph{Implications for systems with short spin-down timescales} In the alternative case where the spin-down timescale is much less than the age of the system (i.e. $\tau_0 < t$), then the radial expansion of the PWN can no longer be expressed analytically and must instead be numerically inferred  \citep{1984ApJ...278..630R,2005ApJ...619..839C}. Such a model has been proposed by \cite{Pandey_2021} to justify the observed early-time photometric evolution of SN 2012au. Specifically their model invokes a millisecond magnetar model with $P_0=18~\mathrm{ms}, B=8\times10^{14}~\mathrm{G}$ and $\tau_0 \approx 72~\mathrm{hr}$. Due to the complexity of this model, we opt to refrain from making any definitive statements regarding its current feasibility and instead defer solving this system for future works. This region is depicted by the ``unconstrained" region in Figure \ref{fig:ppdot1}. However, we do note that our flux decline rate may possibly disfavor this scenario (see Section \ref{sss:flux_evo}, below). 

\subsubsection{When Would we Expect to be Able to Resolve a PWN?}
\begin{figure}
    \centering
    \includegraphics[width=1.0\linewidth]{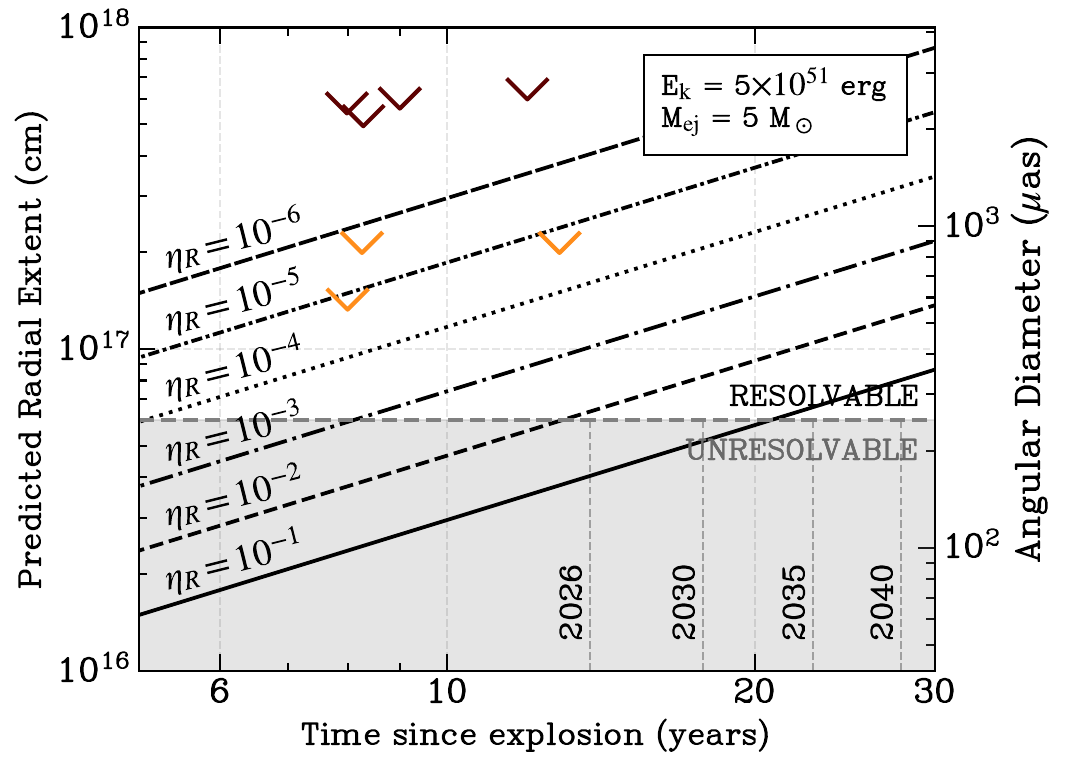}
    \caption{Predicted radial expansion of a putative PWN associated with SN~2012au for varying radio efficiency factors, $\eta_\mathrm{R}$, assuming $\tau_0\gtrsim 8~\mathrm{yr}$. $99.7\%$ confidence interval upper limits imposed by our VLBI observations are overplotted (23 GHz = orange, 5 GHz = maroon; see Table \ref{tab:fitparams}). The dashed, grey horizontal line corresponds to the resolvable limit of the global VLBI network at $22~\mathrm{GHz}$ ($\theta = 250\times250~\upmu\mathrm{as}^{2}$) at a 99.7\% confidence interval upper limit on the distance of $D_A = 32.6~\mathrm{Mpc}$. We assume $E_k=5\times10^{51}~\mathrm{erg}$ and $M_\mathrm{ej} = 5~M_\odot$ inferred from available optical light curve modeling of SN~2012au, leading to a conservative lower-limit on the radial expansion (higher energies and lower masses result in a more rapid radial expansion, becoming resolvable earlier; see text for details). For $\eta_\mathrm{R}\lesssim0.001$, in-line with known Galactic systems, the PWN should now be resolvable ($\sim14~\mathrm{yrs}$ post explosion).}
    \label{fig:res_vs_time}
\end{figure}

Since Equation \ref{eq:dotE} predicts that the PWN should expand with time, \emph{this model should be testable with future VLBI observations}. Under the reasonable assumption that the source remains detectable over the next $\sim15$ years\footnote{Assuming the K-band flux continues to decline as $t^{-1.2}$ (see Section \ref{ss:fluxes}), we expect $F_\nu \approx300~\upmu\mathrm{Jy}$ at $25$ years post-explosion, which is detectable at $\gtrsim6\sigma$ with global VLBI observations assuming $\sim8~\mathrm{hrs}$ on source.}, we compute the predicted expansion for $\eta_\mathrm{R} = 10^{-6}$ to $10^{-1}$ and plot the expectations in Figure \ref{fig:res_vs_time}, assuming $\tau_0\geq8~\mathrm{yr}$, $E_k = 5\times10^{51}~\mathrm{erg}$ and $M_\mathrm{ej} = 5~M_\odot$. We overplot our radial upper limits and indicate the resolvable limit of the global VLBI network (EVN+VLBA+Long Baseline Array+Korean VLBI Network), where we assume an achievable angular resolution of $\theta=250\times250~\mathrm{\upmu as}^2$ at $22~\mathrm{GHz}$ estimated using the EVN observation planner\footnote{\href{https://planobs.jive.eu/}{https://planobs.jive.eu/}}. In comparison to standalone EVN or VLBA observations, global antenna participation achieves more uniform sampling of the $uv$-plane, yielding an improvement in angular resolution by approximately a factor of $2$ at these elevations. To be conservative, we assume the 99.7\% confidence interval upper limit on the angular diameter distance to SN~2012au (e.g. $D_A \leq 32.6~\mathrm{Mpc}$), remaining consistent with our plotted 99.7\% confidence interval radius upper limits. \par
Overall, we find that for $\eta_\mathrm{R}\lesssim10^{-3}$ (i.e., consistent with all known Galactic PWNe), the putative PWN should now be resolvable, $\sim14~\mathrm{yrs}$ post explosion. If the radio source becomes resolved between 2026-2035, then larger efficiency factors would be favored (e.g. $\eta_\mathrm{R}\geq1\times10^{-3}$). As mentioned in Section \ref{sss:spindown}, this would provide a novel opportunity to test whether $\eta_\mathrm{R}$ extends to values beyond what are observed within our own Galaxy. If SN~2012au remains unresolved by $2035$, then a PWN powered by spin-down over the course of $\tau_0\gtrsim8~\mathrm{yr}$ becomes less plausible as it would require that $\eta_\mathrm{R}\gtrsim10^{-1}$, i.e. that greater than $10\%$ of all energy injected into the nebula be converted into radio synchrotron emission, a conclusion that currently lacks a strong theoretical basis. Instead, invoking a magnetar with a short ($P_0\sim10~\mathrm{ms}$) period which rapidly spins-down and no longer continuously contributes to the expansion of the nebula could potentially be a viable alternative, provided that the radio luminosity and evolution could also be explained. This model is worth considering in greater detail: however, as discussed in Section \ref{sss:spindown}, the required analysis will be reserved for future works due to the complexity of the modeling. Moreover, it is also worth noting that the millisecond magnetar model may be inconsistent with the observed the observed flux evolution as we discuss in Section \ref{sss:flux_evo}, below. \par 
Note that these broad conclusions are agnostic to the assumed values of $E_k$ and $M_\mathrm{ej}$: having assumed a lower limit on $E_k$, increasing $E_k$ acts to increase the expansion rate of the PWN in Equation \ref{eq:dotE}, leading it to become resolvable earlier. A similar statement holds for invoking lower values of $M_\mathrm{ej}$.  Consequently, whether an extragalactic PWN resides at the origin of SN 2012au and is resolvable should become clear by $2035$ \emph{at the very latest}. 

\subsubsection{Comparison to PWN Flux Evolution}\label{sss:flux_evo}
We briefly consider here whether the late-time emission is within expectations for a young PWN model: however, we defer comprehensive modeling of SN 2012au's radio lightcurve within the context of a PWN to Paper II.\par

In Section \ref{ss:fluxes}, we measured a steady decay in the flux of $S_\nu\,\propto\, t^{-1.2\pm0.4}$ at $4.93~\mathrm{GHz}$, with a peak flux having occurred before $t_p \leq 2920~\mathrm{days} ~(\sim 8~\mathrm{yr})$ post explosion. At early times, the PWN is originally obscured by both synchrotron self absorption (SSA) and FFA in the surrounding ejecta. However, once the ejecta expands and becomes optically thin to FFA, the PWN emerges as a steadily fading radio source, having reached its peak during the absorption phase \citep{Chevalier1977,1984ApJ...278..630R}. The time for the ejecta to become optically thin to FFA, assuming a spherical and uniform expansion of the ejecta which is homogeneous and expanding with velocity profile $v_\mathrm{ej}\approx R_\mathrm{ej}/t$, is given by \citep{1984ApJ...278..630R}
\begin{align}
\begin{split}\label{eq:tau_ff_time}
    t_\mathrm{ff} \cong~ &470~\left(\frac{T}{10^{4}~\mathrm{K}}\right)^{-3/10}\left(\frac{M_\mathrm{ej}}{2~M_\odot}\right)^{-1/5}\\
    &\times\left(\frac{v_\mathrm{ej}}{300~\mathrm{km~s^{-1}}}\right)^{-1}\left(\frac{\nu}{10^{9}~\mathrm{Hz}}\right)^{-2/5}~\mathrm{yr},
\end{split}
\end{align}
where $T$ is the temperature of the ejecta, $M_\mathrm{ej}$ is the supernova ejecta mass, $v_\mathrm{ej}$ is the velocity of the ejecta and $\nu$ is the observing frequency. For SN~2012au, we assume an ejecta temperature of $T = 10^{4}~\mathrm{K}$, $M_\mathrm{ej} = 5~M_\odot$ and $\nu= 5~\mathrm{GHz}$. With the results of \cite{Kamble_2014} suggesting a low-density, smoothly varying CSM close to the progenitor star (see Section \ref{ss:early_radio} for details) and therefore free-expansion of the ejecta, we approximate $v_\mathrm{ej}\sim \left(2E_k/M_\mathrm{ej}\right)^{1/2}\sim 8000~\mathrm{km~s^{-1}}$, again assuming $E_k=5\times10^{51}~\mathrm{erg}$. Combined, this implies that the ejecta should have become optically thin to FFA after $\sim7.7~\mathrm{years}$ post explosion, in rough agreement with our now-fading flux density measurements which began $\sim8$ years post explosion. We note, however, that uncertainties in model parameters could lead to variations in the timescales over which the PWN becomes optically thin. For example, if the ejecta is sufficiently cooler than $T\sim10^4$ K, the mass of the ejecta is $\ll 5~M_\odot$, or the velocity of the ejecta is $\ll8000~\mathrm{km~s}^{-1}$ (albeit less likely, given the atypically energetic nature of SN~2012au; \citealt{Milisavljevic_2013,Takaki_2013,Pandey_2021,2023ApJ...955...71R}), this timescale could extend to well beyond multiple decades. Additionally, Equation \ref{eq:tau_ff_time} does not account for the increase in ionization by the PWN itself, which could subsequently lead to an increase in free-free absorption \citep{2014MNRAS.437..703M}. For these reasons, the agreement in timescales with our observations should be considered with these caveats in mind. \par

Notably, for $t>\tau_0$, the analytic models in \cite{1984ApJ...278..630R} predict a significantly steeper evolution in the luminosity of $L_\nu \propto t^{-2}$ to $t^{-4}$. The relatively stable and shallow flux evolution associated with SN 2012au ($\propto t^{-1.2\pm0.4}$, Section \ref{ss:fluxes}) therefore possibly disfavors models invoking short spin-down timescales (i.e., millisecond magnetars) which predict a steeper decay in flux as a function of time at this stage of its evolution. This conclusion is in contention with optical analyses which favor energy injection by rapid spin-down of a central magnetar to explain the slow decline in SN~2012au's optical lightcurve \citep{Pandey_2021,DeSoto_2025}. However, models invoking magnetars may be reconciled through variations in the model parameters (e.g., \citealt{2023A&A...673A.107O}). \par
In addition to the temporal evolution, the luminosities of young PWNe are expected to exceed the present-day luminosity of the Crab nebula by factors of $10-1000\times$ \citep{1984ApJ...285..134B,BARTEL20051057}. With a present-day specific luminosity of $\sim 2\times10^{24}~\mathrm{erg~s^{-1}~Hz^{-1}} $  \citep{1968AJ.....73..535T,Perley_2017} and specific luminosity of SN~2012au of $L_\nu = (3\pm1)\times10^{27}~\mathrm{erg~s^{-1}~{Hz}^{-1}}$, both at $4.93~\mathrm{GHz}$, we observe that SN~2012au is $\sim 1500\times$ more luminous, broadly consistent with theoretical expectations.\par

\subsubsection{Comparison to Persistent Radio Sources Associated with Fast Radio Bursts and Their Link to Magnetars}\label{sss:prs-frb}
Some repeating FRBs have been found associated with compact, luminous persistent radio sources (PRSs; often interpreted as young magnetar wind nebulae; \citealt{Margalit18,Chatterjee17,Niu22, Marcote_2017,bhandari2023constraints,bruni2024nebularoriginpersistentradio, moroianu2025milliarcsecondlocalizationassociatesfrb}). Here, we briefly draw comparisons between between the late-time compact radio source identified at the location of SN 2012au and these PRS counterparts. We note, however, that our analysis thus far focuses on a PWN rather than a magnetar-powered wind nebula (the difference being that the former is spin-down powered, while the latter is magnetically powered), and thus this comparison should be treated solely as a qualitative comparison. \par
First, SN 2012au has a measured luminosity of $L_\nu \sim 3\times10^{27}~\mathrm{erg~s^{-1}~Hz^{-1}}$ at $\sim10~\mathrm{yr}$, lying at the lower luminosity range of known FRB-PRS systems which span a luminosity range of $10^{27}-10^{29}~\mathrm{erg~s^{-1}~Hz^{-1}}$ at $\mathrm{GHz}$ frequencies \citep{Marcote_2017,bhandari2023constraints,bruni2024nebularoriginpersistentradio,Bruni_2025, moroianu2025milliarcsecondlocalizationassociatesfrb,snelders2025revisitingfrb20121102amilliarcsecond}\footnote{It should be noted that the compactness on milliarcsecond angular scales of the candidate faint PRS associated with FRB\,20201124A has not yet been confirmed \citep{bruni2024nebularoriginpersistentradio}. Whether PRSs associated with FRBs extend to similar luminosity ranges as SN 2012au is not well understood and is an active area of research.}. Second, the temporal evolution of SN 2012au is not unprecedented among FRB-PRSs over $\lesssim10~\mathrm{yr}$ timescales, with some observed to remain stable over the course of multi-year observations \citep{bhardwaj2025constrainingoriginfrb20121102as}, while others show a steeper flux decay rate of $S_\nu\propto t^{-3.3}$ \citep{balaubramanian2025continuedradioobservationspersistent}. Finally, VLBI observations have constrained the compactness of some of these PRSs to $\lesssim1~\mathrm{pc}$ \citep{Marcote_2017}, consistent with the unresolved nature of SN 2012au on broadly comparable physical scales ($\leq0.045~\mathrm{pc}$ from our VLBI observations). \par
The combination of all three of these remarks suggest that the compact, lower luminosity radio source detected in our VLBI observations of SN 2012au is not unlike the PRSs associated with some repeating FRBs detected to-date. While our search for FRBs yielded non-detections (Section \ref{ss:frb_search}), performing a targeted search once the putative nebula has become optically thin at $\sim1.4~\mathrm{GHz}$ over the coming years (Paper II) may provide a direct avenue to test whether young compact objects formed via CCSNe are a viable formation channel for FRB progenitors. 

\subsection{Interaction with the Circumstellar Medium}\label{s:csm}
The second model for late-time radio re-brightening in CCSNe that we consider in the context of the compact radio source revealed by our VLBI observations is shock interaction with the CSM. In a typical CCSN event, as the rapidly expanding ejecta propagates into the CSM, a shock is formed, giving rise to magnetic fields and accelerated particles at the shock boundary. The shocked material subsequently results in the generation of radio synchrotron emission (e.g., \citealt{chevalier_tauff,2002ARA&A..40..387W,Chevalier_2006}). With the properties of the synchrotron emission being sensitive to both the shock radius and ambient density of the CSM \citep{chevalier1998}, it can provide a probe of the CSM density distribution, and therefore the mass-loss history of the progenitor star.\par
The radio emission of Type Ibc SNe typically peak on a timescale of weeks (e.g., \citealt{2002ARA&A..40..387W}). However, delayed radio emission could potentially occur if either (i) there is dense CSM close to the progenitor from which the radio emission was originally suppressed due to absorption processes, or (ii) the blastwave encounters an overdensity in the CSM at some larger radius (e.g., \citealt{2014c,Bietenholz_2017,Chandra_2020,Stroh_2021,Soria_2025}). In this section, we explore whether the combination of size upper limits and fluxes we measure in this work can be explained by CSM interaction, under the following assumptions about the CSM geometry: (i) an extension of the wind-like environment inferred by radio observations shortly after explosion \citep{Kamble_2014}, (ii) interaction with a dense torus of material surrounding the progenitor star, or (iii) interaction with a dense shell or clump of CSM that was physically separated from the progenitor at the time of explosion.

\subsubsection{Interaction with an extension wind-like environment inferred from early-time observations}\label{ss:early_radio}
The combination of early radio, X-ray and optical follow-up of SN~2012au (within a year of the explosion) constrained a smoothly varying CSM distribution with profile \citep{Kamble_2014}
 \begin{align}
    \rho_\mathrm{CSM} &= \frac{1}{4\pi r_0^2}\left(\frac{\dot{M}}{v_w}\right)\left(\frac{r}{r_0}\right)^{-s} \label{eq:rho_csm}\\
    &\approx 9.2\times10^{-22}\left(\frac{r}{1.4\times10^{16}~\mathrm{cm}}\right)^{-2\pm0.8} ~\mathrm{g~cm}^{-3}.\label{eq:rho_VLA}
\end{align}
A value of $s=2$ is consistent with expectations for a wind-like CSM and \cite{Kamble_2014} inferred a pre-explosion mass loss rate of $\dot{M} = 3.6\times10^{-6}~M_\odot ~\mathrm{yr}^{-1}$ for an assumed wind speed of $v_\mathrm{w}=1000~\mathrm{km~s^{-1}}$.  They further estimated the expansion rate of the shock boundary (assuming a spherical expansion of the blastwave), yielding 
\begin{align}
\begin{split}\label{eq:r_vs_t}
    r &= r_0 \left(\frac{t}{t_0}\right)^m\\
    &\approx1.4\times10^{16}\left(\frac{t}{25~\mathrm{days}}\right)^{0.9\pm 0.1}~\mathrm{cm}.
\end{split}
\end{align}

Could the observed radio emission in our VLBI epochs $8-13$ years post explosion simply be driven by continuous shock interaction with this wind-like profile? First we note that interaction with a smoothly declining density profile, as described by Equation \ref{eq:rho_csm}, cannot lead to a secondary radio re-brightening. Thus, because the 5~GHz radio fluxes measured in Section \ref{ss:fluxes} are a factor of $4$ higher than the final measurements by \cite{Kamble_2014} at 190 days post-explosion (Paper II), it indicates that our VLBI observations must have contributions from a source other than continued expansion into the wind-like CSM measured at early times. Second, our 99.7\% confidence interval VLBI size upper limits require that $R|_{t=2920~\mathrm{days}} \leq 1.4\times10^{17}~\mathrm{cm}$: however, the shock radius at this time (extrapolating from equation \ref{eq:r_vs_t}) must be greater than $r  \geq 1.5\times10^{17}~\mathrm{cm}$ ($99.994\%$ confidence interval lower limit). Given the inconsistency between these results, alongside the observed radio re-brightening, we conclude that the emission cannot be from the original SN shock with a continuation of the wind-like profile. Instead, it must arise from some other compact source of radio emission.

We note, however, that if the assumption of a spherical expansion of the blastwave is incorrect, then one could in principle reconcile the inconsistency between our VLBI radius constraint and the prediction of the blastwave. The extension of a wind-like medium, however, is still unable to explain the presence of a secondary radio re-brightening.

\begin{figure*}
    \centering
    \includegraphics[width=0.953\linewidth]{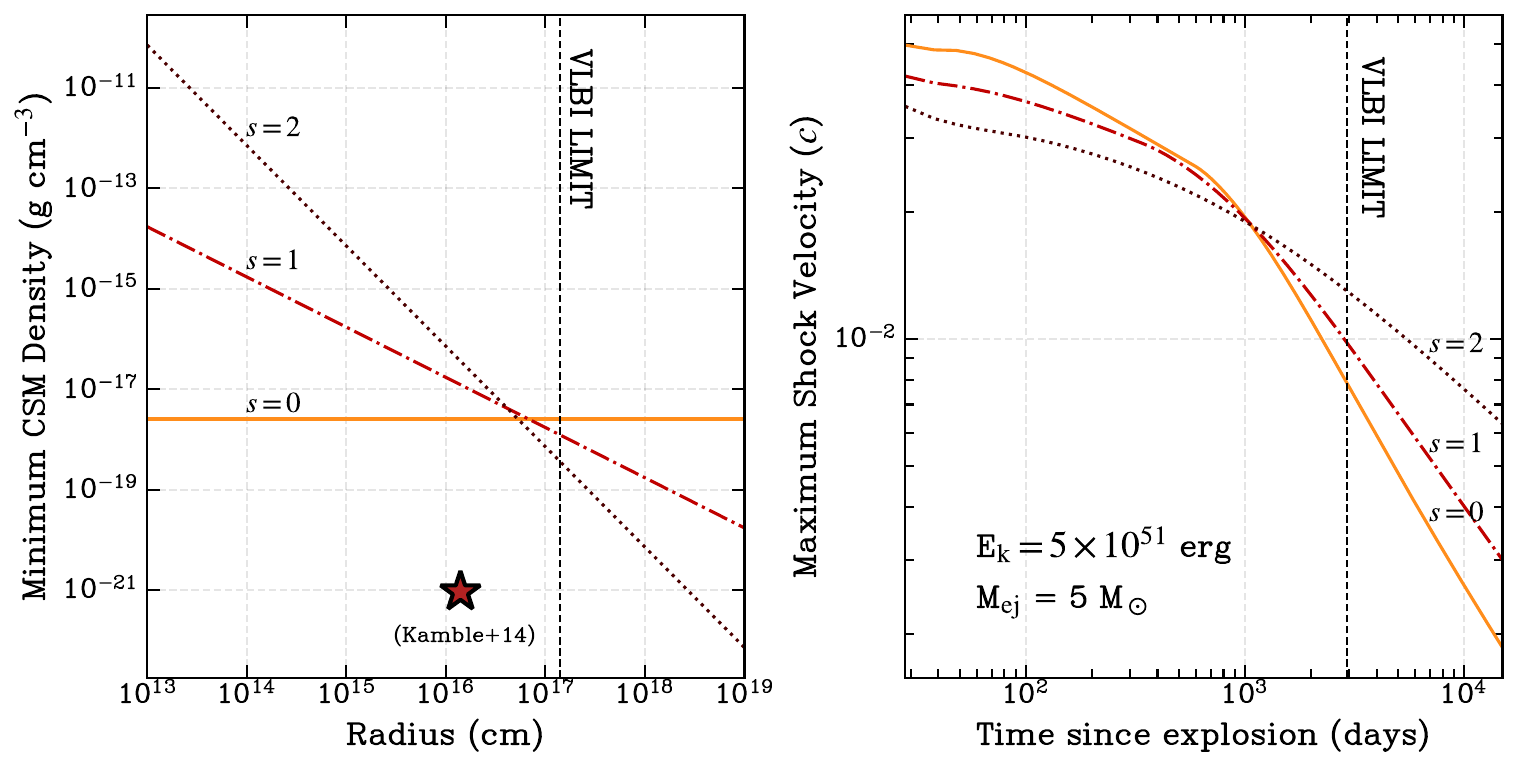}
    \caption{\textit{Left:} Inferred minimum CSM density profiles required to decelerate a shockwave launched on UTC 2012 February 29.5 and remain unresolved by the time of our VLBI observations 8 years post explosion. We assume that the kinetic energy and ejected mass released during the explosion is $E_k = 5\times10^{51}~\mathrm{erg}$ and $M_\mathrm{ej} = 5~M_\odot$ \citep{Pandey_2021}. We plot the scaling for varying CSM density profiles ($\rho_\mathrm{csm}\,\propto\, r^{-s},\,s=0,1,2$). With the assumed CSM profile, all profiles converge to a shock radius of $R_\mathrm{max} = 1.4\times10^{17}~\mathrm{cm}$ at $\Delta t_\mathrm{expl.} = 2925~\mathrm{days}$. The inferred density at $r = 1.4\times10^{14}~\mathrm{cm}$ from \cite{Kamble_2014} is indicated as a red star, highlighting the required difference in density between the polar axis and equatorial axis. \textit{Right}: Evolution of the forward shock velocity as a function of time for a shockwave interacting with the associated density profile. In both cases, we plot the corresponding radius and timestamp from Epoch 2 in Table \ref{tab:fitparams} as a dashed vertical line, respectively. }
    \label{fig:csm_vs}
\end{figure*}

\subsubsection{Interaction with a dense torus of circumstellar material surrounding the progenitor star}\label{ss:torus_csm}
We consider a CSM geometry whereby the progenitor star is surrounded by a dense, equatorial torus. Such a geometry has been proposed in the case of,  e.g., SN~1986J and SN 2014C and is suggested to be the result of a common envelope phase of the progenitor and binary companion, leading to a period of highly anisotropic mass-loss along the equatorial plane \citep{2012ApJ...752L...2C,Bietenholz_2017, Margutti2017,2022ApJ...939..105B}. In this model, we require that there is a component of the CSM surrounding the SN~2012au progenitor that is sufficiently dense to decelerate the shock such that it remains unresolved in our VLBI images. However, in order to justify both the early radio observations in \cite{Kamble_2014} in addition to the results presented in this work, it would be necessary for 
\begin{itemize}
    \item the CSM to be asymmetric in order to have experienced rapid expansion along the lower density polar axis and decelerated expansion along the higher density, equatorial axis (see figure 7 in \cite{2022ApJ...939..105B} for a cartoon depiction),  
    \item for the emission from the dense material to be heavily absorbed at early times (e.g., due to SSA and/or FFA) such that it was not originally detected in the early radio observations; and
    \item for the material to be sufficiently dense and abundant to achieve a detectable late-time radio re-brightening. 
\end{itemize}
With the goal of characterizing the properties of a torus-like CSM that would be required to reproduce the VLBI observations of SN 2012au (and whether or not they are reasonable), we perform a suite of numerical simulations to model the interaction of a shock interacting with varying CSM density profiles, along with the predicted flux evolution. Full details regarding the dynamics and flux evolution modeling are provided in Appendix \ref{s:torus} and our main findings are summarized below.\par   

\paragraph{Density Requirements to Decelerate the Shock} We model the dense torus as a continuous distribution defined by single power law: $\rho_\mathrm{csm}=\rho_0(r/r_0)^{-s}$. Here, $r$ is the radius from the interior of the progenitor, $s$ is the power-law index and $\rho_0$ and $r_0$ are scaling parameters. Throughout this analysis, we fix $r_0 = 1.4\times10^{16}~\mathrm{cm}$ to remain consistent with \cite{Kamble_2014}. A wind-like CSM profile is characterized by $s = 2$ whereas a uniform density profile assumes $s= 0$. Throughout this analysis, we consider three fiducial power-law indices of $s = 0,1$ and $2$. Using the $99.7\%$ confidence interval upper limits imposed by our VLBI observations (Table \ref{tab:fitparams}), we determine the minimum density scaling $\rho_0$ required to sufficiently decelerate the shockwave such that $R_s|_{t=2925~\mathrm{d}} = 1.4\times10^{17}~\mathrm{cm}$  (this is the maximum allowed radius at this time based on the VLBI analysis described above). This is done by numerically integrating the equations for motion for the forward shock following the prescription outlined in Appendix A of \cite{Ibik_2025} which is based on \cite{1994ApJ...420..268C}. \par
When carrying out this analysis, we assume $E_k = 5\times10^{51}~\mathrm{erg}$, $M_\mathrm{ej} = 5~M_\odot$ and an explosion date of MJD 559986.5 based on representative values measured for SN~2012au \citep{Milisavljevic_2013,Pandey_2021}. We further assume that the ejecta density profile is characterized by a broken power law with the inner power-law index of $n_0 = 0$ and outer power-law index of $n_1 = 10$ as is appropriate for a massive stripped-envelope SN progenitor (e.g. \citealt{Chevalier_2006}). However, as we show in Appendix \ref{ss:sensitivity}, our conclusions regarding the properties of the torus CSM are insensitive to the aforementioned assumptions. The inferred density profiles, along with the evolution of the shockwave velocity obtained from these simulations are plotted in Figure \ref{fig:csm_vs} and quoted in Table \ref{tab:torus_csm}. At a reference radius of $r_0=1.4\times10^{16}~\mathrm{cm}$, these results imply a torus that is,\emph{ at a minimum}, $10^3-10^4~\times$ more dense than the densities along the polar axis inferred by \cite{Kamble_2014} (see comparison in Figure \ref{fig:csm_vs}). We emphasize that the density profiles shown in Figure \ref{fig:csm_vs} are those required to decelerate the shock wave enough so that its radial extent is consistent with our VLBI observations. Additionally, given that our observations provide only conservative \emph{upper limits} on the radius, if the true shock location is even smaller this would subsequently lead to an increase in the density scaling parameter $\rho_0$.

\paragraph{Expectation for Flux Suppression at Early Times} 
Next, we consider whether it is reasonable that radio emission from interaction with the aforementioned density profiles could have remained hidden/absorbed within the first $\sim$ year post-explosion. We model the radio flux assuming a standard SSA model for radio supernovae \citep{chevalier1998} with full details regarding the flux modeling provided in Appendix \ref{sss:flux_suppression}.  Where relevant, we assume a luminosity distance of $D_L = 23.5~\mathrm{Mpc}$ \citep{Milisavljevic_2013} and wind velocity of $v_\mathrm{w} = 1000~\mathrm{km~s^{-1}}$. We further assume that equipartition between the relativistic electrons and magnetic field energy density holds, i.e. $\alpha = \epsilon_\mathrm{e}/\epsilon_\mathrm{B} = 1$, adopting the commonly assumed value of $\epsilon_B=\epsilon_\mathrm{e} = 0.1$ \citep{Chandra_2020}. However, we show in Appendix \ref{ss:sensitivity} that our final conclusions are highly sensitive to variations in $\alpha$ and $\epsilon_\mathrm{B}$. We discuss the implications on our interpretation in greater detail in Appendix \ref{ss:sensitivity}. \par
Given the inferred densities, FFA is expected to contribute significantly towards suppressing the radio flux \citep{chevalier_tauff}. To model its contribution, we estimate the line-of-sight optical depth by integrating \citep{1981ApJ...251..259C,rybicky_lightman}
\begin{equation}\label{eq:tauff}
\begin{split}
\tau_{\nu,\mathrm{ff}}
=
\int_{R_\mathrm{sh}}^{+\infty}
&9.8\times10^{-3}\,
n_{e}^{2}\,\nu^{-2}\,T^{-3/2}\,
\\\times&\bigl(7.69 + 1.5\log_{10}T - \log_{10}\nu\bigr)
\;\mathrm{d}r,
\end{split}
\end{equation}
where $n_e(r)$ is the electron density profile assumed to emerge from a He-only shell, $T$ is the temperature of the torus and the equation assumes c.g.s. units\footnote{We note that the integral only converges for $s > 0.5$. For $s < 0.5$, the integral diverges at $+\infty$ and thus a fixed upper bound must be defined. For $s = 0$, we assume an upper bound of $10^{17}~\mathrm{cm}$. This assumption is based on resolved images of the central component of SN 1986J, interpreted as a dense torus, which extends beyond $7\times10^{16}~\mathrm{cm}$ \citep{Bietenholz_2017}.}. We assume a typical CSM temperature of $T \sim 10^5~\mathrm{K}$ \citep{1996ApJ...461..993F}. The optical depth is then evaluated at each timestep in our numerical integration, accounting for the evolution in CSM densities as the shockwave expands into the surrounding environment. \par

For all fiducial values of $s$ that we consider, we find that SSA alone is insufficient to suppress the radio flux from the on-going shock interaction within the dense torus when compared to early VLA observations up to $\sim120~\mathrm{days}$ post explosion. For example, at $13.3~\mathrm{GHz}$, SSA models predict radio flux in excess of $\gtrsim5~\mathrm{mJy}$  nearly three months post explosion, whereas VLA observations at the same time detect only $\sim1~\mathrm{mJy}$ \citep{Kamble_2014}. However, after invoking FFA, the flux levels are easily suppressed to levels below $\lesssim1~\mathrm{mJy}$. We conclude that it is therefore not unreasonable to expect that the radio emission from interaction with such a dense torus could be absorbed on a timescale of $\sim100~\mathrm{days}$, thus far remaining consistent with early VLA observations of SN~2012au.

\paragraph{Flux Modelling to Place Constraints on the Inferred Radio Filling Factor} 
To further investigate if the parameters required by this model are reasonable, we constrain what fraction of the total volume contained within the projected sphere (with radius $R_\mathrm{max} = 1.4\times10^{17}~\mathrm{cm}$, from our VLBI constraints) is actively contributing to the late-time radio emission. This fraction is usually characterized by the radio filling factor, $f$, where $V \equiv \frac{4\pi}{3}fR_\mathrm{max}^3$ and $V$ is the volume. \par
\begin{figure*}
    \centering
    \includegraphics[width=1.0\linewidth]{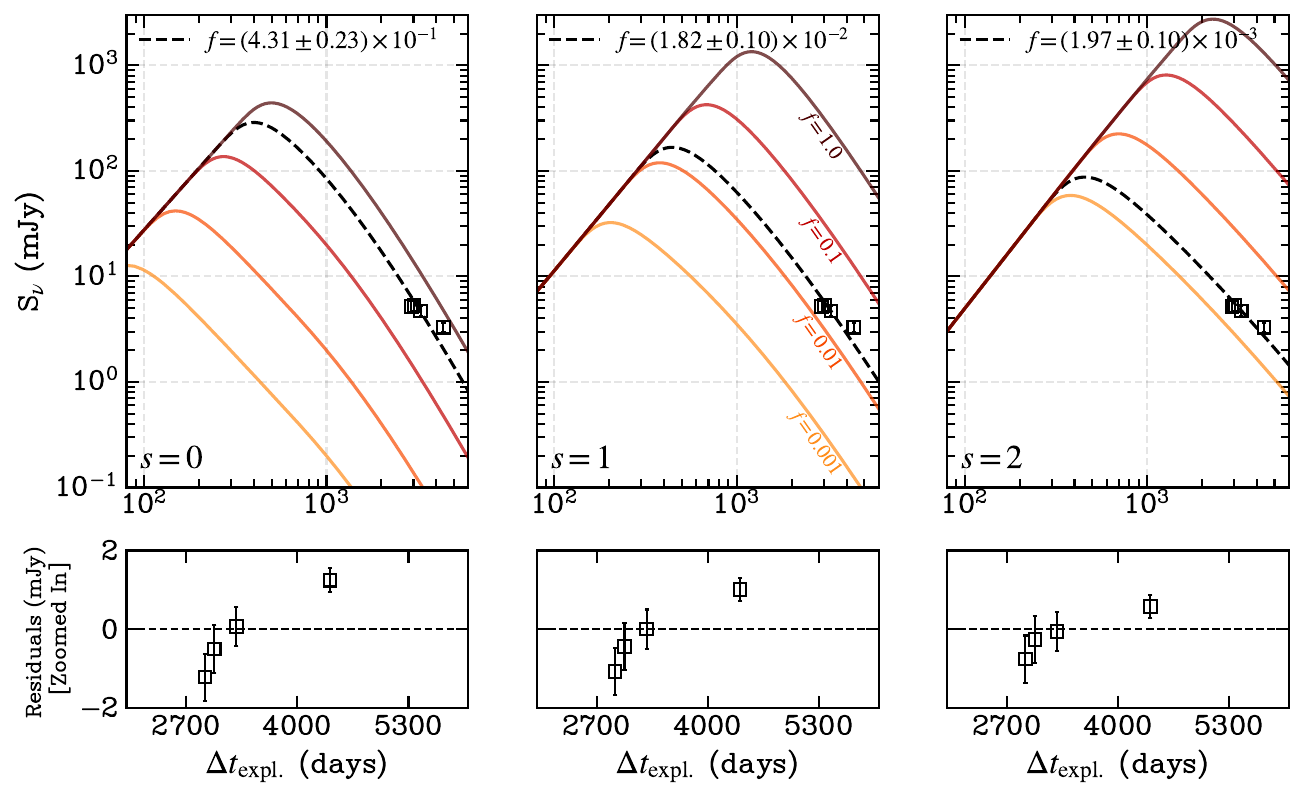}
    \caption{Late-time modeled (SSA-only) flux evolution at $\nu = 4.93~\mathrm{GHz}$ of a shockwave interacting with a torus-like CSM with density profiles characterized by a single power law, $\rho_\mathrm{csm}\,\propto\, r^{-s}$, as shown in Figure \ref{fig:csm_vs}. From left to right, each panel corresponds to the flux evolution for $s = 0,1$ and $2$, highlighting how varying $s$ alters the predicted flux evolution. We plot the late-time VLBI flux measurements at $4.93~\mathrm{GHz}$ from Table \ref{tab:flux-snc} as black squares. The best-fit model, corresponding to the required filling factor to scale the late-time fluxes, is indicated by a dashed black line. The maroon, red, orange and yellow lines on each panel represent filling factors $f = 1.0,0.1,0.01,0.001$, respectively, as labeled in the middle panel. Residuals of each fit are provided in the lower panels, zoomed in around the time of our observations. The figure assumes  $E_k=5\times10^{51}~\mathrm{erg}$, $M_\mathrm{ej} = 5~M_\odot$, $D_L=23.5~\mathrm{Mpc}$, $v_\mathrm{w}=1000~\mathrm{km~s^{-1}}$, $\alpha = 1$, $\epsilon_B=0.1$ and $p=3$.}
    \label{fig:flux_evolution}
\end{figure*}

Applying the same flux modeling framework as in Appendix \ref{sss:flux_suppression}, we determine the best-fit filling factor required to suppress the radio flux to levels measured by our observations in Table \ref{tab:flux-snc} at $4.93~\mathrm{GHz}$. We utilize the fact that we measure a declining flux of $S_\nu \propto t^{-1.2\pm0.4}$ in Section \ref{ss:fluxes} to bypass the need to invoke FFA, noting that while FFA affects the timing at which peak flux is achieved, it ceases to affect the radio flux evolution once the source has become optically thin to FFA and begins fading. In Figure \ref{fig:flux_evolution}, we plot the corresponding SSA flux evolution at $4.93~\mathrm{GHz}$. We observe that in all three cases, the late-time VLBI radio flux evolution can be reasonably explained by shock interaction with a torus-like CSM, provided that a filling factor of $f \sim 0.4,0.02$ or $0.002$ is invoked for $s = 0,1$ and $2$, respectively. \par

\paragraph{Constraining the geometry of the torus} Finally, to translate these results to a constraint on the geometry of the required torus (with the motivation of assessing if it is physically reasonable or not), in Appendix \ref{sss:fillingfactor}, we derive that the filling factor relates to the opening angle $\theta$ of the torus via 
\begin{equation}\label{eq:opening_angle}
    \theta =  \mathrm{arcsin}\left(\frac{f}{1-\gamma^3}\right),
\end{equation}
where $\gamma$ characterizes the ratio of the inner and outer radii actively contributing to the late-time radio emission, which we take to be $0.7$ motivated by available imaging of resolved SNe \citep{2003ApJ...597..374B}. For the quoted filling factors in Table \ref{tab:torus_csm}, we infer opening angles of $\theta \sim40^\circ, 2^\circ$ and $0.2^\circ$ for $s = 0,1$ and $2$, respectively. We summarize the properties of a dense CSM torus that would be required to reproduce the compact radio emission from SN~2012au observed in our VLBI observations in Table \ref{tab:torus_csm}. In Appendix \ref{ss:sensitivity}, we explore how our inferences about the torus change depending on the number of assumptions made throughout this section. We emphasize that while the values in Table \ref{tab:torus_csm} serve as an initial starting point, the large uncertainty on model parameters can lead to significant variation in the torus and thus should be interpreted with caution. \par

\paragraph{Takeaways and caveats of the torus model} We have shown above that CSM material that is dense enough to lead to \emph{compact} radio emission at late times can also (i) reasonably justify the suppression of flux at the time of VLA observations and (ii) predict the gradual late-time decline in flux detected in our observations. Combined, this suggests that the torus model could, in principle, provide an alternative explanation for the late-time radio re-brightening detected in our VLBI observations---provided that the filling factor is relatively small. However, we discuss a few caveats to these conclusions below which could lead to this model being ruled out by future observations. \par

First, given the unresolved nature of the source, it is likely that the radius of the shock is less than the radius quoted from our VLBI upper limit of $R|_{t=2920~\mathrm{d}} \leq 1.4\times10^{17}~\mathrm{cm}$. If the radius of the emitting region is \emph{significantly} smaller than this limit, then our inferences would change dramatically. For example, reducing the radius by a factor of 10 would imply a decrease in the inferred opening angle to $\theta_0 \sim 0.0006^\circ$ for $s = 1$, implying a possibly unphysically-thin torus. As we discuss in Appendix \ref{ss:sensitivity}, we find that the only conceivable way to stop the opening angle from shrinking too dramatically in this case would be to deviate significantly from our assumption of equipartition. \par

Second, given the lack of frequency coverage in our VLBI observations, we could only verify that the observed fading is consistent with expectations at $5~\mathrm{GHz}$. Because of this, our assumptions about the electron energy index ($p$) and/or the contributions of FFA at frequencies other than $5~\mathrm{GHz}$ may be incorrect, potentially resulting in incorrect estimates of the filling factor. One way to validate or reject this model would be to incorporate information from broad-band radio flux and X-ray measurements which provide direct constraints on $p$ and the effects of FFA, and more generally on the properties of the circumstellar material. As such, this model will be revisited in Paper II taking into account these additional considerations. \par

Finally, the lack of detected hydrogen features in the optical spectra of SN 2012au places an additional constraint that this dense torus must be H-poor. While dense H-rich and He/C-rich shells extended to larger radii have been identified in a few instances thus far (see, e.g., SN 2001em, SN 2014C and SN2006jc; \citealt{Chandra_2020,2014c,Anderson_2016,2007ApJ...657L.105F}), this parameter space is only beginning to be explored and a larger sample of H-poor SNe with delayed radio re-brightening is required to make more definitive statements regarding the impacts on stellar evolutionary models. \par
Combined, we conclude that while the torus model may be able to justify our VLBI observations presented in this paper, the model may face challenges in the future if the radius of the emitting region is found to be significantly smaller than our upper limits.

\begin{deluxetable}{cccccc}
\tabletypesize{\small}
\tablecaption{Inferred properties of a torus CSM. \label{tab:torus_csm}}
\tablehead{\colhead{$s^a$} & \colhead{$\rho_0$$^a$} & \colhead{$r_0$$^a$} & \colhead{$f^b$} & \colhead{$\theta^c$} & \colhead{$M_\mathrm{csm}$$^d$} \\
\colhead{} & \colhead{($\mathrm{g~cm^{-3}}$)} & \colhead{(cm)} & \colhead{} & \colhead{(deg)} &\colhead{($M_\odot$)} }
\startdata  
    $0$ & $2.5\times10^{-18}$ & $1.4\times10^{16}$ & $0.4$ & $40$ & $0.3$ \\
    $1$ & $1.2\times10^{-17}$ & $1.4\times10^{16}$ & $0.02$ & $2$ & $0.06$ \\
    $2$ & $3.5\times10^{-17}$ & $1.4\times10^{16}$ & $0.002$ & $0.2$ & $0.02$ \\
\enddata
    \hspace{2mm}$^a$ Density scaling parameters:  $\rho_\mathrm{csm} = \rho_0(r/r_0)^{-s}$\\
    $^b$ Inferred radio filling factor. \\
    $^c$ Opening angle, assuming a torus-like geometry of the CSM. \\
    $^c$ Swept-up CSM mass from $r = 0\longrightarrow R_\mathrm{max}$. \\
\end{deluxetable}

\subsubsection{Interaction with a detached CSM shell located at larger radii}
The final CSM geometry we briefly consider is a shell-like geometry, whereby an overdensity in the CSM exists at larger distances from SN~2012au's progenitor. In this scenario, the shell is expected to be ejected prior to the explosion. Once the ejected mass catches up to the dense CSM overdensity, radio re-brightening is expected as the shockfront interacts with the dense shell (or clump) and subsequently rapidly decelerates. Such a geometry has been proposed for SN~2001em and SN~2014C, both of which were originally classified as Type Ib SNe but later showed narrow hydrogen emission lines and subsequently re-classified as Type IIn \citep{2014c,Chandra_2020}.\par

Given that our VLBI observations constrain the radius of the emitting region to be $\leq1.4\times10^{17}~\mathrm{cm}$, if such a shell were symmetric, it would require that it was located at a radius smaller than this level. In addition, the fact that the shell is not observed in the $\sim4$ months of observations from \cite{Kamble_2014} would require that it was located at a radius $\gtrsim 6\times10^{16}~\mathrm{cm}$. These constraints would imply a shell that is significantly closer to the progenitor than the shell associated with SN 1987A which extends beyond $R\gtrsim4\times10^{17}~\mathrm{cm}$  \citep{1997ApJ...479..845G} and would hence have been resolved by our observations. By contrast, the required shell location remains broadly consistent with inferences of a massive ($\sim1~M_\odot$) H-rich shell component at $R \sim6\times10^{16}~\mathrm{cm}$ associated with SN 2014C \citep{Margutti2017}. Assuming an ejection velocity of the shell of $100~\mathrm{km~s^{-1}}$, our VLBI constraints imply that the shell would have to have been ejected between $\sim200~\mathrm{to}~400~\mathrm{yr}$ prior to core-collapse. The lack of Hydrogen in the optical spectra of SN 2012au would further require that this shell be H-poor. \par
While at present this model remains a viable justification on the basis of our VLBI radial constraints alone, our conclusions are weakened by the fact that we have not considered the expected flux evolution. Further investigation into its feasibility requires detailed broadband flux modeling, which our VLBI observations are ill-suited for, and thus will be presented in Paper~II. Such a shell, if present, would likely need to be very dense to sufficiently decelerate the shock (as in Section~\ref{ss:torus_csm}, above) but also clumpy and/or asymmetric such that the material in the shell does not fully absorb the early-time radio emission observed by \cite{Kamble_2014}.

\subsection{An off-axis, relativistic jet}\label{s:jet}
The final model for late-time radio re-brightening in CCSNe that we examine in the context of the compact and $\sim$stationary radio source revealed by our VLBI images is the successful launch of off-axis, relativistic jets during the collapse of the massive star \citep{1993ApJ...405..273W,1999ApJ...524..262M,2003MNRAS.345..575M}. A subclass of stripped-envelope SNe (SESNe) with high velocity features (Type Ic-BL; for `broad-lined') have been observed to be associated with long-duration gamma-ray bursts (GRBs; see \cite{grb_review} for a review). While SN 2012au was not a Type Ic-BL, it was an atypically energetic SN, and we therefore consider the possibility that it also launched a relativistic jet. In general, these jets are highly collimated, and thus the vast majority of long GRBs should be beamed away from us on Earth. However, as they propagate and decelerate, the radio emission associated with the jet is expected to de-beam enough to become observable irrespective of the viewing angle; thereby leading to a late-time radio rebrightening whose timescale is set by the viewing angle and decelaration timescale (e.g., \citealt{Bietenholz_2014,2014PASA...31...22G,Granot_2018,Leung_2023}).  While a number of searches have been performed, direct confirmation of an off-axis, relativistic jet detected years post-explosion has yet to be established for a SESN (e.g., \citealt{Soderberg_2006,Bietenholz_2014,Stroh_2021,schroeder2025latetimeradiosearchhighly}). \par

Our VLBI images of the late time radio emission for SN~2012au can test the viability of the off-axis jet model for this system by placing constraints on the proper motion and geometry of the emitting region. Similar analysis has been performed for a few on-axis GRBs and tidal disruption events to date (e.g., \citealt{Taylor_2004,Mooley_2018,2019Sci...363..968G,2024A&A...690A..74G,golay2025radioemissioninfraredtidal}). For example, VLBI images of the on-axis GRB 030329 revealed superluminal ($3$-$5c$) expansion over the course of the first 25 to 83 days after the launch of the jet \citep{Taylor_2004}. Here, we consider the implications of our VLBI images of SN~2012au for both (i) a completely off-axis jet and (ii) a partially off-axis jet. \par

 \paragraph{Constraints on a Completely off Axis Jet} In the case of a completely off-axis jet ($\theta_\mathrm{obs}=90^\circ$), the jet head is no longer viewed ``head-on" and thus the apparent expansion and true expansion are equal to one another (e.g. $\beta_\perp = \beta\sin\theta_\mathrm{obs}/[1-\beta\cos\theta_\mathrm{obs}], \theta=\pi/2)$, where $\beta_\perp$ and $\beta$ are the apparent and true expansion rate, respectively; \citealt{Taylor_2004}). As the pair of jets expand, one would eventually expect to resolve two distinct components propagating away relative to one another \citep{Granot_2018}. For SN 2012au, if we assume that the jet is $90^\circ$ off-axis and both jet heads are currently propagating radially outwards, our size limit of $R|_{t=2925~\mathrm{days}}\leq1.4
\times10^{17}~\mathrm{cm}$ requires that $\beta_\perp = \beta\leq r/t \approx 5545~\mathrm{km/s} = 0.02c$ ($99.7\%$ confidence interval; Table \ref{tab:fitparams}). This speed is significantly below those expected for relativistic jets, therefore disfavoring such a scenario. However, a pair of sub-relativistic jets could in principle remain consistent with our VLBI observations.

\paragraph{Constraints on a Partially off Axis Jet}  What if the jet is not completely off axis? In this case, we can compare our VLBI size limits and proper motion constraints to the results of hydrodynamical simulations presented in \cite{Granot_2018} who provide an estimate for the expected physical size of the jet as a function of time for different viewing angles. We emphasize that due to the complexity of these simulations, we do not perform them ourselves and instead make direct comparisons between our VLBI results and the results presented in \cite{Granot_2018}, referencing them throughout this section. Consequently, all of our inferences and statements made here inherit the assumptions made in the aforementioned analysis, which we briefly summarize below.  \par

Our inferences assume an initial GRB jet with a half-opening angle of $\theta_0=0.2~\mathrm{rad}$ and initial Lorentz factor varying between $20\leq\Gamma\leq500$. The jet is assumed to have isotropic equivalent energy of  $E_\mathrm{k,iso} = 10^{53}~\mathrm{erg}$, resulting in a true jet energy of $E_\mathrm{jet} = (1-\mathrm{cos}~\theta_0)E_\mathrm{k,iso}\approx2\times10^{51}~\mathrm{erg}$. We assume that the jet propagates into a constant density external medium with density $\rho_0 = n_\mathrm{ext}m_p = 1.67\times10^{-24}~\mathrm{g~cm^{-3}}$. We further assume that equipartition between the relativistic electrons and magnetic field energy density holds and fix $\epsilon_B=\epsilon_e=0.1$. For specific details regarding the simulation setup, we defer the reader to \cite{Granot_2018} and references therein. \par

Under these assumptions, we can utilize the fact that there was a lack of a radio re-brightening in the first $\sim$190 days post-explosion (\citealt{Kamble_2014}, Paper II) to place a lower limit on the viewing angle of the jet (recall that jets at larger viewing angles should re-brighten later). Based on figure 1 in \citealt{Granot_2018}, we place a conservative lower limit on the viewing angle of $\theta_\mathrm{obs} \gtrsim0.8~\mathrm{rad}$ ($\sim45^\circ$). For angles less than $45^\circ$, the hydrodynamical simulations predict that the jet should have already sufficiently de-beamed and thus a re-brightening should have been detectable by \cite{Kamble_2014}. \par

With this lower limit on the viewing angle for any putative jet, we next compare our VLBI size upper limits to figure 6 in \cite{Granot_2018} which provides estimates for the physical size of the jet as a function of time for different viewing angles. This expected size is characterized by the semi-axes $\sigma_x$ and $\sigma_y$, corresponding to the physical extent of the expanding jet head which is assumed to be shaped as an elliptical Gaussian. We find that at the time of our observations, a jet expanding under the aforementioned assumptions would have expected to expand to a projected size greater than $\sigma_x\approx\sigma_y\gtrsim6\times10^{17}~\mathrm{cm}$. This value extends well beyond our $99.7\%$ confidence interval upper limit of $R\leq1.4\times10^{17}~\mathrm{cm}$. We conclude that under the assumptions described at the beginning of this section, our observations are inconsistent with expectations for an orphan GRB afterglow. \par

In addition to our observational constraints on the size of the emitting region, the expected displacement of the partially off-axis jet head between $3000$ to $4000$ days post explosion is expected to be $ d\sim10^{18}~\mathrm{cm}$ based on figure 6 in \cite{Granot_2018}. This corresponds to an expected proper motion at the time of our VLBI observations of $v_\mathrm{jet}\sim0.39c$. This value is also inconsistent with  our 99.7\% confidence interval upper limit on the proper motion obtained with our VLBI observations of $v_\mathrm{pm}\leq 0.36c$, implying we would have expected greater displacement of the jet head throughout our VLBI imaging campaign. We do note, however, that the models in \cite{Granot_2018} predict that our first set of observations approximately coincide with when the jet entered a period of deceleration. It is possible, then, that the predicted displacement may be overestimated and thus the expected proper motion be less than $v_\mathrm{jet}\sim0.39c$. Despite this caveat, when the size and proper motion constraints are considered together, we deem it unlikely that an orphan, off-axis GRB afterglow represents a realistic justification of the late-time radio emission associated with SN~2012au. We thus currently disfavor models invoking radio re-brightening due to an off-axis, relativistic jet launched during core-collapse.

\section{Summary \& Conclusion} \label{s:conclusion}
In this work, we present the results of a campaign of monitoring the peculiar Type Ib SN~2012au on milliarcsecond angular scales with radio VLBI observations from $8-13~\mathrm{years}$ post explosion. These observations overlap with a time when a radio re-brightening was observed at the location of SN~2012au. We observe that the source remains compact over the course of our campaign. Our most stringent results constrain the projected radius of the emitting region to be $\leq1.4\times10^{17}~\mathrm{cm}$ ($\leq0.045~\mathrm{pc}$; both quoted at the 99.7\% confidence interval) at 2925 days post explosion. We also find that the source is now steadily fading at 5~GHz with a decline rate of $S_\nu \,\propto\, t^{-1.2\pm0.4}$. Using our multi-epoch observations, we constrain the proper motion to be $v_\mathrm{pm}=0.12\pm0.08c$, indicating that it is consistent with being stationary at the $1.5\sigma$ level. Overall, this places a 99.7\% confidence interval upper limit on the proper motion of the source of $v\leq0.36c$. Finally, based on previous work that has invoked a magnetar to explain the early time emission from SN~2012au, we also perform a transient search for FRBs. This search yielded non-detections, placing a fluence upper limit of $7~\mathrm{Jy~ms}$. We examine the implications of the VLBI results (in particular the late-time compactness of the radio source) on three main models that have been put forth to explain late-time re-brightening in core-collapse SNe: the emergence of a pulsar wind nebula, interaction with circumstellar material, and the afterglow of an off-axis relativistic jet. We summarize our main conclusions below.

\paragraph{A young pulsar wind nebula} A young PWN is a particularly intriguing model for the late-time radio emission from SN~2012au because the presence of a PWN was already suggested based on the emergence of sulfur and oxygen emission line features in optical spectra $6.2$ years post-explosion \citep{Milisavljevic_2018}. In Section \ref{s:pwn} we find that our VLBI observations are broadly consistent with the predicted theoretical evolution expected in the case of a young PWN.  However, the details of our observations (e.g. size upper limits, broad flux evolution) place requirements on several properties of the system for this model to remain consistent. In particular, in the case that the initial spin-down timescale of the central pulsar is greater than the age of the system, i.e. $\tau_0\gtrsim8~\mathrm{yr}$, we would require that: 

\begin{itemize}
    \item the initial spin-down luminosity fall between $10^{36}~\mathrm{erg~s^{-1}}\leq\dot{E}_0\leq {4\times10^{42}}~\mathrm{erg~s^{-1}}$ ($99.7\%$ confidence interval); consistent with the independently inferred value of $\dot{E}_0 \approx10^{40}~\mathrm{erg~s}^{-1}$ from late-time optical spectroscopy of SN~2012au \citep{Milisavljevic_2018},
    \item the radio efficiency factor be greater than $\eta_\mathrm{R}\geq{3\times10^{-7}}$ (99.7\% confidence interval lower limit); broadly consistent with known Galactic PWNe with $10^{-6}\leq\eta_\mathrm{R}\leq10^{-3}$ \citep{1997ApJ...480..364F,Gaensler_2006}, 
    \item and that the initial formation properties of the central NS lie in a region broadly consistent with the youngest, known sample of Galactic PWNe (see, e.g., the P-Pdot diagram in Figure~\ref{fig:ppdot1}).
\end{itemize}

Additionally, we predict that the putative nebula should expand to become resolvable at the highest angular resolution achievable by the global VLBI network as early as 2026 and by 2035 at the latest. Lastly, the gradual observed flux evolution possibly disfavors models in which $\tau_0\ll8~\mathrm{yr}$ (i.e. models invoking millisecond magnetars): however, further analysis is required to explore this parameter space in greater detail and will be presented in future works.  

\paragraph{Circumstellar medium interaction} 
In Section \ref{s:csm} we also investigate whether our VLBI observations could be consistent with expectations for radio emission due to the interaction of the SN shock with circumstellar material. We consider multiple possible geometries for the CSM. We find that our size constraints rule out that the emission present in the VLBI observations is due to continued interaction of the forward shock with the wind-like medium that was inferred from radio observations in the first $\sim4~\mathrm{months}$ post-explosion \citep{Kamble_2014}; as this should have been resolvable by our observations if it remained sufficiently bright. However, we find that either a (i) dense torus of CSM close to the progenitor star or (ii) a detached and `clumpy' shell located between $(0.6-1.4)\times10^{17}~\mathrm{cm}$ could in principle decelerate a portion of the shock sufficiently such that the emission remains unresolved in our late-time observations. However, both models face significant constraints on their densities and geometries such that they neither are visible in the first $\sim$ year post-explosion nor fully absorb radio emission from the interaction with a wind-like medium close to the progenitor observed by \cite{Kamble_2014}. In addition, all models which invoke shock interaction with the CSM \emph{require} that the CSM be hydrogen-poor to remain consistent with the optical evolution of SN 2012au \citep{Milisavljevic_2018}. Finally, these models may face further challenges in the future when incorporating information from broad-band radio and X-ray flux measurements, both of which will be presented in Paper II.

\paragraph{An off-axis, relativistic jet} In Section \ref{s:jet}, under the standard assumption of a jet launched during core-collapse with isotropic equivalent kinetic energy of $E_\mathrm{k,iso}=10^{53}~\mathrm{erg}$, we find that it would have expanded to a size well beyond our $99.7\%$ confidence interval size upper limit, thus inconsistent with our VLBI observations. When further considering the proper motion of the source in comparison to theoretical expectations, we  disfavor models invoking an off-axis, orphan GRB afterglow as a justification for the late-time radio re-brightening. 
\\

\indent Overall, we conclude that our current VLBI observations are consistent with the hypothesis that a young, extragalactic PWN has emerged at the site of SN~2012au, adding to the multiwavelength evidence in favor of the scenario originally suggested by \cite{Milisavljevic_2018}. Over the next 5 to 10 years, assuming SN 2012au maintains its stable flux evolution, planned monitoring of the radio source will allow us to further constrain its evolution. If the source becomes resolved, then the required formation properties of the central NS in the PWN model will be refined, providing insight into the early evolution of a young NS. If the source remains unresolved, then models invoking a magnetar-powered nebula possibly become favored.\par

Future observations will also play a critical role in further constraining the complex CSM geometries explored throughout this analysis. If future observations resolve a clumpy ring-like or torus structure, interaction with a dense CSM become the favored model to justify the late-time radio evolution. Such a confirmation would itself be notable, marking the first resolved instance of shock interaction with an H-poor CSM. \par

Finally, our analysis presented in this work provides a general framework for interpreting the radio emission associated with a CCSN on milliarcsecond angular scales decades post explosion. As the sample of CCSNe with confirmed late-time radio re-brightening grows (e.g., \citealt{Stroh_2021,rose2024latetimesupernovaeradiorebrightening}), our analysis will serve as a general guide on how VLBI observations can be used to disentangle between the competing models discussed in this work. In doing so, we can aim to improve our understanding of the mass-loss history of massive progenitors, the energy budgets of young NSs and jet-mechanisms by a central engine. An on-going effort to extend this work to a broader sample of CCSNe is underway under the VLBA large program VLBA/24B-252 (PI: Margutti \& Drout) and will be presented in future works.

\software{Astropy \citep{astropy:2013,astropy:2018,astropy:2022},
Matplotlib \citep{Hunter:2007}, 
NumPy \citep{numpy2020},
SciPy \citep{2020SciPy-NMeth}.}

\section*{Acknowledgments}
We thank the anonymous referee for their insightful comments which led to improvement of the manuscript. M.L. thanks Amanda M. Cook for insightful discussions on the statistical handling of the data. M.L. acknowledges the support of the Natural Sciences and Engineering Research Council of Canada (NSERC-CGSD). K.N. acknowledges support by NASA through the NASA Hubble
Fellowship grant \# HST-HF2-51582.001-A awarded by the Space Telescope
Science Institute, which is operated by the Association of Universities for
Research in Astronomy, Incorporated, under NASA contract NAS5-
26555. M.R.D. acknowledges support from the NSERC through grant RGPIN-2019-06186, the Canada Research Chairs Program, and the Dunlap Institute at the University of Toronto. J.W.T.H., B.M., O.O.-B., and the AstroFlash research group acknowledge support from a Canada Excellence Research Chair in Transient Astrophysics (CERC-2022-00009); an Advanced Grant from the European Research Council (ERC) under the European Union’s Horizon 2020 research and innovation programme (`EuroFlash'; Grant agreement No. 101098079); and an NWO-Vici grant (`AstroFlash'; VI.C.192.045). J.K.L. acknowledges support from the University of Toronto and Hebrew University of Jerusalem through the University of Toronto - Hebrew University of Jerusalem Research and Training Alliance program. 
The Dunlap Institute is funded through an endowment established by the David Dunlap family and the University of Toronto. D.L.C. acknowledges support from the Science and Technology Facilities Council (STFC) grant number ST/X001121/1. B.M. acknowledges financial support from the State Agency for Research of the Spanish Ministry of Science and Innovation, and FEDER, UE, under grant PID2022-136828NB-C41/MICIU/AEI/10.13039/501100011033, and through the Unit of Excellence Mar\'ia de Maeztu 2020--2023 award to the Institute of Cosmos Sciences (CEX2019- 000918-M). D.M. acknowledges support from the National Science Foundation through grants PHY-2209451 and AST-2206532. J.M.P. acknowledges the support of an NSERC Discovery Grant (RGPIN-2023-05373). D.J.P. acknowledges support from the Chandra X-ray Center, which is operated by the Smithsonian Institution under NASA contract NAS8-03060.

The European VLBI Network is a joint facility of independent European, African, Asian, and North American radio astronomy institutes.  Scientific results from data presented in this publication are derived from the following EVN project code: EN006 \& EL071. 

The National Radio Astronomy Observatory is a facility of the National Science Foundation operated under cooperative agreement by Associated Universities, Inc. This work made use of the Swinburne University of Technology software correlator, developed as part of the Australian Major National Research Facilities Programme and operated under licence.

\newpage

\appendix

\section{SN 2012au Morphology}\label{appendix:resolved?}
We investigate in greater whether SN~2012au exhibits evidence in favor of (a) being marginally resolved and (b) having extension at larger radii. 
\subsection{Is the source marginally resolved?}
In our final set of images in Figures \ref{fig:clean_2012au_kband} and \ref{fig:clean_2012au_cband}, we observe that in some instances (see images taken at e.g., $3030,3289,4382~\mathrm{days}$), the synthesized beam is marginally misaligned with position angle of the Gaussian source. Could this be evidence that SN~2012au is marginally resolved? To test this, we combined the calibrated visibilities from Epochs $1,4$ and $5$ into a single dataset using \texttt{AIPS} and exported for imaging in \texttt{DIFMAP}. We adopt the same imaging technique as used for the individual epochs, as discussed in Section \ref{s:observations}. We exclude Epoch $6$ when stacking the observations due to the large temporal separation between the observation and the remaining epochs at $4.93~\mathrm{GHz}$, in addition to the observed astrometric offset in the check source discussed in Section \ref{ss:astrometry}. Note that we do not perform a similar analysis with the first two $22.24~\mathrm{GHz}$ images for a similar reason: The large differences in astrometry between the two epochs, driven by dominant tropospheric variations at low elevations (as discussed in Section \ref{ss:astrometry}), made it such that combining the two datasets did not improve the individual noise properties of the image. The same flags that were applied on a per-epoch basis were again applied to the final dataset. By combining the individual datasets, an increase in the amount of data achieves a more uniform sampling of the $uv-$plane. To verify that combining the data did not introduce any additional morphology to the visibilities, we performed a similar test using the check source. \par
In Figure \ref{fig:stacked_combined}, we plot the clean, combined images of J1303$-$1051 and SN~2012au, respectively, along with their residuals in the lower panel. The properties of the images (e.g., the chosen color scaling) are identical to those discussed in Section \ref{s:results}. In comparison to the individual images, we find that by increasing our sampling of the $uv$ plane, we no longer observe any evidence for misalignment with the synthesized beam. We conclude that images where marginal misalignment is identified can thus be attributed to reduced $uv$-sampling driven by excess flagging to compensate for the low elevations, rather than evidence for the source being marginally resolved.
\subsection{Is there extended emission at larger radii?}
\begin{figure*}
    \centering
\includegraphics[width=0.95\linewidth]{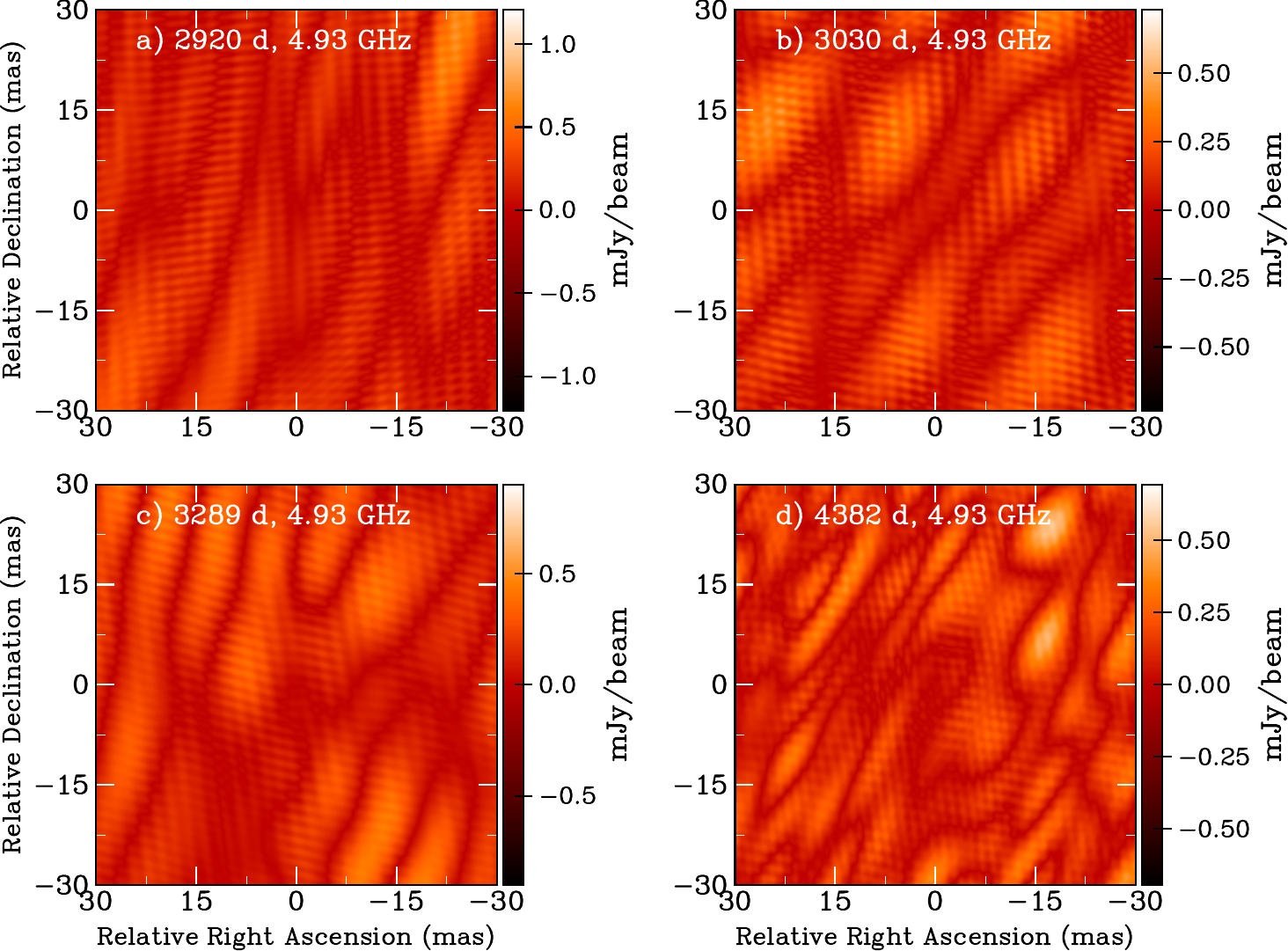}
    \caption{Residuals associated with CLEAN images plotted in Figure \ref{fig:clean_2012au_cband}. Images are zoomed out to $\pm30~\mathrm{mas}$ and saturated to $\pm5\sigma$. No significant peaks are identified in the residuals above the $>5\sigma$ level in the vicinity of the target.}
    \label{fig:residuals_2012au_C-band}
\end{figure*}

\begin{figure*}
    \centering
\includegraphics[width=0.95\linewidth]{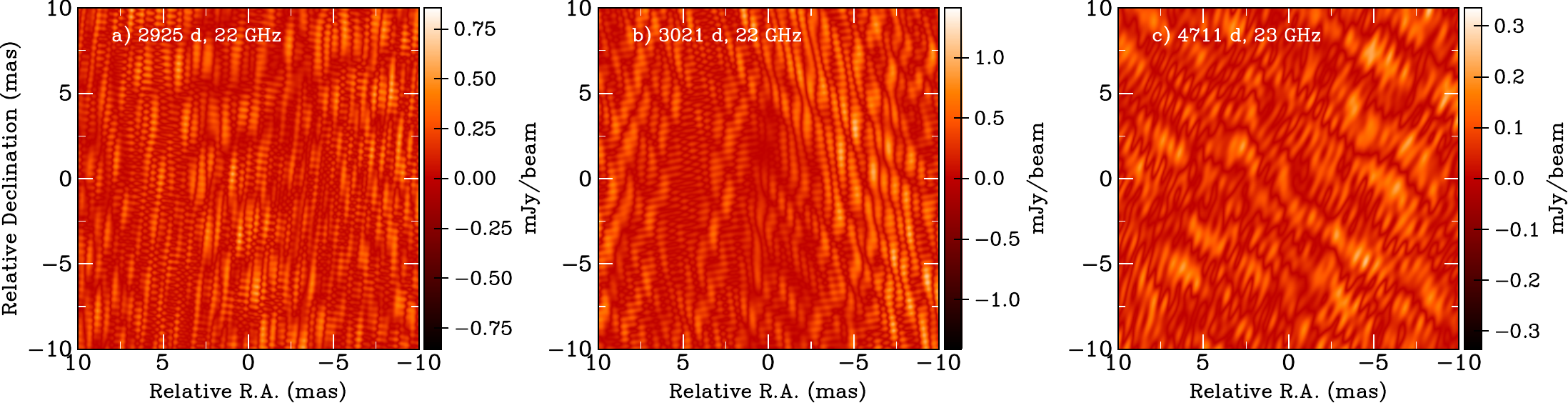}
    \caption{Residuals associated with CLEAN images plotted in Figure \ref{fig:clean_2012au_kband}. Images are zoomed out to $\pm10~\mathrm{mas}$ and saturated to $\pm5\sigma$. No significant peaks are identified in the residuals above the $>5\sigma$ level in the vicinity of the target.}
    \label{fig:residuals_2012au_K-band}
\end{figure*}

\begin{figure*}[htbp]
  \centering
  \begin{minipage}{0.48\linewidth}
    \centering
    \includegraphics[width=\linewidth]{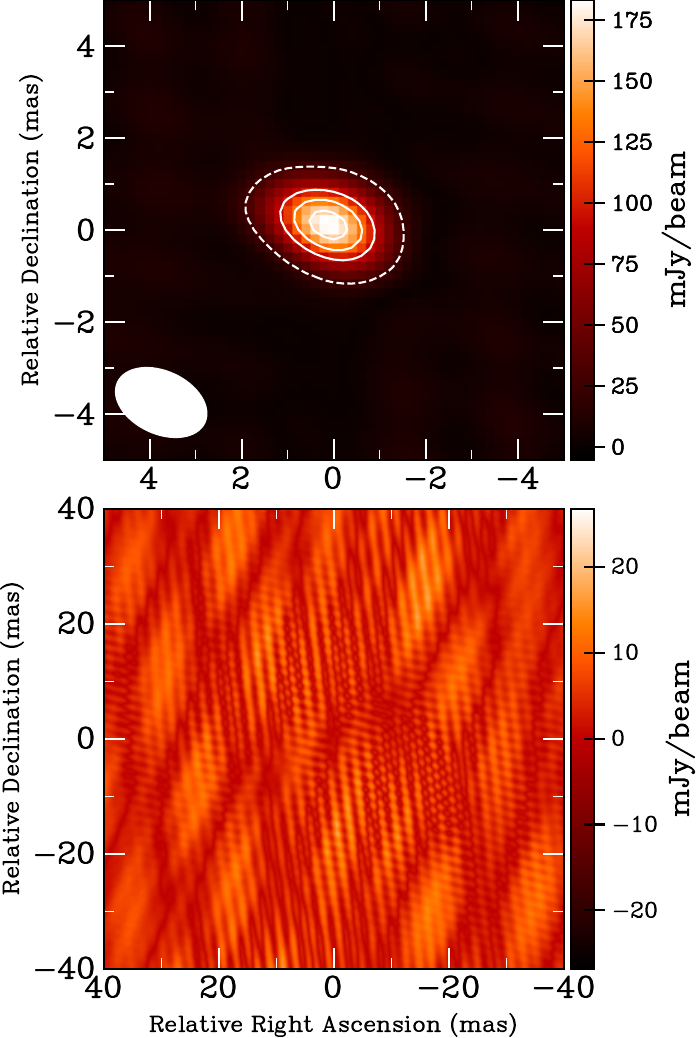}
  \end{minipage}%
  \hfill
  \begin{minipage}{0.48\linewidth}
    \centering
    \includegraphics[width=\linewidth]{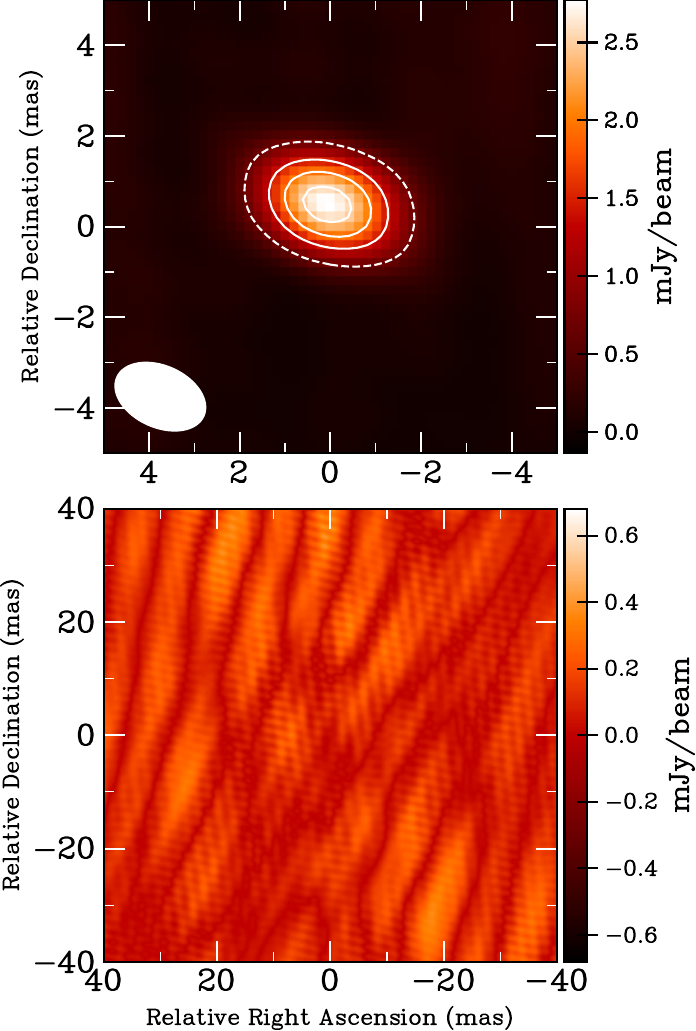}
  \end{minipage}

  \caption{Images of J1303$-$1051 (left) and SN~2012au (right), showing the final clean images (top row) and residuals (bottom row) after combining epochs 1, 4, and 5 at $4.93\,$GHz.  The $5\sigma$ contour is dashed; solid contours are at 50\%, 70\%, and 90\% of peak.  Colorscale is in mJy beam$^{-1}$, and the synthesized beam is shown in the lower‐left of each clean image.  Residuals are saturated to $\pm5\sigma$ to highlight systematic calibration errors common to both datasets.}
  \label{fig:stacked_combined}
\end{figure*}

In Figures \ref{fig:residuals_2012au_C-band} and \ref{fig:residuals_2012au_K-band}, we plot the final set of residuals of our cleaned images presented in Figure \ref{fig:clean_2012au_kband} and \ref{fig:clean_2012au_cband}. We identify peaks in our residual dirty images at the $3-4\sigma$ level, but none exceeding the $\ge5\sigma$ level. Are these peaks astrophysical in origin, or simply residual calibration errors driven by the low elevation observations? To answer this question, we draw the reader's attention back to the residuals of the combined images in Figure \ref{fig:stacked_combined}. Under the reasonable assumption that a point source model is a representative description of the check source, and in the absence of strong systematic residual calibration contributions, the residuals of J1303$-$1051 in Figure \ref{fig:stacked_combined} should be approximately Gaussian distributed. If, on the other hand, dominant residual calibration errors persist throughout the visibilities of the check source, they should manifest as structure in the residual dirty image. Opting to include these structures in the model while imaging either target would consequently lead to a biased model of the true sky. \par

In the residuals presented in Figure \ref{fig:stacked_combined}, we observe evident structure, suggesting that residual calibration errors persist throughout the first three epochs at $4.93~\mathrm{GHz}$ of J1303-1051. Comparing the structure to the observed residuals for SN~2012au in Figure \ref{fig:stacked_combined}, we observe a strong qualitative correlation between the structure of both sets of residuals, suggesting that the same residual calibration errors persist throughout both sets of visibilities. For this reason, we argue that it is highly unlikely that the peaks in SN~2012au's residuals at the $3-4\sigma$ confidence level are astrophysical, given that we observe the same peaks in the residuals of J1303$-$1051. \par
The aforementioned exercise led us to compare the residuals between the check source and SN~2012au on a per-epoch basis in which we found that nearly all extension in the residuals of SN~2012au could be mapped to an equivalent component in J1303$-$1051's residuals. For these reasons, at the sensitivity limits provided by our observations dominated by systematic uncertainties, we rule out the presence of any dominant extended emission in the field of SN~2012au and claim that SN~2012au is an unresolved point source in all of our images across all epochs.

\section{DYNAMICAL MODELING OF SHOCK INTERACTION WITH A TORUS-LIKE CIRCUMSTELLAR MEDIUM}\label{s:torus}
\subsection{Blastwave modeling and constraints on the CSM densities required to decelerate forward shock}\label{sss:csm_density}
We model the evolution of the SN blastwave expanding into a dense CSM following \cite{Ibik_2025}. This model assumes a broken power law for the SN ejecta and here we assume it expands into a CSM described by a single power law, e.g., $\rho_\mathrm{csm}\,\propto\, r^{-s}$.  We consider only the interaction with the equatorial component and ignore the secondary, low density component as it is expected to contribute minimally to the late-time radio flux, as motivated by our discussion in Section \ref{ss:early_radio}. The ejecta profile is defined as 
 \begin{equation}\label{eq:rho_ej}
\rho_{\text{ej}} = \rho_0 \left( \frac{t}{t_0} \right)^{-3} \left( \frac{v_0 t}{r} \right)^n 
\begin{cases} 
n = n_0 & \text{for } \frac{r}{t} > v_0 \\
n = n_1 & \text{for } \frac{r}{t} < v_0
\end{cases}, 
\end{equation}
where the density of the ejecta is computed at a given time $t$ and radius $r$. The inner (slow moving) ejecta has a power-law index of $n_0$, while the outer (fast moving) ejecta has a power-law index of $n_1$. Here, $\rho_0$ and $t_0$ are scaling parameters and $v_0$ is a transition velocity between the inner and outer ejecta profiles, estimated following the prescription provided in \cite{1994ApJ...420..268C}. \par
Following \cite{Ibik_2025,1994ApJ...420..268C}, we forward model the evolution of the blastwave by implementing a fourth-order Runge-Kutta scheme which numerically solves the 1-dimensional dynamical interaction between the ejected mass propagating into the surrounding CSM. For a given CSM density profile, ejecta profile and explosion parameters ($M_\mathrm{ej}$ and $E_k$), the simulations output the radius and radial velocity of the forward shock as a function of time. Asserting that the forward shock must obey our radial upper limits at a given time in Table \ref{tab:fitparams}, we determine the minimum required density scaling of the CSM profile to sufficiently decelerate the shock front. For the explosion parameters, we assume $E_k = 5\times10^{51}~\mathrm{erg}$ and $M_\mathrm{ej} = 5~M_\odot$ \citep{Pandey_2021}. We further assume power law indices for the ejecta profile of $n_0 = 0$ and $n_1 = 10$. We discuss the sensitivity of our results to these assumptions in Appendix \ref{ss:sensitivity}.

\subsection{Expectations of flux suppression at early times}\label{sss:flux_suppression}
We wish to understand whether the inferred distributions in Figure \ref{fig:csm_vs} would be sufficiently self-absorbed at early times to suppress the radio emission such that it only emerges at late times. To answer this, we model the radio flux assuming a standard SSA model for radio supernovae \citep{chevalier1998}, self-consistently verifying that the spectrum peaks at $\nu_\mathrm{SSA}$. At each time step, we evaluate the instantaneous flux density at frequency $\nu$ given by \citep{chevalier1998,2015ApJ...806..106A}
\begin{align}\label{eq:ssa_simple}
\begin{split}
    F_\nu = &1.582~F_{\nu_p}\left(\frac{\nu}{\nu_p}\right)^{5/2}\\
    &\times\left\{1 - \mathrm{exp}\left[-\left(\frac{\nu}{\nu_p}\right)^{-(p+4)/2}\right]\right\} ,
\end{split}
\end{align}
where $p$ corresponds to the electron energy index ($N(E)dE\,\propto\, E^{-p}dE$) and $F_{\nu_p}$ is the peak flux of the SSA spectrum occurring at frequency $\nu_p$.  We calculate $F_{\nu_p}$ and $\nu_p$ at each time step by inverting Equations 2 and 5 in \cite{2015ApJ...806..106A} where we have assumed $p=3$, typical for SNe Ib/c \citep{Chevalier_2006}. The peak flux and peak frequency of the SSA spectrum at a given timestep is then given by: 
\begin{align}\label{eq:fnup}
\begin{split}
    {F_{\nu_p}}=
~\alpha^{\tfrac{5}{7}}\left(\frac{R}{4.0\times10^{14}\,\mathrm{cm}}\right)^{\tfrac{19}{7}}\left(\frac{\dot{M}}{9.25\times10^{-6}\,M_\odot\,\mathrm{yr}^{-1}}\right)^{\tfrac{19}{14}}\left(\frac{\epsilon_{B}}{0.1}\right)^{\tfrac{19}{14}}\\
\times\left(\frac{D}{\mathrm{Mpc}}\right)^{-2}
\left(\frac{t}{10\,\mathrm{days}}\right)^{-\tfrac{19}{7}}\left(\frac{v_{w}}{1000\,\mathrm{km\,s}^{-1}}\right)^{-\tfrac{19}{14}}\left(\frac{f}{0.5}\right)^{\tfrac{5}{7}}~\mathrm{mJy},
\end{split}
\end{align}
and 
\begin{align}\label{eq:nup}
\begin{split}
    {{\nu_p}}
=  5~\alpha^{\tfrac{2}{7}}\left(\frac{R}{4.0\times10^{14}\,\mathrm{cm}}\right)^{\tfrac{2}{7}}\left(\frac{\dot{M}}{9.25\times10^{-6}\,M_{\odot}\,\mathrm{yr}^{-1}}\right)^{\tfrac{9}{14}} \left(\frac{f}{0.5}\right)^{\tfrac{2}{7}}\left(\frac{\epsilon_{B}}{0.1}\right)^{\tfrac{9}{14}} \left(\frac{t}{10\,\mathrm{days}}\right)^{-\tfrac{9}{7}}\left(\frac{v_{w}}{1000\,\mathrm{km\,s}^{-1}}\right)^{-\tfrac{9}{14}}~\mathrm{GHz}.
\end{split}
\end{align}
Here, $R$ is the radius of the shockfront at time $t$, $\dot{M}$ is the equivalent mass-loss rate related to the density of the CSM at radius $R$ via Equation \ref{eq:rho_csm}, $v_\mathrm{w}=1000~\mathrm{km~s}^{-1}$ is the assumed wind velocity, $D = 23.5~\mathrm{Mpc}$ is the luminosity distance to the source \citep{Milisavljevic_2013}, $f=0.5$ is the filling factor, $\alpha=\epsilon_\mathrm{e}/\epsilon_\mathrm{B}$ is the ratio of the relativistic electron energy density to the magnetic energy density and $\epsilon_\mathrm{B}$ is the fraction of post-shock energy density that is converted into magnetic energy density in the shocked region \citep{2015ApJ...806..106A}. Throughout this analysis, we assume default values of $\epsilon_\mathrm{B} = 0.1$ and $\alpha = 1$, again addressing the sensitivity of our conclusions to these assumptions in Section \ref{ss:sensitivity}. 

\subsection{Flux Modeling to Place Constraints on the Inferred Radio Filling Factor}\label{sss:fillingfactor}
In Section \ref{ss:torus_csm}, we are interested in determining the fraction of the total volume contained within a sphere imposed by our VLBI observations is actively contributing to powering the late-time radio emission. To do this, we require a constraint the radio filling factor, $f$, which relates to the SSA flux evolution via Equations \ref{eq:ssa_simple}, \ref{eq:fnup} and \ref{eq:nup}. We obtain the best-fit filling factors by performing a least-squares fit of Equation \ref{eq:ssa_simple} to our flux measurements, utilizing SciPy's \texttt{curve\_fit} routine. While performing the fit, we use the same assumptions as in Section \ref{sss:flux_suppression}, and allow $f$ to be the only free parameter.\par
With a constraint on $f$, the volume of the emitting region can be estimated via $V \approx \frac{4\pi}{3}fR_\mathrm{max}^3$. Once the volume is known, we can use it to make a statement regarding the geometry of the torus. In particular, the volume of CSM actively driving the late-time emission can be related to the opening angle $\theta$ of the torus via 
\begin{equation}
    V = \frac{4\pi}{3}fR_\mathrm{max}^3 = \int_0^{2\pi}\int_{\pi/2 - \theta}^{\pi/2 + \theta}\int_{R_\star}^{R_\mathrm{max}} r^2\mathrm{sin}\theta drd\theta d\phi,
\end{equation}
where $R_\star = \gamma R_\mathrm{max}$ and $\gamma$ is the ratio of outer shock radius to inner shock boundary actively contributing to powering the light curve which we assume to be $\gamma = 0.7$. This implies that only the outer $30\%$ of the torus is radiating. Solving the above integral and isolating the opening angle, we obtain the equation for the opening angle given by Equation \ref{eq:opening_angle}. The total amount of CSM mass that has been swept up is $M_\mathrm{csm} \sim \int\rho_\mathrm{csm}dV$, where $dV = r^2\mathrm{sin}\zeta drd\zeta d\phi$, $r\in[0,R_\mathrm{max}], \phi\in[0,2\pi]$ and $\zeta\in [\pi/2 - \theta, \pi/2 + \theta]$.

\subsection{Impact of parameter variation}\label{ss:sensitivity}
Here, we breakdown the sensitivity of our results to the numerous assumptions that we have made to infer the radio filling factor in Section \ref{ss:torus_csm}. 

\paragraph{$E_k$ \& $M_\mathrm{ej}$} Throughout our analysis, we have adopted conservative values for the explosion parameters of $E_k = 5\times10^{51}~\mathrm{erg}$ and $M_\mathrm{ej} = 6~M_\odot$ based off of representative values obtained from modeling of SN~2012au's optical lightcurve \citep{Pandey_2021}. However, \cite{Milisavljevic_2013} suggests a more energetic event with  $E_k\sim10^{52}~\mathrm{erg}$ and $M_\mathrm{ej}\sim3-5~M_\odot$. Increasing $E_k$ to $10^{52}~\mathrm{erg}$ decreases the filling factor by $\sim60\%$, implying a decrease in the opening angle. This is consistent with expectations: increasing the kinetic energy requires an even denser torus to decelerate the increasingly energetic blastwave, thus implying an even smaller filling factor to suppress the radio emission. While the opposite is true when decreasing the kinetic energy, we do not consider $E_k< 5\times10^{51}~\mathrm{erg}$ as it would contradict the inferred kinetic energies from optical observations \citep{Milisavljevic_2013,Takaki_2013,Pandey_2021}.  For the expected range of $3~M_\odot\lesssim M_\mathrm{ej} \lesssim 8M_\odot$ \citep{Milisavljevic_2013,Pandey_2021}, decreasing $M_\mathrm{ej}$ to $\sim3~M_\odot$ increases the filling factor by $\sim50\%$ while increasing to $M_\mathrm{ej}\sim8~M_\odot$ decreases the filling factor by $\sim10\%$. 

\paragraph{$n_0$ \& $n_1$} We assume a broken power law of the ejecta with $n_0 = 0$ and $n_1 = 10$. Typical values in the literature assume $n_0\in[0,2]$ and $n_1 \in [8,12]$ (e.g. \citealt{1994ApJ...420..268C,2005ApJ...619..839C}). Increasing the inner ejecta profile to $n_0 = 2$ increases the filling factor by $\sim25\%$ while varying the the outer ejecta profile scaling has a minimal effect, varying the filling factor by $\pm 5\%$. Both assumptions therefore have a relatively minimal effect on our inferred filling factors. 

\paragraph{$\alpha$ \& $\epsilon_\mathrm{B}$} We assumed that equipartition between the relativistic electron  and magnetic field energy density holds, i.e. $\epsilon_\mathrm{B} = \epsilon_\mathrm{e} = 0.1$. However, as \cite{Ibik_2025} also notes, these parameters are highly uncertain, with values for some SNe in radio achieving $\alpha = 10-100$ (e.g., \citealt{2013MNRAS.436.1258H,2014ApJ...782...30Y,2020ApJ...903..132H}). To assess the sensitivity of our results to the choice of $\alpha$ and $\epsilon_\mathrm{B}$, we fix $\alpha = 10$ and $\epsilon_\mathrm{B} = 0.01$. We observe a significant increase in the radio filling factor by a factor of $\sim10\times$ across all density profiles, significantly inflating the opening angles across all density profiles. The assumption of $\alpha$ and $\epsilon_\mathrm{B}$ therefore dominates the uncertainty budget and independent constraints on these parameters is necessary to accurately model a torus-like CSM. Alternatively, improvement on the modeling could further be refined if the torus geometry was resolved with higher angular resolution observations. 

\paragraph{$R_\mathrm{max}$} Our VLBI observations place only upper limits on the radius of the forward shock. However, it is also possible the shock radius is at an even smaller radius at the same time of our observations. Decreasing the radius of the shock by $10\%$, we observe an decrease in the filling factor of $\sim20-50\%$ depending on the assumed density profile, decreasing the opening angle of the torus by similar amounts. The strong dependence on $R_\mathrm{max}$ suggests that if the source remains unresolved, it will continue to push $\theta$ to smaller and smaller opening angles, potentially disfavoring the torus model over the next few years. 

\newpage

\bibliography{main}{}

@ARTICLE{2022ApJ...939..105B,
       author = {{Brethauer}, Daniel and {Margutti}, Raffaella and {Milisavljevic}, Dan and {Bietenholz}, Michael F. and {Chornock}, Ryan and {Coppejans}, Deanne L. and {De Colle}, Fabio and {Hajela}, Aprajita and {Terreran}, Giacomo and {Vargas}, Felipe and {DeMarchi}, Lindsay and {Harris}, Chelsea and {Jacobson-Gal{\'a}n}, Wynn V. and {Kamble}, Atish and {Patnaude}, Daniel and {Stroh}, Michael C.},
        title = "{Seven Years of Coordinated Chandra-NuSTAR Observations of SN 2014C Unfold the Extreme Mass-loss History of Its Stellar Progenitor}",
      journal = {\apj},
         year = 2022,
        month = nov,
       volume = {939},
       number = {2},
          eid = {105},
        pages = {105},
          doi = {10.3847/1538-4357/ac8b14},
archivePrefix = {arXiv},
       eprint = {2206.00842},
 primaryClass = {astro-ph.HE},
       adsurl = {https://ui.adsabs.harvard.edu/abs/2022ApJ...939..105B}
}

@ARTICLE{2024Natur.628..733R,
       author = {{Rodr{\'\i}guez}, {\'O}smar and {Nakar}, Ehud and {Maoz}, Dan},
        title = "{Stripped-envelope supernova light curves argue for central engine activity}",
      journal = {\nat},
         year = 2024,
        month = apr,
       volume = {628},
       number = {8009},
        pages = {733-735},
          doi = {10.1038/s41586-024-07262-x},
archivePrefix = {arXiv},
       eprint = {2404.10846},
 primaryClass = {astro-ph.HE},
       adsurl = {https://ui.adsabs.harvard.edu/abs/2024Natur.628..733R}
}

@ARTICLE{2012CBET.3052....1H,
       author = {{Howerton}, S. and {Drake}, A.~J. and {Djorgovski}, S.~G. and {Mahabal}, A. and {Graham}, M.~J. and {Williams}, R. and {Roy}, R. and {Mohan}, V. and {Prieto}, J.~L. and {Catelan}, M. and {Beshore}, E.~C. and {Larson}, S.~M. and {Christensen}, E. and {Elenin}, L. and {Molotov}, I. and {Koff}, R.~A. and {Silverman}, J.~M. and {Cenko}, S.~B. and {Miller}, A.~A. and {Nugent}, P.~E. and {Filippenko}, A.~V.},
        title = "{Supernova 2012au in NGC 4790 = Psn J12545218-1014502}",
      journal = {Central Bureau Electronic Telegrams},
         year = 2012,
        month = mar,
       volume = {3052},
        pages = {1},
       adsurl = {https://ui.adsabs.harvard.edu/abs/2012CBET.3052....1H}
}

@ARTICLE{2014MNRAS.437..703M,
       author = {{Metzger}, Brian D. and {Vurm}, Indrek and {Hasco{\"e}t}, Romain and {Beloborodov}, Andrei M.},
        title = "{Ionization break-out from millisecond pulsar wind nebulae: an X-ray probe of the origin of superluminous supernovae}",
      journal = {\mnras},
         year = 2014,
        month = jan,
       volume = {437},
       number = {1},
        pages = {703-720},
          doi = {10.1093/mnras/stt1922},
archivePrefix = {arXiv},
       eprint = {1307.8115},
 primaryClass = {astro-ph.HE},
       adsurl = {https://ui.adsabs.harvard.edu/abs/2014MNRAS.437..703M}
}

@ARTICLE{2023ApJ...955...71R,
       author = {{Rodr{\'\i}guez}, {\'O}smar and {Maoz}, Dan and {Nakar}, Ehud},
        title = "{The Iron Yield of Core-collapse Supernovae}",
      journal = {\apj},
         year = 2023,
        month = sep,
       volume = {955},
       number = {1},
          eid = {71},
        pages = {71},
          doi = {10.3847/1538-4357/ace2bd},
archivePrefix = {arXiv},
       eprint = {2209.05552},
 primaryClass = {astro-ph.HE},
       adsurl = {https://ui.adsabs.harvard.edu/abs/2023ApJ...955...71R}
}

@article{Kourkchi_2020,
   title={Cosmicflows-4: The Catalog of ∼10,000 Tully–Fisher Distances},
   volume={902},
   ISSN={1538-4357},
   url={http://dx.doi.org/10.3847/1538-4357/abb66b},
   DOI={10.3847/1538-4357/abb66b},
   number={2},
   journal={The Astrophysical Journal},
   publisher={American Astronomical Society},
   author={Kourkchi, Ehsan and Tully, R. Brent and Eftekharzadeh, Sarah and Llop, Jordan and Courtois, Hélène M. and Guinet, Daniel and Dupuy, Alexandra and Neill, James D. and Seibert, Mark and Andrews, Michael and Chuang, Juana and Danesh, Arash and Gonzalez, Randy and Holthaus, Alexandria and Mokelke, Amber and Schoen, Devin and Urasaki, Chase},
   year={2020},
   month=oct, pages={145} }

@misc{snelders2025revisitingfrb20121102amilliarcsecond,
      title={Revisiting FRB 20121102A: milliarcsecond localisation and a decreasing dispersion measure}, 
      author={M. P. Snelders and J. W. T. Hessels and J. Huang and N. Sridhar and B. Marcote and A. M. Moroianu and O. S. Ould-Boukattine and F. Kirsten and S. Bhandari and D. M. Hewitt and D. Pelliciari and L. Rhodes and R. Anna-Thomas and U. Bach and E. K. Bempong-Manful and V. Bezrukovs and J. D. Bray and S. Buttaccio and I. Cognard and A. Corongiu and R. Feiler and M. P. Gawroński and M. Giroletti and L. Guillemot and R. Karuppusamy and M. Lindqvist and K. Nimmo and A. Possenti and W. Puchalska and D. Williams-Baldwin},
      year={2025},
      eprint={2510.11352},
      archivePrefix={arXiv},
      primaryClass={astro-ph.HE},
      url={https://arxiv.org/abs/2510.11352}, 
}

@ARTICLE{1985A&AS...59...43B,
       author = {{Bottinelli}, L. and {Gouguenheim}, L. and {Paturel}, G. and {de Vaucouleurs}, G.},
        title = "{HI line studies of galaxies. IV. Distance moduli of 468 disk galaxies.}",
      journal = {\aaps},
         year = 1985,
        month = jan,
       volume = {59},
        pages = {43-58},
       adsurl = {https://ui.adsabs.harvard.edu/abs/1985A&AS...59...43B}
}

@ARTICLE{2007A&A...465...71T,
       author = {{Theureau}, G. and {Hanski}, M.~O. and {Coudreau}, N. and {Hallet}, N. and {Martin}, J. -M.},
        title = "{Kinematics of the Local Universe. XIII. 21-cm line measurements of 452 galaxies with the Nan{\c{c}}ay radiotelescope, JHK Tully-Fisher relation, and preliminary maps of the peculiar velocity field}",
      journal = {\aap},
         year = 2007,
        month = apr,
       volume = {465},
       number = {1},
        pages = {71-85},
          doi = {10.1051/0004-6361:20066187},
archivePrefix = {arXiv},
       eprint = {astro-ph/0611626},
 primaryClass = {astro-ph},
       adsurl = {https://ui.adsabs.harvard.edu/abs/2007A&A...465...71T}
}

@ARTICLE{2015AJ....150...58F,
       author = {{Fey}, A.~L. and {Gordon}, D. and {Jacobs}, C.~S. and {Ma}, C. and {Gaume}, R.~A. and {Arias}, E.~F. and {Bianco}, G. and {Boboltz}, D.~A. and {B{\"o}ckmann}, S. and {Bolotin}, S. and {Charlot}, P. and {Collioud}, A. and {Engelhardt}, G. and {Gipson}, J. and {Gontier}, A. -M. and {Heinkelmann}, R. and {Kurdubov}, S. and {Lambert}, S. and {Lytvyn}, S. and {MacMillan}, D.~S. and {Malkin}, Z. and {Nothnagel}, A. and {Ojha}, R. and {Skurikhina}, E. and {Sokolova}, J. and {Souchay}, J. and {Sovers}, O.~J. and {Tesmer}, V. and {Titov}, O. and {Wang}, G. and {Zharov}, V.},
        title = "{The Second Realization of the International Celestial Reference Frame by Very Long Baseline Interferometry}",
      journal = {\aj},
         year = 2015,
        month = aug,
       volume = {150},
       number = {2},
          eid = {58},
        pages = {58},
          doi = {10.1088/0004-6256/150/2/58},
       adsurl = {https://ui.adsabs.harvard.edu/abs/2015AJ....150...58F}
}

@Article{Hunter:2007,
  Author    = {Hunter, J. D.},
  Title     = {Matplotlib: A 2D graphics environment},
  Journal   = {Computing in Science \& Engineering},
  Volume    = {9},
  Number    = {3},
  Pages     = {90--95},
  publisher = {IEEE COMPUTER SOC},
  doi       = {10.1109/MCSE.2007.55},
  year      = 2007
}

@Article{numpy2020,
 title         = {Array programming with {NumPy}},
 author        = {Charles R. Harris and K. Jarrod Millman and St{\'{e}}fan J.
                 van der Walt and Ralf Gommers and Pauli Virtanen and David
                 Cournapeau and Eric Wieser and Julian Taylor and Sebastian
                 Berg and Nathaniel J. Smith and Robert Kern and Matti Picus
                 and Stephan Hoyer and Marten H. van Kerkwijk and Matthew
                 Brett and Allan Haldane and Jaime Fern{\'{a}}ndez del
                 R{\'{i}}o and Mark Wiebe and Pearu Peterson and Pierre
                 G{\'{e}}rard-Marchant and Kevin Sheppard and Tyler Reddy and
                 Warren Weckesser and Hameer Abbasi and Christoph Gohlke and
                 Travis E. Oliphant},
 year          = {2020},
 month         = sep,
 journal       = {Nature},
 volume        = {585},
 number        = {7825},
 pages         = {357--362},
 doi           = {10.1038/s41586-020-2649-2},
 publisher     = {Springer Science and Business Media {LLC}},
 url           = {https://doi.org/10.1038/s41586-020-2649-2}
}

@ARTICLE{1994ApJ...420..268C,
       author = {{Chevalier}, Roger A. and {Fransson}, Claes},
        title = "{Emission from Circumstellar Interaction in Normal Type II Supernovae}",
      journal = {\apj},
         year = 1994,
        month = jan,
       volume = {420},
        pages = {268},
          doi = {10.1086/173557},
       adsurl = {https://ui.adsabs.harvard.edu/abs/1994ApJ...420..268C}
}

@article{astropy:2013,
Adsurl = {http://adsabs.harvard.edu/abs/2013A%26A...558A..33A},
Archiveprefix = {arXiv},
Author = {{Astropy Collaboration} and {Robitaille}, T.~P. and {Tollerud}, E.~J. and {Greenfield}, P. and {Droettboom}, M. and {Bray}, E. and {Aldcroft}, T. and {Davis}, M. and {Ginsburg}, A. and {Price-Whelan}, A.~M. and {Kerzendorf}, W.~E. and {Conley}, A. and {Crighton}, N. and {Barbary}, K. and {Muna}, D. and {Ferguson}, H. and {Grollier}, F. and {Parikh}, M.~M. and {Nair}, P.~H. and {Unther}, H.~M. and {Deil}, C. and {Woillez}, J. and {Conseil}, S. and {Kramer}, R. and {Turner}, J.~E.~H. and {Singer}, L. and {Fox}, R. and {Weaver}, B.~A. and {Zabalza}, V. and {Edwards}, Z.~I. and {Azalee Bostroem}, K. and {Burke}, D.~J. and {Casey}, A.~R. and {Crawford}, S.~M. and {Dencheva}, N. and {Ely}, J. and {Jenness}, T. and {Labrie}, K. and {Lim}, P.~L. and {Pierfederici}, F. and {Pontzen}, A. and {Ptak}, A. and {Refsdal}, B. and {Servillat}, M. and {Streicher}, O.},
Doi = {10.1051/0004-6361/201322068},
Eid = {A33},
Eprint = {1307.6212},
Journal = {\aap},
Month = oct,
Pages = {A33},
Primaryclass = {astro-ph.IM},
Title = {{Astropy: A community Python package for astronomy}},
Volume = 558,
Year = 2013}

@ARTICLE{astropy:2018,
       author = {{Astropy Collaboration} and {Price-Whelan}, A.~M. and
         {Sip{\H{o}}cz}, B.~M. and {G{\"u}nther}, H.~M. and {Lim}, P.~L. and
         {Crawford}, S.~M. and {Conseil}, S. and {Shupe}, D.~L. and
         {Craig}, M.~W. and {Dencheva}, N. and {Ginsburg}, A. and {Vand
        erPlas}, J.~T. and {Bradley}, L.~D. and {P{\'e}rez-Su{\'a}rez}, D. and
         {de Val-Borro}, M. and {Aldcroft}, T.~L. and {Cruz}, K.~L. and
         {Robitaille}, T.~P. and {Tollerud}, E.~J. and {Ardelean}, C. and
         {Babej}, T. and {Bach}, Y.~P. and {Bachetti}, M. and {Bakanov}, A.~V. and
         {Bamford}, S.~P. and {Barentsen}, G. and {Barmby}, P. and
         {Baumbach}, A. and {Berry}, K.~L. and {Biscani}, F. and {Boquien}, M. and
         {Bostroem}, K.~A. and {Bouma}, L.~G. and {Brammer}, G.~B. and
         {Bray}, E.~M. and {Breytenbach}, H. and {Buddelmeijer}, H. and
         {Burke}, D.~J. and {Calderone}, G. and {Cano Rodr{\'\i}guez}, J.~L. and
         {Cara}, M. and {Cardoso}, J.~V.~M. and {Cheedella}, S. and {Copin}, Y. and
         {Corrales}, L. and {Crichton}, D. and {D'Avella}, D. and {Deil}, C. and
         {Depagne}, {\'E}. and {Dietrich}, J.~P. and {Donath}, A. and
         {Droettboom}, M. and {Earl}, N. and {Erben}, T. and {Fabbro}, S. and
         {Ferreira}, L.~A. and {Finethy}, T. and {Fox}, R.~T. and
         {Garrison}, L.~H. and {Gibbons}, S.~L.~J. and {Goldstein}, D.~A. and
         {Gommers}, R. and {Greco}, J.~P. and {Greenfield}, P. and
         {Groener}, A.~M. and {Grollier}, F. and {Hagen}, A. and {Hirst}, P. and
         {Homeier}, D. and {Horton}, A.~J. and {Hosseinzadeh}, G. and {Hu}, L. and
         {Hunkeler}, J.~S. and {Ivezi{\'c}}, {\v{Z}}. and {Jain}, A. and
         {Jenness}, T. and {Kanarek}, G. and {Kendrew}, S. and {Kern}, N.~S. and
         {Kerzendorf}, W.~E. and {Khvalko}, A. and {King}, J. and {Kirkby}, D. and
         {Kulkarni}, A.~M. and {Kumar}, A. and {Lee}, A. and {Lenz}, D. and
         {Littlefair}, S.~P. and {Ma}, Z. and {Macleod}, D.~M. and
         {Mastropietro}, M. and {McCully}, C. and {Montagnac}, S. and
         {Morris}, B.~M. and {Mueller}, M. and {Mumford}, S.~J. and {Muna}, D. and
         {Murphy}, N.~A. and {Nelson}, S. and {Nguyen}, G.~H. and
         {Ninan}, J.~P. and {N{\"o}the}, M. and {Ogaz}, S. and {Oh}, S. and
         {Parejko}, J.~K. and {Parley}, N. and {Pascual}, S. and {Patil}, R. and
         {Patil}, A.~A. and {Plunkett}, A.~L. and {Prochaska}, J.~X. and
         {Rastogi}, T. and {Reddy Janga}, V. and {Sabater}, J. and
         {Sakurikar}, P. and {Seifert}, M. and {Sherbert}, L.~E. and
         {Sherwood-Taylor}, H. and {Shih}, A.~Y. and {Sick}, J. and
         {Silbiger}, M.~T. and {Singanamalla}, S. and {Singer}, L.~P. and
         {Sladen}, P.~H. and {Sooley}, K.~A. and {Sornarajah}, S. and
         {Streicher}, O. and {Teuben}, P. and {Thomas}, S.~W. and
         {Tremblay}, G.~R. and {Turner}, J.~E.~H. and {Terr{\'o}n}, V. and
         {van Kerkwijk}, M.~H. and {de la Vega}, A. and {Watkins}, L.~L. and
         {Weaver}, B.~A. and {Whitmore}, J.~B. and {Woillez}, J. and
         {Zabalza}, V. and {Astropy Contributors}},
        title = "{The Astropy Project: Building an Open-science Project and Status of the v2.0 Core Package}",
      journal = {\aj},
         year = 2018,
        month = sep,
       volume = {156},
       number = {3},
          eid = {123},
        pages = {123},
          doi = {10.3847/1538-3881/aabc4f},
archivePrefix = {arXiv},
       eprint = {1801.02634},
 primaryClass = {astro-ph.IM},
       adsurl = {https://ui.adsabs.harvard.edu/abs/2018AJ....156..123A}
}

@ARTICLE{astropy:2022,
       author = {{Astropy Collaboration} and {Price-Whelan}, Adrian M. and {Lim}, Pey Lian and {Earl}, Nicholas and {Starkman}, Nathaniel and {Bradley}, Larry and {Shupe}, David L. and {Patil}, Aarya A. and {Corrales}, Lia and {Brasseur}, C.~E. and {N{"o}the}, Maximilian and {Donath}, Axel and {Tollerud}, Erik and {Morris}, Brett M. and {Ginsburg}, Adam and {Vaher}, Eero and {Weaver}, Benjamin A. and {Tocknell}, James and {Jamieson}, William and {van Kerkwijk}, Marten H. and {Robitaille}, Thomas P. and {Merry}, Bruce and {Bachetti}, Matteo and {G{"u}nther}, H. Moritz and {Aldcroft}, Thomas L. and {Alvarado-Montes}, Jaime A. and {Archibald}, Anne M. and {B{'o}di}, Attila and {Bapat}, Shreyas and {Barentsen}, Geert and {Baz{'a}n}, Juanjo and {Biswas}, Manish and {Boquien}, M{'e}d{'e}ric and {Burke}, D.~J. and {Cara}, Daria and {Cara}, Mihai and {Conroy}, Kyle E. and {Conseil}, Simon and {Craig}, Matthew W. and {Cross}, Robert M. and {Cruz}, Kelle L. and {D'Eugenio}, Francesco and {Dencheva}, Nadia and {Devillepoix}, Hadrien A.~R. and {Dietrich}, J{"o}rg P. and {Eigenbrot}, Arthur Davis and {Erben}, Thomas and {Ferreira}, Leonardo and {Foreman-Mackey}, Daniel and {Fox}, Ryan and {Freij}, Nabil and {Garg}, Suyog and {Geda}, Robel and {Glattly}, Lauren and {Gondhalekar}, Yash and {Gordon}, Karl D. and {Grant}, David and {Greenfield}, Perry and {Groener}, Austen M. and {Guest}, Steve and {Gurovich}, Sebastian and {Handberg}, Rasmus and {Hart}, Akeem and {Hatfield-Dodds}, Zac and {Homeier}, Derek and {Hosseinzadeh}, Griffin and {Jenness}, Tim and {Jones}, Craig K. and {Joseph}, Prajwel and {Kalmbach}, J. Bryce and {Karamehmetoglu}, Emir and {Ka{l}uszy{'n}ski}, Miko{l}aj and {Kelley}, Michael S.~P. and {Kern}, Nicholas and {Kerzendorf}, Wolfgang E. and {Koch}, Eric W. and {Kulumani}, Shankar and {Lee}, Antony and {Ly}, Chun and {Ma}, Zhiyuan and {MacBride}, Conor and {Maljaars}, Jakob M. and {Muna}, Demitri and {Murphy}, N.~A. and {Norman}, Henrik and {O'Steen}, Richard and {Oman}, Kyle A. and {Pacifici}, Camilla and {Pascual}, Sergio and {Pascual-Granado}, J. and {Patil}, Rohit R. and {Perren}, Gabriel I. and {Pickering}, Timothy E. and {Rastogi}, Tanuj and {Roulston}, Benjamin R. and {Ryan}, Daniel F. and {Rykoff}, Eli S. and {Sabater}, Jose and {Sakurikar}, Parikshit and {Salgado}, Jes{'u}s and {Sanghi}, Aniket and {Saunders}, Nicholas and {Savchenko}, Volodymyr and {Schwardt}, Ludwig and {Seifert-Eckert}, Michael and {Shih}, Albert Y. and {Jain}, Anany Shrey and {Shukla}, Gyanendra and {Sick}, Jonathan and {Simpson}, Chris and {Singanamalla}, Sudheesh and {Singer}, Leo P. and {Singhal}, Jaladh and {Sinha}, Manodeep and {Sip{H{o}}cz}, Brigitta M. and {Spitler}, Lee R. and {Stansby}, David and {Streicher}, Ole and {{{S}}umak}, Jani and {Swinbank}, John D. and {Taranu}, Dan S. and {Tewary}, Nikita and {Tremblay}, Grant R. and {Val-Borro}, Miguel de and {Van Kooten}, Samuel J. and {Vasovi{'c}}, Zlatan and {Verma}, Shresth and {de Miranda Cardoso}, Jos{'e} Vin{'i}cius and {Williams}, Peter K.~G. and {Wilson}, Tom J. and {Winkel}, Benjamin and {Wood-Vasey}, W.~M. and {Xue}, Rui and {Yoachim}, Peter and {Zhang}, Chen and {Zonca}, Andrea and {Astropy Project Contributors}},
        title = "{The Astropy Project: Sustaining and Growing a Community-oriented Open-source Project and the Latest Major Release (v5.0) of the Core Package}",
      journal = {\apj},
         year = 2022,
        month = aug,
       volume = {935},
       number = {2},
          eid = {167},
        pages = {167},
          doi = {10.3847/1538-4357/ac7c74},
archivePrefix = {arXiv},
       eprint = {2206.14220},
 primaryClass = {astro-ph.IM},
       adsurl = {https://ui.adsabs.harvard.edu/abs/2022ApJ...935..167A}
}

@ARTICLE{2020ApJ...903..132H,
       author = {{Horesh}, Assaf and {Sfaradi}, Itai and {Ergon}, Mattias and {Barbarino}, Cristina and {Sollerman}, Jesper and {Moldon}, Javier and {Dobie}, Dougal and {Schulze}, Steve and {P{\'e}rez-Torres}, Miguel and {Williams}, David R.~A. and {Fremling}, Christoffer and {Gal-Yam}, Avishay and {Kulkarni}, Shrinivas R. and {O'Brien}, Andrew and {Lundqvist}, Peter and {Murphy}, Tara and {Fender}, Rob and {Anand}, Shreya and {Belicki}, Justin and {Bellm}, Eric C. and {Coughlin}, Michael W. and {De}, Kishalay and {Golkhou}, V. Zach and {Graham}, Matthew J. and {Green}, Dave A. and {Hankins}, Matt and {Kasliwal}, Mansi and {Kupfer}, Thomas and {Laher}, Russ R. and {Masci}, Frank J. and {Miller}, A.~A. and {Neill}, James D. and {Ofek}, Eran O. and {Perrott}, Yvette and {Porter}, Michael and {Reiley}, Daniel J. and {Rigault}, Mickael and {Rodriguez}, Hector and {Rusholme}, Ben and {Shupe}, David L. and {Titterington}, David},
        title = "{A Non-equipartition Shock Wave Traveling in a Dense Circumstellar Environment around SN 2020oi}",
      journal = {\apj},
         year = 2020,
        month = nov,
       volume = {903},
       number = {2},
          eid = {132},
        pages = {132},
          doi = {10.3847/1538-4357/abbd38},
archivePrefix = {arXiv},
       eprint = {2006.13952},
 primaryClass = {astro-ph.HE},
       adsurl = {https://ui.adsabs.harvard.edu/abs/2020ApJ...903..132H}
}

@ARTICLE{2014ApJ...782...30Y,
       author = {{Yadav}, Naveen and {Ray}, Alak and {Chakraborti}, Sayan and {Stockdale}, Christopher and {Chandra}, Poonam and {Smith}, Randall and {Roy}, Rupak and {Bose}, Subhash and {Dwarkadas}, Vikram and {Sutaria}, Firoza and {Pooley}, David},
        title = "{Electron Cooling in a Young Radio Supernova: SN 2012aw}",
      journal = {\apj},
         year = 2014,
        month = feb,
       volume = {782},
       number = {1},
          eid = {30},
        pages = {30},
          doi = {10.1088/0004-637X/782/1/30},
archivePrefix = {arXiv},
       eprint = {1311.3568},
 primaryClass = {astro-ph.HE},
       adsurl = {https://ui.adsabs.harvard.edu/abs/2014ApJ...782...30Y}
}

@ARTICLE{2013MNRAS.436.1258H,
       author = {{Horesh}, Assaf and {Stockdale}, Christopher and {Fox}, Derek B. and {Frail}, Dale A. and {Carpenter}, John and {Kulkarni}, S.~R. and {Ofek}, Eran O. and {Gal-Yam}, Avishay and {Kasliwal}, Mansi M. and {Arcavi}, Iair and {Quimby}, Robert and {Cenko}, S. Bradley and {Nugent}, Peter E. and {Bloom}, Joshua S. and {Law}, Nicholas M. and {Poznanski}, Dovi and {Gorbikov}, Evgeny and {Polishook}, David and {Yaron}, Ofer and {Ryder}, Stuart and {Weiler}, Kurt W. and {Bauer}, Franz and {Van Dyk}, Schuyler D. and {Immler}, Stefan and {Panagia}, Nino and {Pooley}, Dave and {Kassim}, Namir},
        title = "{An early and comprehensive millimetre and centimetre wave and X-ray study of SN 2011dh: a non-equipartition blast wave expanding into a massive stellar wind}",
      journal = {\mnras},
         year = 2013,
        month = dec,
       volume = {436},
       number = {2},
        pages = {1258-1267},
          doi = {10.1093/mnras/stt1645},
archivePrefix = {arXiv},
       eprint = {1209.1102},
 primaryClass = {astro-ph.CO},
       adsurl = {https://ui.adsabs.harvard.edu/abs/2013MNRAS.436.1258H}
}

@ARTICLE{2022ApJS..263...24A,
       author = {{Ajello}, M. and {Baldini}, L. and {Ballet}, J. and {Bastieri}, D. and {Becerra Gonzalez}, J. and {Bellazzini}, R. and {Berretta}, A. and {Bissaldi}, E. and {Bonino}, R. and {Brill}, A. and {Bruel}, P. and {Buson}, S. and {Caputo}, R. and {Caraveo}, P.~A. and {Cheung}, C.~C. and {Chiaro}, G. and {Cibrario}, N. and {Ciprini}, S. and {Crnogorcevic}, M. and {Cutini}, S. and {D'Ammando}, F. and {De Gaetano}, S. and {Di Lalla}, N. and {Di Venere}, L. and {Dom{\'\i}nguez}, A. and {Ramazani}, V. Fallah and {Ferrara}, E.~C. and {Fiori}, A. and {Fukazawa}, Y. and {Funk}, S. and {Fusco}, P. and {Gammaldi}, V. and {Gargano}, F. and {Garrappa}, S. and {Gasparrini}, D. and {Giglietto}, N. and {Giordano}, F. and {Giroletti}, M. and {Green}, D. and {Grenier}, I.~A. and {Guiriec}, S. and {Horan}, D. and {Hou}, X. and {Kayanoki}, T. and {Kuss}, M. and {Larsson}, S. and {Latronico}, L. and {Lewis}, T. and {Li}, J. and {Liodakis}, I. and {Longo}, F. and {Loparco}, F. and {Lott}, B. and {Lovellette}, M.~N. and {Lubrano}, P. and {Madejski}, G.~M. and {Maldera}, S. and {Manfreda}, A. and {Mart{\'\i}-Devesa}, G. and {Mazziotta}, M.~N. and {Mereu}, I. and {Michelson}, P.~F. and {Mirabal}, N. and {Mitthumsiri}, W. and {Mizuno}, T. and {Monzani}, M.~E. and {Morselli}, A. and {Moskalenko}, I.~V. and {Negro}, M. and {Ojha}, R. and {Orienti}, M. and {Orlando}, E. and {Ormes}, J.~F. and {Pei}, Z. and {Pe{\~n}a-Herazo}, H. and {Persic}, M. and {Pesce-Rollins}, M. and {Petrosian}, V. and {Pillera}, R. and {Poon}, H. and {Porter}, T.~A. and {Principe}, G. and {Rain{\`o}}, S. and {Rando}, R. and {Rani}, B. and {Razzano}, M. and {Razzaque}, S. and {Reimer}, A. and {Reimer}, O. and {Scotton}, L. and {Serini}, D. and {Sgr{\`o}}, C. and {Siskind}, E.~J. and {Spandre}, G. and {Spinelli}, P. and {Suson}, D.~J. and {Tajima}, H. and {Torres}, D.~F. and {Valverde}, J. and {Yassin}, H. and {Zaharijas}, G.},
        title = "{The Fourth Catalog of Active Galactic Nuclei Detected by the Fermi Large Area Telescope: Data Release 3}",
      journal = {\apjs},
         year = 2022,
        month = dec,
       volume = {263},
       number = {2},
          eid = {24},
        pages = {24},
          doi = {10.3847/1538-4365/ac9523},
archivePrefix = {arXiv},
       eprint = {2209.12070},
 primaryClass = {astro-ph.HE},
       adsurl = {https://ui.adsabs.harvard.edu/abs/2022ApJS..263...24A}
}

@article{Carli_2024,
   title={The TRAPUM Small Magellanic Cloud pulsar survey with MeerKAT – I. Discovery of seven new pulsars and two Pulsar Wind Nebula associations},
   volume={531},
   ISSN={1365-2966},
   url={http://dx.doi.org/10.1093/mnras/stae1310},
   DOI={10.1093/mnras/stae1310},
   number={2},
   journal={Monthly Notices of the Royal Astronomical Society},
   publisher={Oxford University Press (OUP)},
   author={Carli, E and Levin, L and Stappers, B W and Barr, E D and Breton, R P and Buchner, S and Burgay, M and Geyer, M and Kramer, M and Padmanabh, P V and Possenti, A and Venkatraman Krishnan, V and Becker, W and Filipović, M D and Maitra, C and Behrend, J and Champion, D J and Chen, W and Men, Y P and Ridolfi, A},
   year={2024},
   month=may, pages={2835–2863} }

@article{Haberl_2012,
   title={Multi-frequency observations of SNR J0453–6829 in the LMC: A composite supernova remnant with a pulsar wind nebula},
   volume={543},
   ISSN={1432-0746},
   url={http://dx.doi.org/10.1051/0004-6361/201218971},
   DOI={10.1051/0004-6361/201218971},
   journal={Astronomy \&; Astrophysics},
   publisher={EDP Sciences},
   author={Haberl, F. and Filipović, M. D. and Bozzetto, L. M. and Crawford, E. J. and Points, S. D. and Pietsch, W. and De Horta, A. Y. and Tothill, N. and Payne, J. L. and Sasaki, M.},
   year={2012},
   month=jul, pages={A154} }

@ARTICLE{2013MNRAS.429.2677K,
       author = {{Karachentsev}, I.~D. and {Nasonova}, O.~G.},
        title = "{Intense look at Virgo Southern Extension}",
      journal = {\mnras},
         year = 2013,
        month = mar,
       volume = {429},
       number = {3},
        pages = {2677-2686},
          doi = {10.1093/mnras/sts557},
archivePrefix = {arXiv},
       eprint = {1212.0840},
 primaryClass = {astro-ph.CO},
       adsurl = {https://ui.adsabs.harvard.edu/abs/2013MNRAS.429.2677K}
}

@ARTICLE{1996ApJ...461..993F,
       author = {{Fransson}, Claes and {Lundqvist}, Peter and {Chevalier}, Roger A.},
        title = "{Circumstellar Interaction in SN 1993J}",
      journal = {\apj},
         year = 1996,
        month = apr,
       volume = {461},
        pages = {993},
          doi = {10.1086/177119},
       adsurl = {https://ui.adsabs.harvard.edu/abs/1996ApJ...461..993F}
}

@misc{moroianu2025milliarcsecondlocalizationassociatesfrb,
      title={A milliarcsecond localization associates FRB 20190417A with a compact, luminous persistent radio source and an extreme magneto-ionic environment}, 
      author={Alexandra M. Moroianu and Shivani Bhandari and Maria R. Drout and Jason W. T. Hessels and Danté M. Hewitt and Franz Kirsten and Benito Marcote and Ziggy Pleunis and Mark P. Snelders and Navin Sridhar and Uwe Bach and Emmanuel K. Bempong-Manful and Vladislavs Bezrukovs and Richard Blaauw and Justin D. Bray and Salvatore Buttaccio and Shami Chatterjee and Alessandro Corongiu and Roman Feiler and Bryan M. Gaensler and Marcin P. Gawroński and Marcello Giroletti and Adaeze L. Ibik and Ramesh Karuppusamy and Mattias Lazda and Calvin Leung and Michael Lindqvist and Kiyoshi W. Masui and Daniele Michilli and Kenzie Nimmo and Omar S. Ould-Boukattine and Ayush Pandhi and Zsolt Paragi and Aaron B. Pearlman and Weronika Puchalska and Paul Scholz and Kaitlyn Shin and Jurjen J. Sluman and Matteo Trudu and David Williams-Baldwin and Jun Yang},
      year={2025},
      eprint={2509.05174},
      archivePrefix={arXiv},
      primaryClass={astro-ph.HE},
      url={https://arxiv.org/abs/2509.05174}, 
}

@ARTICLE{2008ARA&A..46..127H,
       author = {{Hester}, J.~J.},
        title = "{The Crab Nebula : an astrophysical chimera.}",
      journal = {\araa},
         year = 2008,
        month = sep,
       volume = {46},
        pages = {127-155},
          doi = {10.1146/annurev.astro.45.051806.110608},
       adsurl = {https://ui.adsabs.harvard.edu/abs/2008ARA&A..46..127H}
}

@ARTICLE{2020SciPy-NMeth,
  author  = {Virtanen, Pauli and Gommers, Ralf and Oliphant, Travis E. and
            Haberland, Matt and Reddy, Tyler and Cournapeau, David and
            Burovski, Evgeni and Peterson, Pearu and Weckesser, Warren and
            Bright, Jonathan and {van der Walt}, St{\'e}fan J. and
            Brett, Matthew and Wilson, Joshua and Millman, K. Jarrod and
            Mayorov, Nikolay and Nelson, Andrew R. J. and Jones, Eric and
            Kern, Robert and Larson, Eric and Carey, C J and
            Polat, {\.I}lhan and Feng, Yu and Moore, Eric W. and
            {VanderPlas}, Jake and Laxalde, Denis and Perktold, Josef and
            Cimrman, Robert and Henriksen, Ian and Quintero, E. A. and
            Harris, Charles R. and Archibald, Anne M. and
            Ribeiro, Ant{\^o}nio H. and Pedregosa, Fabian and
            {van Mulbregt}, Paul and {SciPy 1.0 Contributors}},
  title   = {{{SciPy} 1.0: Fundamental Algorithms for Scientific
            Computing in Python}},
  journal = {Nature Methods},
  year    = {2020},
  volume  = {17},
  pages   = {261--272},
  adsurl  = {https://rdcu.be/b08Wh},
  doi     = {10.1038/s41592-019-0686-2},
}

@ARTICLE{2010ApJ...719L.204W,
       author = {{Woosley}, S.~E.},
        title = "{Bright Supernovae from Magnetar Birth}",
      journal = {\apjl},
         year = 2010,
        month = aug,
       volume = {719},
       number = {2},
        pages = {L204-L207},
          doi = {10.1088/2041-8205/719/2/L204},
archivePrefix = {arXiv},
       eprint = {0911.0698},
 primaryClass = {astro-ph.HE},
       adsurl = {https://ui.adsabs.harvard.edu/abs/2010ApJ...719L.204W}
}

@ARTICLE{1998Natur.395..670G,
       author = {{Galama}, T.~J. and {Vreeswijk}, P.~M. and {van Paradijs}, J. and {Kouveliotou}, C. and {Augusteijn}, T. and {B{\"o}hnhardt}, H. and {Brewer}, J.~P. and {Doublier}, V. and {Gonzalez}, J. -F. and {Leibundgut}, B. and {Lidman}, C. and {Hainaut}, O.~R. and {Patat}, F. and {Heise}, J. and {in't Zand}, J. and {Hurley}, K. and {Groot}, P.~J. and {Strom}, R.~G. and {Mazzali}, P.~A. and {Iwamoto}, K. and {Nomoto}, K. and {Umeda}, H. and {Nakamura}, T. and {Young}, T.~R. and {Suzuki}, T. and {Shigeyama}, T. and {Koshut}, T. and {Kippen}, M. and {Robinson}, C. and {de Wildt}, P. and {Wijers}, R.~A.~M.~J. and {Tanvir}, N. and {Greiner}, J. and {Pian}, E. and {Palazzi}, E. and {Frontera}, F. and {Masetti}, N. and {Nicastro}, L. and {Feroci}, M. and {Costa}, E. and {Piro}, L. and {Peterson}, B.~A. and {Tinney}, C. and {Boyle}, B. and {Cannon}, R. and {Stathakis}, R. and {Sadler}, E. and {Begam}, M.~C. and {Ianna}, P.},
        title = "{An unusual supernova in the error box of the {\ensuremath{\gamma}}-ray burst of 25 April 1998}",
      journal = {\nat},
         year = 1998,
        month = oct,
       volume = {395},
       number = {6703},
        pages = {670-672},
          doi = {10.1038/27150},
archivePrefix = {arXiv},
       eprint = {astro-ph/9806175},
 primaryClass = {astro-ph},
       adsurl = {https://ui.adsabs.harvard.edu/abs/1998Natur.395..670G}
}

@article{Petrov_2025,
   title={The Radio Fundamental Catalog. I. Astrometry},
   volume={276},
   ISSN={1538-4365},
   url={http://dx.doi.org/10.3847/1538-4365/ad8c36},
   DOI={10.3847/1538-4365/ad8c36},
   number={2},
   journal={The Astrophysical Journal Supplement Series},
   publisher={American Astronomical Society},
   author={Petrov, L. Y. and Kovalev, Y. Y.},
   year={2025},
   month=jan, pages={38} }

@ARTICLE{2005AJ....129.1993M,
       author = {{Manchester}, R.~N. and {Hobbs}, G.~B. and {Teoh}, A. and {Hobbs}, M.},
        title = "{The Australia Telescope National Facility Pulsar Catalogue}",
      journal = {\aj},
         year = 2005,
        month = apr,
       volume = {129},
       number = {4},
        pages = {1993-2006},
          doi = {10.1086/428488},
archivePrefix = {arXiv},
       eprint = {astro-ph/0412641},
 primaryClass = {astro-ph},
       adsurl = {https://ui.adsabs.harvard.edu/abs/2005AJ....129.1993M}
}

@ARTICLE{2019Sci...363..968G,
       author = {{Ghirlanda}, G. and {Salafia}, O.~S. and {Paragi}, Z. and {Giroletti}, M. and {Yang}, J. and {Marcote}, B. and {Blanchard}, J. and {Agudo}, I. and {An}, T. and {Bernardini}, M.~G. and {Beswick}, R. and {Branchesi}, M. and {Campana}, S. and {Casadio}, C. and {Chassande-Mottin}, E. and {Colpi}, M. and {Covino}, S. and {D'Avanzo}, P. and {D'Elia}, V. and {Frey}, S. and {Gawronski}, M. and {Ghisellini}, G. and {Gurvits}, L.~I. and {Jonker}, P.~G. and {van Langevelde}, H.~J. and {Melandri}, A. and {Moldon}, J. and {Nava}, L. and {Perego}, A. and {Perez-Torres}, M.~A. and {Reynolds}, C. and {Salvaterra}, R. and {Tagliaferri}, G. and {Venturi}, T. and {Vergani}, S.~D. and {Zhang}, M.},
        title = "{Compact radio emission indicates a structured jet was produced by a binary neutron star merger}",
      journal = {Science},
         year = 2019,
        month = mar,
       volume = {363},
       number = {6430},
        pages = {968-971},
          doi = {10.1126/science.aau8815},
archivePrefix = {arXiv},
       eprint = {1808.00469},
 primaryClass = {astro-ph.HE},
       adsurl = {https://ui.adsabs.harvard.edu/abs/2019Sci...363..968G}
}

@ARTICLE{1997ApJ...479..845G,
       author = {{Gaensler}, B.~M. and {Manchester}, R.~N. and {Staveley-Smith}, L. and {Tzioumis}, A.~K. and {Reynolds}, J.~E. and {Kesteven}, M.~J.},
        title = "{The Asymmetric Radio Remnant of SN 1987A}",
      journal = {\apj},
         year = 1997,
        month = apr,
       volume = {479},
       number = {2},
        pages = {845-858},
          doi = {10.1086/303917},
archivePrefix = {arXiv},
       eprint = {astro-ph/9612234},
 primaryClass = {astro-ph},
       adsurl = {https://ui.adsabs.harvard.edu/abs/1997ApJ...479..845G}
}

@article{Taylor_2004,
   title={The Angular Size and Proper Motion of the Afterglow of GRB 030329},
   volume={609},
   ISSN={1538-4357},
   url={http://dx.doi.org/10.1086/422554},
   DOI={10.1086/422554},
   number={1},
   journal={The Astrophysical Journal},
   publisher={American Astronomical Society},
   author={Taylor, G. B. and Frail, D. A. and Berger, E. and Kulkarni, S. R.},
   year={2004},
   month=may, pages={L1–L4} }

@misc{schroeder2025latetimeradiosearchhighly,
      title={A Late-time Radio Search for Highly Off-axis Jets from PTF Broad-lined Ic Supernovae in GRB-like Host Galaxy Environments}, 
      author={Genevieve Schroeder and Anna Y. Q. Ho and Ranadeep G. Dastidar and Maryam Modjaz and Alessandra Corsi and Paul C. Duffell},
      year={2025},
      eprint={2507.15928},
      archivePrefix={arXiv},
      primaryClass={astro-ph.HE},
      url={https://arxiv.org/abs/2507.15928}, 
}

@ARTICLE{2024A&A...690A..74G,
       author = {{Giarratana}, S. and {Salafia}, O.~S. and {Giroletti}, M. and {Ghirlanda}, G. and {Rhodes}, L. and {Atri}, P. and {Marcote}, B. and {Yang}, J. and {An}, T. and {Anderson}, G. and {Bright}, J.~S. and {Farah}, W. and {Fender}, R. and {Leung}, J.~K. and {Motta}, S.~E. and {P{\'e}rez-Torres}, M. and {van der Horst}, A.~J.},
        title = "{The expansion of the GRB 221009A afterglow}",
      journal = {\aap},
         year = 2024,
        month = oct,
       volume = {690},
          eid = {A74},
        pages = {A74},
          doi = {10.1051/0004-6361/202348524},
archivePrefix = {arXiv},
       eprint = {2311.05527},
 primaryClass = {astro-ph.HE},
       adsurl = {https://ui.adsabs.harvard.edu/abs/2024A&A...690A..74G}
}

@ARTICLE{grb_review,
       author = {{Woosley}, S.~E. and {Bloom}, J.~S.},
        title = "{The Supernova Gamma-Ray Burst Connection}",
      journal = {\araa},
         year = 2006,
        month = sep,
       volume = {44},
       number = {1},
        pages = {507-556},
          doi = {10.1146/annurev.astro.43.072103.150558},
archivePrefix = {arXiv},
       eprint = {astro-ph/0609142},
 primaryClass = {astro-ph},
       adsurl = {https://ui.adsabs.harvard.edu/abs/2006ARA&A..44..507W}
}

@article{Granot_2018,
   title={Off-axis afterglow light curves and images from 2D hydrodynamic simulations of double-sided GRB jets in a stratified external medium},
   volume={481},
   ISSN={1365-2966},
   url={http://dx.doi.org/10.1093/mnras/sty2454},
   DOI={10.1093/mnras/sty2454},
   number={2},
   journal={Monthly Notices of the Royal Astronomical Society},
   publisher={Oxford University Press (OUP)},
   author={Granot, Jonathan and De Colle, Fabio and Ramirez-Ruiz, Enrico},
   year={2018},
   month=sep, pages={2711–2720} }

@article{Bietenholz_2014,
   title={Radio limits on off-axis GRB afterglows and VLBI observations of SN 2003gk},
   volume={440},
   ISSN={0035-8711},
   url={http://dx.doi.org/10.1093/mnras/stu246},
   DOI={10.1093/mnras/stu246},
   number={1},
   journal={Monthly Notices of the Royal Astronomical Society},
   publisher={Oxford University Press (OUP)},
   author={Bietenholz, M. F. and De Colle, F. and Granot, J. and Bartel, N. and Soderberg, A. M.},
   year={2014},
   month=mar, pages={821–832} }

@ARTICLE{1999ApJ...524..262M,
       author = {{MacFadyen}, A.~I. and {Woosley}, S.~E.},
        title = "{Collapsars: Gamma-Ray Bursts and Explosions in ``Failed Supernovae''}",
      journal = {\apj},
         year = 1999,
        month = oct,
       volume = {524},
       number = {1},
        pages = {262-289},
          doi = {10.1086/307790},
archivePrefix = {arXiv},
       eprint = {astro-ph/9810274},
 primaryClass = {astro-ph},
       adsurl = {https://ui.adsabs.harvard.edu/abs/1999ApJ...524..262M}
}

@ARTICLE{2018JOSS....3..538P,
       author = {{Pitkin}, Matthew},
        title = "{psrqpy: a python interface for querying the ATNF pulsar catalogue}",
      journal = {The Journal of Open Source Software},
         year = 2018,
        month = feb,
       volume = {3},
       number = {22},
        pages = {538},
          doi = {10.21105/joss.00538},
archivePrefix = {arXiv},
       eprint = {1806.07809},
 primaryClass = {astro-ph.IM},
       adsurl = {https://ui.adsabs.harvard.edu/abs/2018JOSS....3..538P}
}

@ARTICLE{2020A&A...644A.159C,
       author = {{Charlot}, P. and {Jacobs}, C.~S. and {Gordon}, D. and {Lambert}, S. and {de Witt}, A. and {B{\"o}hm}, J. and {Fey}, A.~L. and {Heinkelmann}, R. and {Skurikhina}, E. and {Titov}, O. and {Arias}, E.~F. and {Bolotin}, S. and {Bourda}, G. and {Ma}, C. and {Malkin}, Z. and {Nothnagel}, A. and {Mayer}, D. and {MacMillan}, D.~S. and {Nilsson}, T. and {Gaume}, R.},
        title = "{The third realization of the International Celestial Reference Frame by very long baseline interferometry}",
      journal = {\aap},
         year = 2020,
        month = dec,
       volume = {644},
          eid = {A159},
        pages = {A159},
          doi = {10.1051/0004-6361/202038368},
archivePrefix = {arXiv},
       eprint = {2010.13625},
 primaryClass = {astro-ph.GA},
       adsurl = {https://ui.adsabs.harvard.edu/abs/2020A&A...644A.159C}
}

@ARTICLE{2002ARA&A..40..387W,
       author = {{Weiler}, Kurt W. and {Panagia}, Nino and {Montes}, Marcos J. and {Sramek}, Richard A.},
        title = "{Radio Emission from Supernovae and Gamma-Ray Bursters}",
      journal = {\araa},
         year = 2002,
        month = jan,
       volume = {40},
        pages = {387-438},
          doi = {10.1146/annurev.astro.40.060401.093744},
       adsurl = {https://ui.adsabs.harvard.edu/abs/2002ARA&A..40..387W}
}

@article{Ng_2011,
   title={FIRST VLBI DETECTION OF THE RADIO REMNANT OF SUPERNOVA 1987A: EVIDENCE FOR SMALL-SCALE FEATURES},
   volume={728},
   ISSN={2041-8213},
   url={http://dx.doi.org/10.1088/2041-8205/728/1/L15},
   DOI={10.1088/2041-8205/728/1/l15},
   number={1},
   journal={The Astrophysical Journal},
   publisher={American Astronomical Society},
   author={Ng, C.-Y. and Potter, T. M. and Staveley-Smith, L. and Tingay, S. and Gaensler, B. M. and Phillips, C. and Tzioumis, A. K. and Zanardo, G.},
   year={2011},
   month=jan, pages={L15} }

@ARTICLE{smith_2014_review,
       author = {{Smith}, Nathan},
        title = "{Mass Loss: Its Effect on the Evolution and Fate of High-Mass Stars}",
      journal = {\araa},
         year = 2014,
        month = aug,
       volume = {52},
        pages = {487-528},
          doi = {10.1146/annurev-astro-081913-040025},
archivePrefix = {arXiv},
       eprint = {1402.1237},
 primaryClass = {astro-ph.SR},
       adsurl = {https://ui.adsabs.harvard.edu/abs/2014ARA&A..52..487S}
}

@ARTICLE{2012ApJ...751..125B,
       author = {{Bietenholz}, M.~F. and {Brunthaler}, A. and {Soderberg}, A.~M. and {Krauss}, M. and {Zauderer}, B. and {Bartel}, N. and {Chomiuk}, L. and {Rupen}, M.~P.},
        title = "{VLBI Observations of the Nearby Type IIb Supernova 2011dh}",
      journal = {\apj},
         year = 2012,
        month = jun,
       volume = {751},
       number = {2},
          eid = {125},
        pages = {125},
          doi = {10.1088/0004-637X/751/2/125},
archivePrefix = {arXiv},
       eprint = {1201.0771},
 primaryClass = {astro-ph.HE},
       adsurl = {https://ui.adsabs.harvard.edu/abs/2012ApJ...751..125B}
}

@ARTICLE{2001em_vlbi,
       author = {{Bietenholz}, M.~F. and {Bartel}, N.},
        title = "{SN 2001em: No Jet-driven Gamma-Ray Burst Event}",
      journal = {\apjl},
         year = 2007,
        month = aug,
       volume = {665},
       number = {1},
        pages = {L47-L50},
          doi = {10.1086/521048},
archivePrefix = {arXiv},
       eprint = {0706.3344},
 primaryClass = {astro-ph},
       adsurl = {https://ui.adsabs.harvard.edu/abs/2007ApJ...665L..47B}
}

@ARTICLE{1979c_vlbi,
       author = {{Bartel}, Norbert and {Bietenholz}, Michael F.},
        title = "{SN 1979C VLBI: 22 Years of Almost Free Expansion}",
      journal = {\apj},
         year = 2003,
        month = jul,
       volume = {591},
       number = {1},
        pages = {301-315},
          doi = {10.1086/375267},
       adsurl = {https://ui.adsabs.harvard.edu/abs/2003ApJ...591..301B}
}

@ARTICLE{1996cr_vlbi,
       author = {{Bauer}, F.~E. and {Dwarkadas}, V.~V. and {Brandt}, W.~N. and {Immler}, S. and {Smartt}, S. and {Bartel}, N. and {Bietenholz}, M.~F.},
        title = "{Supernova 1996cr: SN 1987A's Wild Cousin?}",
      journal = {\apj},
         year = 2008,
        month = dec,
       volume = {688},
       number = {2},
        pages = {1210-1234},
          doi = {10.1086/589761},
archivePrefix = {arXiv},
       eprint = {0804.3597},
 primaryClass = {astro-ph},
       adsurl = {https://ui.adsabs.harvard.edu/abs/2008ApJ...688.1210B}
}

@article{Soderberg_2006,
   title={Late‐Time Radio Observations of 68 Type Ibc Supernovae: Strong Constraints on Off‐Axis Gamma‐Ray Bursts},
   volume={638},
   ISSN={1538-4357},
   url={http://dx.doi.org/10.1086/499121},
   DOI={10.1086/499121},
   number={2},
   journal={The Astrophysical Journal},
   publisher={American Astronomical Society},
   author={Soderberg, A. M. and Nakar, E. and Berger, E. and Kulkarni, S. R.},
   year={2006},
   month=feb, pages={930–937} }

@article{Bietenholz_2017,
   title={SN 1986J VLBI. IV. The Nature of the Central Component},
   volume={851},
   ISSN={1538-4357},
   url={http://dx.doi.org/10.3847/1538-4357/aa960b},
   DOI={10.3847/1538-4357/aa960b},
   number={1},
   journal={The Astrophysical Journal},
   publisher={American Astronomical Society},
   author={Bietenholz, Michael F. and Bartel, Norbert},
   year={2017},
   month=dec, pages={7} }

@ARTICLE{Margutti2017,
       author = {{Margutti}, Raffaella and {Kamble}, A. and {Milisavljevic}, D. and {Zapartas}, E. and {de Mink}, S.~E. and {Drout}, M. and {Chornock}, R. and {Risaliti}, G. and {Zauderer}, B.~A. and {Bietenholz}, M. and {Cantiello}, M. and {Chakraborti}, S. and {Chomiuk}, L. and {Fong}, W. and {Grefenstette}, B. and {Guidorzi}, C. and {Kirshner}, R. and {Parrent}, J.~T. and {Patnaude}, D. and {Soderberg}, A.~M. and {Gehrels}, N.~C. and {Harrison}, F.},
        title = "{Ejection of the Massive Hydrogen-rich Envelope Timed with the Collapse of the Stripped SN 2014C}",
      journal = {\apj},
         year = 2017,
        month = feb,
       volume = {835},
       number = {2},
          eid = {140},
        pages = {140},
          doi = {10.3847/1538-4357/835/2/140},
archivePrefix = {arXiv},
       eprint = {1601.06806},
 primaryClass = {astro-ph.HE},
       adsurl = {https://ui.adsabs.harvard.edu/abs/2017ApJ...835..140M}
}

@ARTICLE{Petroff22,
       author = {{Petroff}, E. and {Hessels}, J.~W.~T. and {Lorimer}, D.~R.},
        title = "{Fast radio bursts at the dawn of the 2020s}",
      journal = {\aapr},
         year = 2022,
        month = dec,
       volume = {30},
       number = {1},
          eid = {2},
        pages = {2},
          doi = {10.1007/s00159-022-00139-w},
archivePrefix = {arXiv},
       eprint = {2107.10113},
 primaryClass = {astro-ph.HE},
       adsurl = {https://ui.adsabs.harvard.edu/abs/2022A&ARv..30....2P}
}

@ARTICLE{Chatterjee17,
       author = {{Chatterjee}, S. and {Law}, C.~J. and {Wharton}, R.~S. and {Burke-Spolaor}, S. and {Hessels}, J.~W.~T. and {Bower}, G.~C. and {Cordes}, J.~M. and {Tendulkar}, S.~P. and {Bassa}, C.~G. and {Demorest}, P. and {Butler}, B.~J. and {Seymour}, A. and {Scholz}, P. and {Abruzzo}, M.~W. and {Bogdanov}, S. and {Kaspi}, V.~M. and {Keimpema}, A. and {Lazio}, T.~J.~W. and {Marcote}, B. and {McLaughlin}, M.~A. and {Paragi}, Z. and {Ransom}, S.~M. and {Rupen}, M. and {Spitler}, L.~G. and {van Langevelde}, H.~J.},
        title = "{A direct localization of a fast radio burst and its host}",
      journal = {\nat},
         year = 2017,
        month = jan,
       volume = {541},
       number = {7635},
        pages = {58-61},
          doi = {10.1038/nature20797},
archivePrefix = {arXiv},
       eprint = {1701.01098},
 primaryClass = {astro-ph.HE},
       adsurl = {https://ui.adsabs.harvard.edu/abs/2017Natur.541...58C}
}

@ARTICLE{Niu22,
       author = {{Niu}, C. -H. and {Aggarwal}, K. and {Li}, D. and {Zhang}, X. and {Chatterjee}, S. and {Tsai}, C. -W. and {Yu}, W. and {Law}, C.~J. and {Burke-Spolaor}, S. and {Cordes}, J.~M. and {Zhang}, Y. -K. and {Ocker}, S.~K. and {Yao}, J. -M. and {Wang}, P. and {Feng}, Y. and {Niino}, Y. and {Bochenek}, C. and {Cruces}, M. and {Connor}, L. and {Jiang}, J. -A. and {Dai}, S. and {Luo}, R. and {Li}, G. -D. and {Miao}, C. -C. and {Niu}, J. -R. and {Anna-Thomas}, R. and {Sydnor}, J. and {Stern}, D. and {Wang}, W. -Y. and {Yuan}, M. and {Yue}, Y. -L. and {Zhou}, D. -J. and {Yan}, Z. and {Zhu}, W. -W. and {Zhang}, B.},
        title = "{A repeating fast radio burst associated with a persistent radio source}",
      journal = {\nat},
         year = 2022,
        month = jun,
       volume = {606},
       number = {7916},
        pages = {873-877},
          doi = {10.1038/s41586-022-04755-5},
archivePrefix = {arXiv},
       eprint = {2110.07418},
 primaryClass = {astro-ph.HE},
       adsurl = {https://ui.adsabs.harvard.edu/abs/2022Natur.606..873N}
}

@ARTICLE{Margalit18,
       author = {{Margalit}, Ben and {Metzger}, Brian D.},
        title = "{A Concordance Picture of FRB 121102 as a Flaring Magnetar Embedded in a Magnetized Ion-Electron Wind Nebula}",
      journal = {\apjl},
         year = 2018,
        month = nov,
       volume = {868},
       number = {1},
          eid = {L4},
        pages = {L4},
          doi = {10.3847/2041-8213/aaedad},
archivePrefix = {arXiv},
       eprint = {1808.09969},
 primaryClass = {astro-ph.HE},
       adsurl = {https://ui.adsabs.harvard.edu/abs/2018ApJ...868L...4M}
}

@article{Marcote_2017,
   title={The Repeating Fast Radio Burst FRB 121102 as Seen on Milliarcsecond Angular Scales},
   volume={834},
   ISSN={2041-8213},
   url={http://dx.doi.org/10.3847/2041-8213/834/2/L8},
   DOI={10.3847/2041-8213/834/2/l8},
   number={2},
   journal={ApJL},
   publisher={American Astronomical Society},
   author={Marcote, B. and Paragi, Z. and Hessels, J. W. T. and Keimpema, A. and Langevelde, H. J. van and Huang, Y. and Bassa, C. G. and Bogdanov, S. and Bower, G. C. and Burke-Spolaor, S. and Butler, B. J. and Campbell, R. M. and Chatterjee, S. and Cordes, J. M. and Demorest, P. and Garrett, M. A. and Ghosh, T. and Kaspi, V. M. and Law, C. J. and Lazio, T. J. W. and McLaughlin, M. A. and Ransom, S. M. and Salter, C. J. and Scholz, P. and Seymour, A. and Siemion, A. and Spitler, L. G. and Tendulkar, S. P. and Wharton, R. S.},
   year={2017},
   month=jan, pages={L8} }

@article{Kamble_2014,
doi = {10.1088/0004-637X/797/1/2},
url = {https://dx.doi.org/10.1088/0004-637X/797/1/2},
year = {2014},
month = {nov},
publisher = {The American Astronomical Society},
volume = {797},
number = {1},
pages = {2},
author = {Atish Kamble and Alicia M. Soderberg and Laura Chomiuk and Raffaella Margutti and Mikhail Medvedev and Dan Milisavljevic and Sayan Chakraborti and Roger Chevalier and Nikolai Chugai and Jason Dittmann and Maria Drout and Claes Fransson and Ehud Nakar and Nathan Sanders},
title = {RADIO OBSERVATIONS REVEAL A SMOOTH CIRCUMSTELLAR ENVIRONMENT AROUND THE EXTRAORDINARY TYPE Ib SUPERNOVA 2012au},
journal = {The Astrophysical Journal}
}

@article{Milisavljevic_2018,
doi = {10.3847/2041-8213/aadd4e},
url = {https://dx.doi.org/10.3847/2041-8213/aadd4e},
year = {2018},
month = {sep},
publisher = {The American Astronomical Society},
volume = {864},
number = {2},
pages = {L36},
author = {Dan Milisavljevic and Daniel J. Patnaude and Roger A. Chevalier and John C. Raymond and Robert A. Fesen and Raffaella Margutti and Brody Conner and John Banovetz},
title = {Evidence for a Pulsar Wind Nebula in the Type Ib Peculiar Supernova SN 2012au},
journal = {The Astrophysical Journal Letters}
}

@article{Milisavljevic_2013,
doi = {10.1088/2041-8205/770/2/L38},
url = {https://dx.doi.org/10.1088/2041-8205/770/2/L38},
year = {2013},
month = {jun},
publisher = {The American Astronomical Society},
volume = {770},
number = {2},
pages = {L38},
author = {Dan Milisavljevic and Alicia M. Soderberg and Raffaella Margutti and Maria R. Drout and G. Howie Marion and Nathan E. Sanders and Eric Y. Hsiao and Ragnhild Lunnan and Ryan Chornock and Robert A. Fesen and Jerod T. Parrent and Emily M. Levesque and Edo Berger and Ryan J. Foley and Pete Challis and Robert P. Kirshner and Jason Dittmann and Allyson Bieryla and Atish Kamble and Sayan Chakraborti and Gisella De Rosa and Michael Fausnaugh and Kevin N. Hainline and Chien-Ting Chen and Ryan C. Hickox and Nidia Morrell and Mark M. Phillips and Maximilian Stritzinger},
title = {SN 2012au: A GOLDEN LINK BETWEEN SUPERLUMINOUS SUPERNOVAE AND THEIR LOWER-LUMINOSITY COUNTERPARTS},
journal = {The Astrophysical Journal Letters}
}

@article{Pandey_2021,
	doi = {10.1093/mnras/stab1889},
  
	url = {https://doi.org/10.1093%2Fmnras%2Fstab1889},
  
	year = 2021,
	month = {aug},
  
	publisher = {Oxford University Press ({OUP})},
  
	volume = {507},
  
	number = {1},
  
	pages = {1229--1253},
  
	author = {S B Pandey and Amit Kumar and Brajesh Kumar and G C Anupama and S Srivastav and D K Sahu and J Vinko and A Aryan and A Pastorello and S Benetti and L Tomasella and Avinash Singh and A S Moskvitin and V V Sokolov and R Gupta and K Misra and P Ochner and S Valenti},
  
	title = {Photometric, polarimetric, and spectroscopic studies of the luminous, slow-decaying Type Ib {SN}~2012au},
  
	journal = {Monthly Notices of the Royal Astronomical Society}
}

@article{Takaki_2013,
doi = {10.1088/2041-8205/772/2/L17},
url = {https://dx.doi.org/10.1088/2041-8205/772/2/L17},
year = {2013},
month = {jul},
publisher = {The American Astronomical Society},
volume = {772},
number = {2},
pages = {L17},
author = {Katsutoshi Takaki and Koji S. Kawabata and Masayuki Yamanaka and Keiichi Maeda and Masaomi Tanaka and Hiroshi Akitaya and Yasushi Fukazawa and Ryosuke Itoh and Kenzo Kinugasa and Yuki Moritani and Takashi Ohsugi and Mahito Sasada and Makoto Uemura and Issei Ueno and Takahiro Ui and Takeshi Urano and Michitoshi Yoshida and Ken'ichi Nomoto},
title = {A LUMINOUS AND FAST-EXPANDING TYPE Ib SUPERNOVA SN 2012au},
journal = {The Astrophysical Journal Letters}
}

@ARTICLE{1997_filippenko,
       author = {{Filippenko}, Alexei V.},
        title = "{Optical Spectra of Supernovae}",
      journal = {\araa},
         year = 1997,
        month = jan,
       volume = {35},
        pages = {309-355},
          doi = {10.1146/annurev.astro.35.1.309},
       adsurl = {https://ui.adsabs.harvard.edu/abs/1997ARA&A..35..309F}
}

@misc{bhandari2023constraints,
      title={Constraints on the persistent radio source associated with FRB 20190520B using the European VLBI Network}, 
      author={Shivani Bhandari and Benito Marcote and Navin Sridhar and Tarraneh Eftekhari and Jason W. T. Hessels and Danté M. Hewitt and Franz Kirsten and Omar S. Ould-Boukattine and Zsolt Paragi and Mark P. Snelders},
      year={2023},
      eprint={2308.12801},
      archivePrefix={arXiv},
      primaryClass={astro-ph.HE}
}

@ARTICLE{1993ApJ...405..273W,
       author = {{Woosley}, S.~E.},
        title = "{Gamma-Ray Bursts from Stellar Mass Accretion Disks around Black Holes}",
      journal = {\apj},
         year = 1993,
        month = mar,
       volume = {405},
        pages = {273},
          doi = {10.1086/172359},
       adsurl = {https://ui.adsabs.harvard.edu/abs/1993ApJ...405..273W}
}

@ARTICLE{2015ExA....39..259K,
       author = {{Keimpema}, A. and {Kettenis}, M.~M. and {Pogrebenko}, S.~V. and {Campbell}, R.~M. and {Cim{\'o}}, G. and {Duev}, D.~A. and {Eldering}, B. and {Kruithof}, N. and {van Langevelde}, H.~J. and {Marchal}, D. and {Molera Calv{\'e}s}, G. and {Ozdemir}, H. and {Paragi}, Z. and {Pidopryhora}, Y. and {Szomoru}, A. and {Yang}, J.},
        title = "{The SFXC software correlator for very long baseline interferometry: algorithms and implementation}",
      journal = {Experimental Astronomy},
         year = 2015,
        month = jun,
       volume = {39},
       number = {2},
        pages = {259-279},
          doi = {10.1007/s10686-015-9446-1},
archivePrefix = {arXiv},
       eprint = {1502.00467},
 primaryClass = {astro-ph.IM},
       adsurl = {https://ui.adsabs.harvard.edu/abs/2015ExA....39..259K}
}

@INPROCEEDINGS{1994BAAS...26..987S,
       author = {{Shepherd}, M.~C. and {Pearson}, T.~J. and {Taylor}, G.~B.},
        title = "{DIFMAP: an interactive program for synthesis imaging.}",
    booktitle = {Bulletin of the American Astronomical Society},
         year = 1994,
       volume = {26},
        month = may,
        pages = {987-989},
       adsurl = {https://ui.adsabs.harvard.edu/abs/1994BAAS...26..987S}
}

@article{Chevalier_2006,
   title={Circumstellar Emission from Type Ib and Ic Supernovae},
   volume={651},
   ISSN={1538-4357},
   url={http://dx.doi.org/10.1086/507606},
   DOI={10.1086/507606},
   number={1},
   journal={The Astrophysical Journal},
   publisher={American Astronomical Society},
   author={Chevalier, Roger A. and Fransson, Claes},
   year={2006},
   month=nov, pages={381–391} }

@article{Gaensler_2006,
   title={The Evolution and Structure of Pulsar Wind Nebulae},
   volume={44},
   ISSN={1545-4282},
   url={http://dx.doi.org/10.1146/annurev.astro.44.051905.092528},
   DOI={10.1146/annurev.astro.44.051905.092528},
   number={1},
   journal={Annual Review of Astronomy and Astrophysics},
   publisher={Annual Reviews},
   author={Gaensler, Bryan M. and Slane, Patrick O.},
   year={2006},
   month=sep, pages={17–47} }

@article{Dong_2023,
   title={A Flat-spectrum Radio Transient at 122 Mpc Consistent with an Emerging Pulsar Wind Nebula},
   volume={948},
   ISSN={1538-4357},
   url={http://dx.doi.org/10.3847/1538-4357/acc06c},
   DOI={10.3847/1538-4357/acc06c},
   number={2},
   journal={The Astrophysical Journal},
   publisher={American Astronomical Society},
   author={Dong, Dillon Z. and Hallinan, Gregg},
   year={2023},
   month=may, pages={119} }

@ARTICLE{1984ApJ...285..134B,
       author = {{Bandiera}, R. and {Pacini}, F. and {Salvati}, M.},
        title = "{The evolution of nonthermal supernova remnants. II. Can radio supernovae become plerions ?}",
      journal = {\apj},
         year = 1984,
        month = oct,
       volume = {285},
        pages = {134-140},
          doi = {10.1086/162484},
       adsurl = {https://ui.adsabs.harvard.edu/abs/1984ApJ...285..134B}
}

@ARTICLE{2009Natur.462..624G,
       author = {{Gal-Yam}, A. and {Mazzali}, P. and {Ofek}, E.~O. and {Nugent}, P.~E. and {Kulkarni}, S.~R. and {Kasliwal}, M.~M. and {Quimby}, R.~M. and {Filippenko}, A.~V. and {Cenko}, S.~B. and {Chornock}, R. and {Waldman}, R. and {Kasen}, D. and {Sullivan}, M. and {Beshore}, E.~C. and {Drake}, A.~J. and {Thomas}, R.~C. and {Bloom}, J.~S. and {Poznanski}, D. and {Miller}, A.~A. and {Foley}, R.~J. and {Silverman}, J.~M. and {Arcavi}, I. and {Ellis}, R.~S. and {Deng}, J.},
        title = "{Supernova 2007bi as a pair-instability explosion}",
      journal = {\nat},
         year = 2009,
        month = dec,
       volume = {462},
       number = {7273},
        pages = {624-627},
          doi = {10.1038/nature08579},
archivePrefix = {arXiv},
       eprint = {1001.1156},
 primaryClass = {astro-ph.CO},
       adsurl = {https://ui.adsabs.harvard.edu/abs/2009Natur.462..624G}
}

@ARTICLE{2014PASA...31...22G,
       author = {{Ghirlanda}, G. and {Burlon}, D. and {Ghisellini}, G. and {Salvaterra}, R. and {Bernardini}, M.~G. and {Campana}, S. and {Covino}, S. and {D'Avanzo}, P. and {D'Elia}, V. and {Melandri}, A. and {Murphy}, T. and {Nava}, L. and {Vergani}, S.~D. and {Tagliaferri}, G.},
        title = "{GRB Orphan Afterglows in Present and Future Radio Transient Surveys}",
      journal = {\pasa},
         year = 2014,
        month = may,
       volume = {31},
          eid = {e022},
        pages = {e022},
          doi = {10.1017/pasa.2014.14},
archivePrefix = {arXiv},
       eprint = {1402.6338},
 primaryClass = {astro-ph.HE},
       adsurl = {https://ui.adsabs.harvard.edu/abs/2014PASA...31...22G}
}

@article{Leung_2023,
   title={A matched-filter approach to radio variability and transients: searching for orphan afterglows in the VAST Pilot Survey},
   volume={523},
   ISSN={1365-2966},
   url={http://dx.doi.org/10.1093/mnras/stad1670},
   DOI={10.1093/mnras/stad1670},
   number={3},
   journal={Monthly Notices of the Royal Astronomical Society},
   publisher={Oxford University Press (OUP)},
   author={Leung, James K and Murphy, Tara and Lenc, Emil and Edwards, Philip G and Ghirlanda, Giancarlo and Kaplan, David L and O’Brien, Andrew and Wang, Ziteng},
   year={2023},
   month=jun, pages={4029–4048} }

@ARTICLE{2015ApJ...815..120M,
       author = {{Milisavljevic}, D. and {Margutti}, R. and {Kamble}, A. and {Patnaude}, D.~J. and {Raymond}, J.~C. and {Eldridge}, J.~J. and {Fong}, W. and {Bietenholz}, M. and {Challis}, P. and {Chornock}, R. and {Drout}, M.~R. and {Fransson}, C. and {Fesen}, R.~A. and {Grindlay}, J.~E. and {Kirshner}, R.~P. and {Lunnan}, R. and {Mackey}, J. and {Miller}, G.~F. and {Parrent}, J.~T. and {Sanders}, N.~E. and {Soderberg}, A.~M. and {Zauderer}, B.~A.},
        title = "{Metamorphosis of SN 2014C: Delayed Interaction between a Hydrogen Poor Core-collapse Supernova and a Nearby Circumstellar Shell}",
      journal = {\apj},
         year = 2015,
        month = dec,
       volume = {815},
       number = {2},
          eid = {120},
        pages = {120},
          doi = {10.1088/0004-637X/815/2/120},
archivePrefix = {arXiv},
       eprint = {1511.01907},
 primaryClass = {astro-ph.HE},
       adsurl = {https://ui.adsabs.harvard.edu/abs/2015ApJ...815..120M}
}

@article{Soria_2025,
   title={The radio re-brightening of the Type IIb SN 2001ig},
   volume={42},
   ISSN={1448-6083},
   url={http://dx.doi.org/10.1017/pasa.2025.38},
   DOI={10.1017/pasa.2025.38},
   journal={Publications of the Astronomical Society of Australia},
   publisher={Cambridge University Press (CUP)},
   author={Soria, Roberto and Russell, Thomas D. and Wiston, Eli and Cheng, Siying and Margutti, Raffaella and Rose, Kovi and Ryder, Stuart and Terreran, Giacomo},
   year={2025} }

@ARTICLE{2023A&A...673A.107O,
       author = {{Omand}, C.~M.~B. and {Jerkstrand}, A.},
        title = "{Toward nebular spectral modeling of magnetar-powered supernovae}",
      journal = {\aap},
         year = 2023,
        month = may,
       volume = {673},
          eid = {A107},
        pages = {A107},
          doi = {10.1051/0004-6361/202245406},
archivePrefix = {arXiv},
       eprint = {2211.04502},
 primaryClass = {astro-ph.HE},
       adsurl = {https://ui.adsabs.harvard.edu/abs/2023A&A...673A.107O}
}

@article{Bietenholz_2002,
   title={SN 1986J VLBI: The Evolution and Deceleration of the Complex Source and a Search for a Pulsar Nebula},
   volume={581},
   ISSN={1538-4357},
   url={http://dx.doi.org/10.1086/344251},
   DOI={10.1086/344251},
   number={2},
   journal={The Astrophysical Journal},
   publisher={American Astronomical Society},
   author={Bietenholz, M. F. and Bartel, N. and Rupen, M. P.},
   year={2002},
   month=dec, pages={1132–1147} }

@ARTICLE{2019arXiv190506690B,
       author = {{Bietenholz}, Michael F. and {Bartel}, Norbert},
        title = "{Recent VLBI Results on SN 1986J and the Possibility of FRBs Originating from Inside the Expanding Ejecta of Supernovae}",
      journal = {arXiv e-prints},
         year = 2019,
        month = may,
          eid = {arXiv:1905.06690},
        pages = {arXiv:1905.06690},
          doi = {10.48550/arXiv.1905.06690},
archivePrefix = {arXiv},
       eprint = {1905.06690},
 primaryClass = {astro-ph.HE},
       adsurl = {https://ui.adsabs.harvard.edu/abs/2019arXiv190506690B}
}

@ARTICLE{2006ApJ...641.1051C,
       author = {{Chugai}, Nikolai N. and {Chevalier}, Roger A.},
        title = "{Late Emission from the Type Ib/c SN 2001em: Overtaking the Hydrogen Envelope}",
      journal = {\apj},
         year = 2006,
        month = apr,
       volume = {641},
       number = {2},
        pages = {1051-1059},
          doi = {10.1086/500539},
archivePrefix = {arXiv},
       eprint = {astro-ph/0510362},
 primaryClass = {astro-ph},
       adsurl = {https://ui.adsabs.harvard.edu/abs/2006ApJ...641.1051C}
}

@article{DeSoto_2025,
   title={Spectropolarimetric Evolution Reveals Dual-axis Ejecta in the Atypical Magnetar-powered SN 2012au},
   volume={995},
   ISSN={1538-4357},
   url={http://dx.doi.org/10.3847/1538-4357/ae0e6f},
   DOI={10.3847/1538-4357/ae0e6f},
   number={1},
   journal={The Astrophysical Journal},
   publisher={American Astronomical Society},
   author={DeSoto, Sabrina and Hoffman, Jennifer L. and Williams, G. Grant and Bilinski, Christopher and Leonard, Douglas C. and Milne, Peter A. and Pickens, Christopher and Shrestha, Manisha and Smith, Nathan and Smith, Paul S.},
   year={2025},
   month=dec, pages={5} }

@ARTICLE{1992Natur.355..147S,
       author = {{Staveley-Smith}, L. and {Manchester}, R.~N. and {Kesteven}, M.~J. and {Reynolds}, J.~E. and {Tzioumis}, A.~K. and {Killeen}, N.~E.~B. and {Jauncey}, D.~L. and {Campbell-Wilson}, D. and {Crawford}, D.~F. and {Turtle}, A.~J.},
        title = "{Birth of a radio supernova remnant in supernova 1987A}",
      journal = {\nat},
         year = 1992,
        month = jan,
       volume = {355},
       number = {6356},
        pages = {147-149},
          doi = {10.1038/355147a0},
       adsurl = {https://ui.adsabs.harvard.edu/abs/1992Natur.355..147S}
}

@article{Anderson_2016,
   title={The peculiar mass-loss history of SN 2014C as revealed through AMI radio observations},
   volume={466},
   ISSN={1365-2966},
   url={http://dx.doi.org/10.1093/mnras/stw3310},
   DOI={10.1093/mnras/stw3310},
   number={3},
   journal={Monthly Notices of the Royal Astronomical Society},
   publisher={Oxford University Press (OUP)},
   author={Anderson, G. E. and Horesh, A. and Mooley, K. P. and Rushton, A. P. and Fender, R. P. and Staley, T. D. and Argo, M. K. and Beswick, R. J. and Hancock, P. J. and Pérez-Torres, M. A. and Perrott, Y. C. and Plotkin, R. M. and Pretorius, M. L. and Rumsey, C. and Titterington, D. J.},
   year={2016},
   month=dec, pages={3648–3662} }

@misc{rose2024latetimesupernovaeradiorebrightening,
      title={Late-Time Supernovae Radio Re-brightening in the VAST Pilot Survey}, 
      author={Kovi Rose and Assaf Horesh and Tara Murphy and David L. Kaplan and Itai Sfaradi and Stuart D. Ryder and Robert J. Aloisi and Dougal Dobie and Laura Driessen and Rob Fender and David A. Green and James K. Leung and Emil Lenc and Hao Qiu and David Williams-Baldwin},
      year={2024},
      eprint={2410.01375},
      archivePrefix={arXiv},
      primaryClass={astro-ph.HE},
      url={https://arxiv.org/abs/2410.01375}, 
}

@ARTICLE{2001AJ....121.1648M,
       author = {{Matheson}, Thomas and {Filippenko}, Alexei V. and {Li}, Weidong and {Leonard}, Douglas C. and {Shields}, Joseph C.},
        title = "{Optical Spectroscopy of Type IB/C Supernovae}",
      journal = {\aj},
         year = 2001,
        month = mar,
       volume = {121},
       number = {3},
        pages = {1648-1675},
          doi = {10.1086/319390},
archivePrefix = {arXiv},
       eprint = {astro-ph/0101119},
 primaryClass = {astro-ph},
       adsurl = {https://ui.adsabs.harvard.edu/abs/2001AJ....121.1648M}
}

@article{Deller_2007,
doi = {10.1086/513572},
url = {https://dx.doi.org/10.1086/513572},
year = {2007},
month = {feb},
publisher = {The University of Chicago Press},
volume = {119},
number = {853},
pages = {318},
author = {Deller, A. T. and Tingay, S. J. and Bailes, M. and West, C.},
title = {DiFX: A Software Correlator for Very Long Baseline Interferometry Using Multiprocessor Computing Environments},
journal = {Publications of the Astronomical Society of the Pacific}
}

@ARTICLE{2010ApJ...717..245K,
       author = {{Kasen}, Daniel and {Bildsten}, Lars},
        title = "{Supernova Light Curves Powered by Young Magnetars}",
      journal = {\apj},
         year = 2010,
        month = jul,
       volume = {717},
       number = {1},
        pages = {245-249},
          doi = {10.1088/0004-637X/717/1/245},
archivePrefix = {arXiv},
       eprint = {0911.0680},
 primaryClass = {astro-ph.HE},
       adsurl = {https://ui.adsabs.harvard.edu/abs/2010ApJ...717..245K}
}

@article{Mooley_2018,
   title={Superluminal motion of a relativistic jet in the neutron-star merger GW170817},
   volume={561},
   ISSN={1476-4687},
   url={http://dx.doi.org/10.1038/s41586-018-0486-3},
   DOI={10.1038/s41586-018-0486-3},
   number={7723},
   journal={Nature},
   publisher={Springer Science and Business Media LLC},
   author={Mooley, K. P. and Deller, A. T. and Gottlieb, O. and Nakar, E. and Hallinan, G. and Bourke, S. and Frail, D. A. and Horesh, A. and Corsi, A. and Hotokezaka, K.},
   year={2018},
   month=sep, pages={355–359} }

@ARTICLE{2014c,
       author = {{Margutti}, Raffaella and {Kamble}, A. and {Milisavljevic}, D. and {Zapartas}, E. and {de Mink}, S.~E. and {Drout}, M. and {Chornock}, R. and {Risaliti}, G. and {Zauderer}, B.~A. and {Bietenholz}, M. and {Cantiello}, M. and {Chakraborti}, S. and {Chomiuk}, L. and {Fong}, W. and {Grefenstette}, B. and {Guidorzi}, C. and {Kirshner}, R. and {Parrent}, J.~T. and {Patnaude}, D. and {Soderberg}, A.~M. and {Gehrels}, N.~C. and {Harrison}, F.},
        title = "{Ejection of the Massive Hydrogen-rich Envelope Timed with the Collapse of the Stripped SN 2014C}",
      journal = {\apj},
         year = 2017,
        month = feb,
       volume = {835},
       number = {2},
          eid = {140},
        pages = {140},
          doi = {10.3847/1538-4357/835/2/140},
archivePrefix = {arXiv},
       eprint = {1601.06806},
 primaryClass = {astro-ph.HE},
       adsurl = {https://ui.adsabs.harvard.edu/abs/2017ApJ...835..140M}
}

@ARTICLE{CLEANalgo,
       author = {{H{\"o}gbom}, J.~A.},
        title = "{Aperture Synthesis with a Non-Regular Distribution of Interferometer Baselines}",
      journal = {\aaps},
         year = 1974,
        month = jun,
       volume = {15},
        pages = {417},
       adsurl = {https://ui.adsabs.harvard.edu/abs/1974A&AS...15..417H}
}

@misc{bietenholz2011radioimagingsn1993j,
      title={Radio Imaging of SN 1993J: The Story Continues}, 
      author={M. F. Bietenholz and N. Bartel and M. P. Rupen and V. V. Dwarkadas and A. J. Beasley and D. A. Graham and T. Venturi and G. Umana and W. Cannon and J. Conway},
      year={2011},
      eprint={1103.1783},
      archivePrefix={arXiv},
      primaryClass={astro-ph.CO},
      url={https://arxiv.org/abs/1103.1783}, 
}

@article{pradel2006astrometric-774, 
  year     = {2006}, 
  title    = {Astrometric accuracy of phase-referenced observations with the {VLBA} and {EVN}}, 
  author   = {Pradel, N. and Charlot, P. and Lestrade, J.-F.}, 
  journal  = {Astronomy \& Astrophysics}, 
  issn     = {0004-6361}, 
  doi      = {10.1051/0004-6361:20053021}, 
  eprint   = {astro-ph/0603015}, 
  pages    = {1099--1106}, 
  number   = {3}, 
  volume   = {452}
}

@misc{eiden2025dynamicsobservationalsignaturescorecollapse,
      title={Dynamics and Observational Signatures of Core-Collapse Supernovae with Central Engines: Hydrodynamics Simulations with Monte Carlo Post-Processing}, 
      author={Kiran Eiden and Daniel Kasen},
      year={2025},
      eprint={2510.13741},
      archivePrefix={arXiv},
      primaryClass={astro-ph.HE},
      url={https://arxiv.org/abs/2510.13741}, 
}

@misc{bhardwaj2025constrainingoriginfrb20121102as,
      title={Constraining the Origin of FRB 20121102A's Persistent Radio Source with Long-Term Radio Observations}, 
      author={Mohit Bhardwaj and Arvind Balasubramanian and Yasha Kaushal and Shriharsh P. Tendulkar},
      year={2025},
      eprint={2506.23861},
      archivePrefix={arXiv},
      primaryClass={astro-ph.HE},
      url={https://arxiv.org/abs/2506.23861}, 
}

@ARTICLE{2018ApJ...867...65C,
       author = {{Cendes}, Y. and {Gaensler}, B.~M. and {Ng}, C. -Y. and {Zanardo}, G. and {Staveley-Smith}, L. and {Tzioumis}, A.~K.},
        title = "{The Reacceleration of the Shock Wave in the Radio Remnant of SN 1987A}",
      journal = {\apj},
         year = 2018,
        month = nov,
       volume = {867},
       number = {1},
          eid = {65},
        pages = {65},
          doi = {10.3847/1538-4357/aae261},
archivePrefix = {arXiv},
       eprint = {1809.02364},
 primaryClass = {astro-ph.SR},
       adsurl = {https://ui.adsabs.harvard.edu/abs/2018ApJ...867...65C}
}

@ARTICLE{2025arXiv251013629B,
       author = {{Barlow}, Michael J.},
        title = "{The Supernova 1987A system and its recent evolution - a review}",
      journal = {arXiv e-prints},
         year = 2025,
        month = oct,
          eid = {arXiv:2510.13629},
        pages = {arXiv:2510.13629},
          doi = {10.48550/arXiv.2510.13629},
archivePrefix = {arXiv},
       eprint = {2510.13629},
 primaryClass = {astro-ph.SR},
       adsurl = {https://ui.adsabs.harvard.edu/abs/2025arXiv251013629B}
}

@ARTICLE{2013ApJ...773..139V,
       author = {{Vorster}, M.~J. and {Tibolla}, O. and {Ferreira}, S.~E.~S. and {Kaufmann}, S.},
        title = "{Time-dependent Modeling of Pulsar Wind Nebulae}",
      journal = {\apj},
         year = 2013,
        month = aug,
       volume = {773},
       number = {2},
          eid = {139},
        pages = {139},
          doi = {10.1088/0004-637X/773/2/139},
archivePrefix = {arXiv},
       eprint = {1309.7137},
 primaryClass = {astro-ph.HE},
       adsurl = {https://ui.adsabs.harvard.edu/abs/2013ApJ...773..139V}
}

@article{Stroh_2021,
   title={Luminous Late-time Radio Emission from Supernovae Detected by the Karl G. Jansky Very Large Array Sky Survey (VLASS)},
   volume={923},
   ISSN={2041-8213},
   url={http://dx.doi.org/10.3847/2041-8213/ac375e},
   DOI={10.3847/2041-8213/ac375e},
   number={2},
   journal={The Astrophysical Journal Letters},
   publisher={American Astronomical Society},
   author={Stroh, Michael C. and Terreran, Giacomo and Coppejans, Deanne L. and Bright, Joe S. and Margutti, Raffaella and Bietenholz, Michael F. and De Colle, Fabio and DeMarchi, Lindsay and Duran, Rodolfo Barniol and Milisavljevic, Danny and Murase, Kohta and Paterson, Kerry and Williams, Wendy L.},
   year={2021},
   month=dec, pages={L24} }

@misc{balaubramanian2025continuedradioobservationspersistent,
      title={Continued radio observations of the persistent radio source associated with FRB20190520B provides insights into its origin}, 
      author={Arvind Balaubramanian and Mohit Bhardwaj and Shriharsh P. Tendulkar},
      year={2025},
      eprint={2507.03113},
      archivePrefix={arXiv},
      primaryClass={astro-ph.HE},
      url={https://arxiv.org/abs/2507.03113}, 
}

@article{Fransson_2024,
   title={Emission lines due to ionizing radiation from a compact object in the remnant of Supernova 1987A},
   volume={383},
   ISSN={1095-9203},
   url={http://dx.doi.org/10.1126/science.adj5796},
   DOI={10.1126/science.adj5796},
   number={6685},
   journal={Science},
   publisher={American Association for the Advancement of Science (AAAS)},
   author={Fransson, C. and Barlow, M. J. and Kavanagh, P. J. and Larsson, J. and Jones, O. C. and Sargent, B. and Meixner, M. and Bouchet, P. and Temim, T. and Wright, G. S. and Blommaert, J. A. D. L. and Habel, N. and Hirschauer, A. S. and Hjorth, J. and Lenkić, L. and Tikkanen, T. and Wesson, R. and Coulais, A. and Fox, O. D. and Gastaud, R. and Glasse, A. and Jaspers, J. and Krause, O. and Lau, R. M. and Nayak, O. and Rest, A. and Colina, L. and van Dishoeck, E. F. and Güdel, M. and Henning, Th. and Lagage, P.-O. and Östlin, G. and Ray, T. P. and Vandenbussche, B.},
   year={2024},
   month=feb, pages={898–903} }

@misc{golay2025radioemissioninfraredtidal,
      title={Radio Emission from the Infrared Tidal Disruption Event WTP14adeqka: The First Directly Resolved Delayed Outflow from a TDE}, 
      author={Walter W. Golay and Edo Berger and Yvette Cendes and Megan Masterson and Emil Polisensky and Robert L. Mutel and Peter K. Blanchard and Harsh Kumar and Raffaella Margutti and Maria Drout and Christos Panagiotou and Kishalay De and Erin Kara},
      year={2025},
      eprint={2508.16756},
      archivePrefix={arXiv},
      primaryClass={astro-ph.HE},
      url={https://arxiv.org/abs/2508.16756}, 
}

@ARTICLE{2007ApJ...657L.105F,
       author = {{Foley}, Ryan J. and {Smith}, Nathan and {Ganeshalingam}, Mohan and {Li}, Weidong and {Chornock}, Ryan and {Filippenko}, Alexei V.},
        title = "{SN 2006jc: A Wolf-Rayet Star Exploding in a Dense He-rich Circumstellar Medium}",
      journal = {\apjl},
         year = 2007,
        month = mar,
       volume = {657},
       number = {2},
        pages = {L105-L108},
          doi = {10.1086/513145},
archivePrefix = {arXiv},
       eprint = {astro-ph/0612711},
 primaryClass = {astro-ph},
       adsurl = {https://ui.adsabs.harvard.edu/abs/2007ApJ...657L.105F}
}

@ARTICLE{2012ApJ...752L...2C,
       author = {{Chevalier}, Roger A.},
        title = "{Common Envelope Evolution Leading to Supernovae with Dense Interaction}",
      journal = {\apjl},
         year = 2012,
        month = jun,
       volume = {752},
       number = {1},
          eid = {L2},
        pages = {L2},
          doi = {10.1088/2041-8205/752/1/L2},
archivePrefix = {arXiv},
       eprint = {1204.3300},
 primaryClass = {astro-ph.HE},
       adsurl = {https://ui.adsabs.harvard.edu/abs/2012ApJ...752L...2C}
}

@ARTICLE{2003MNRAS.345..575M,
       author = {{Matzner}, Christopher D.},
        title = "{Supernova hosts for gamma-ray burst jets: dynamical constraints}",
      journal = {\mnras},
         year = 2003,
        month = oct,
       volume = {345},
       number = {2},
        pages = {575-589},
          doi = {10.1046/j.1365-8711.2003.06969.x},
archivePrefix = {arXiv},
       eprint = {astro-ph/0203085},
 primaryClass = {astro-ph},
       adsurl = {https://ui.adsabs.harvard.edu/abs/2003MNRAS.345..575M}
}

@ARTICLE{chevalier_tauff,
       author = {{Chevalier}, R.~A.},
        title = "{The radio and X-ray emission from type II supernovae.}",
      journal = {\apj},
         year = 1982,
        month = aug,
       volume = {259},
        pages = {302-310},
          doi = {10.1086/160167},
       adsurl = {https://ui.adsabs.harvard.edu/abs/1982ApJ...259..302C}
}

@ARTICLE{2003ApJ...597..374B,
       author = {{Bietenholz}, M.~F. and {Bartel}, N. and {Rupen}, M.~P.},
        title = "{SN 1993J VLBI. III. The Evolution of the Radio Shell}",
      journal = {\apj},
         year = 2003,
        month = nov,
       volume = {597},
       number = {1},
        pages = {374-398},
          doi = {10.1086/378265},
archivePrefix = {arXiv},
       eprint = {astro-ph/0307382},
 primaryClass = {astro-ph},
       adsurl = {https://ui.adsabs.harvard.edu/abs/2003ApJ...597..374B}
}

@ARTICLE{2021A&A...655A..41Z,
       author = {{Zhu}, Bo-Tao and {Lu}, Fang-Wu and {Zhou}, Bing and {Zhang}, Li},
        title = "{Multiband nonthermal radiative model of pulsar wind nebulae: Study of the effects of advection and diffusion}",
      journal = {\aap},
         year = 2021,
        month = nov,
       volume = {655},
          eid = {A41},
        pages = {A41},
          doi = {10.1051/0004-6361/202141042},
       adsurl = {https://ui.adsabs.harvard.edu/abs/2021A&A...655A..41Z}
}

@article{10.1111/j.1365-2966.2012.22014.x,
    author = {Martín, Jonatan and Torres, Diego F. and Rea, Nanda},
    title = {Time-dependent modelling of pulsar wind nebulae: study on the impact of the diffusion-loss approximations},
    journal = {Monthly Notices of the Royal Astronomical Society},
    volume = {427},
    number = {1},
    pages = {415-427},
    year = {2012},
    month = {11},
    issn = {0035-8711},
    doi = {10.1111/j.1365-2966.2012.22014.x},
    url = {https://doi.org/10.1111/j.1365-2966.2012.22014.x},
    eprint = {https://academic.oup.com/mnras/article-pdf/427/1/415/18235481/427-1-415.pdf},
}

@BOOK{rybicky_lightman,
       author = {{Rybicki}, George B. and {Lightman}, Alan P.},
        title = "{Radiative Processes in Astrophysics}",
         year = 1986,
       adsurl = {https://ui.adsabs.harvard.edu/abs/1986rpa..book.....R},
  publisher = {Wiley-VCH}
}

@ARTICLE{1981ApJ...251..259C,
       author = {{Chevalier}, R.~A.},
        title = "{The interaction of the radiation from a type II supernova with a circumstellar shell.}",
      journal = {\apj},
         year = 1981,
        month = dec,
       volume = {251},
        pages = {259-265},
          doi = {10.1086/159460},
       adsurl = {https://ui.adsabs.harvard.edu/abs/1981ApJ...251..259C}
}

@article{Chandra_2020,
   title={Supernova Interaction with a Dense Detached Shell in SN 2001em},
   volume={902},
   ISSN={1538-4357},
   url={http://dx.doi.org/10.3847/1538-4357/abb460},
   DOI={10.3847/1538-4357/abb460},
   number={1},
   journal={The Astrophysical Journal},
   publisher={American Astronomical Society},
   author={Chandra, Poonam and Chevalier, Roger A. and Chugai, Nikolai and Milisavljevic, Dan and Fransson, Claes},
   year={2020},
   month=oct, pages={55} }

@article{Ibik_2025,
doi = {10.3847/1538-4357/ad9336},
url = {https://dx.doi.org/10.3847/1538-4357/ad9336},
year = {2025},
month = {jan},
publisher = {The American Astronomical Society},
volume = {979},
number = {1},
pages = {16},
author = {Ibik, Adaeze L. and Drout, Maria R. and Margutti, Raffaella and Matthews, David and Villar, V. Ashley and Berger, Edo and Chornock, Ryan and Alexander, Kate D. and Eftekhari, Tarraneh and Laskar, Tanmoy and Lunnan, Ragnhild and Foley, Ryan J. and Jones, David and Milisavljevic, Dan and Rest, Armin and Scolnic, Daniel and Williams, Peter K. G.},
title = {PS1-11aop: Probing the Mass-loss History of a Luminous Interacting Supernova Prior to Its Final Eruption with Multiwavelength Observations},
journal = {The Astrophysical Journal}
}

@ARTICLE{2015ApJ...806..106A,
       author = {{Alexander}, Kate D. and {Soderberg}, Alicia M. and {Chomiuk}, Laura B.},
        title = "{A New Model for the Radio Emission from SN 1994I and an Associated Search for Radio Transients in M51}",
      journal = {\apj},
         year = 2015,
        month = jun,
       volume = {806},
       number = {1},
          eid = {106},
        pages = {106},
          doi = {10.1088/0004-637X/806/1/106},
archivePrefix = {arXiv},
       eprint = {1405.0228},
 primaryClass = {astro-ph.HE},
       adsurl = {https://ui.adsabs.harvard.edu/abs/2015ApJ...806..106A}
}

@ARTICLE{chevalier1998,
       author = {{Chevalier}, Roger A.},
        title = "{Synchrotron Self-Absorption in Radio Supernovae}",
      journal = {\apj},
         year = 1998,
        month = may,
       volume = {499},
       number = {2},
        pages = {810-819},
          doi = {10.1086/305676},
       adsurl = {https://ui.adsabs.harvard.edu/abs/1998ApJ...499..810C}
}

@INPROCEEDINGS{greisen_2003,
       author = {{Greisen}, E.~W.},
        title = "{AIPS, the VLA, and the VLBA}",
    booktitle = {Information Handling in Astronomy - Historical Vistas},
         year = 2003,
       editor = {{Heck}, Andr{\'e}},
       series = {Astrophysics and Space Science Library},
       volume = {285},
        month = mar,
        pages = {109},
          doi = {10.1007/0-306-48080-8_7},
       adsurl = {https://ui.adsabs.harvard.edu/abs/2003ASSL..285..109G}
}

@ARTICLE{ouldboukattine_2024_arxiv,
       author = {{Ould-Boukattine}, O.~S. and {Chawla}, P. and {Hessels}, J.~W.~T. and {Cooper}, A.~J. and {Gawro{\'n}ski}, M.~P. and {Herrmann}, W. and {Kirsten}, F. and {Hewitt}, D.~M. and {Konijn}, D.~C. and {Nimmo}, K. and {Pleunis}, Z. and {Puchalska}, W. and {Snelders}, M.~P.},
        title = "{A probe of the maximum energetics of fast radio bursts through a prolific repeating source}",
      journal = {arXiv e-prints},
         year = 2024,
        month = oct,
          eid = {arXiv:2410.17024},
        pages = {arXiv:2410.17024},
          doi = {10.48550/arXiv.2410.17024},
archivePrefix = {arXiv},
       eprint = {2410.17024},
 primaryClass = {astro-ph.HE},
       adsurl = {https://ui.adsabs.harvard.edu/abs/2024arXiv241017024O}
}

@ARTICLE{kirsten_2024_natas,
       author = {{Kirsten}, F. and {Ould-Boukattine}, O.~S. and {Herrmann}, W. and {Gawro{\'n}ski}, M.~P. and {Hessels}, J.~W.~T. and {Lu}, W. and {Snelders}, M.~P. and {Chawla}, P. and {Yang}, J. and {Blaauw}, R. and {Nimmo}, K. and {Puchalska}, W. and {Wolak}, P. and {van Ruiten}, R.},
        title = "{A link between repeating and non-repeating fast radio bursts through their energy distributions}",
      journal = {Nature Astronomy},
         year = 2024,
        month = mar,
       volume = {8},
        pages = {337-346},
          doi = {10.1038/s41550-023-02153-z},
archivePrefix = {arXiv},
       eprint = {2306.15505},
 primaryClass = {astro-ph.HE},
       adsurl = {https://ui.adsabs.harvard.edu/abs/2024NatAs...8..337K}
}

@INPROCEEDINGS{whitney_2009_evlb,
       author = {{Whitney}, A. and {Kettenis}, M. and {Phillips}, C. and {Sekido}, M.},
        title = "{VLBI Data Interchange Format (VDIF) (invited)}",
    booktitle = {8th International e-VLBI Workshop},
         year = 2009,
        month = jan,
          eid = {42},
        pages = {42},
          doi = {10.22323/1.082.0042},
       adsurl = {https://ui.adsabs.harvard.edu/abs/2009evlb.confE..42W}
}

@incollection{Chevalier1977,
  author       = {Chevalier, R. A.},
  title        = {—},      
  booktitle    = {Supernovae},
  editor       = {Schramm, D. N.},
  pages        = {53--61},
  year         = {1977},
  publisher    = {Reidel},
  address      = {Dordrecht},
}

@ARTICLE{1997ApJ...480..364F,
       author = {{Frail}, D.~A. and {Scharringhausen}, B.~R.},
        title = "{A Radio Survey for Pulsar Wind Nebulae}",
      journal = {\apj},
         year = 1997,
        month = may,
       volume = {480},
       number = {1},
        pages = {364-370},
          doi = {10.1086/303943},
       adsurl = {https://ui.adsabs.harvard.edu/abs/1997ApJ...480..364F}
}

@ARTICLE{1973ApJ...186..249P,
       author = {{Pacini}, F. and {Salvati}, M.},
        title = "{On the Evolution of Supernova Remnants. Evolution of the Magnetic Field, Particles, Content, and Luminosity}",
      journal = {\apj},
         year = 1973,
        month = nov,
       volume = {186},
        pages = {249-266},
          doi = {10.1086/152495},
       adsurl = {https://ui.adsabs.harvard.edu/abs/1973ApJ...186..249P}
}

@misc{bruni2024nebularoriginpersistentradio,
      title={A nebular origin for the persistent radio emission of fast radio bursts}, 
      author={Gabriele Bruni and Luigi Piro and Yuan-Pei Yang and Salvatore Quai and Bing Zhang and Eliana Palazzi and Luciano Nicastro and Chiara Feruglio and Roberta Tripodi and Brendan O'Connor and Angela Gardini and Sandra Savaglio and Andrea Rossi and A. M. Nicuesa Guelbenzu and Rosita Paladino},
      year={2024},
      eprint={2312.15296},
      archivePrefix={arXiv},
      primaryClass={astro-ph.HE},
      url={https://arxiv.org/abs/2312.15296}, 
}

@article{BARTEL20051057,
title = {A VLBI search for pulsar wind nebulae in supernovae},
journal = {Advances in Space Research},
volume = {35},
number = {6},
pages = {1057-1061},
year = {2005},
note = {Young Neutron Stars and Supernova Remnants},
issn = {0273-1177},
doi = {https://doi.org/10.1016/j.asr.2005.05.060},
url = {https://www.sciencedirect.com/science/article/pii/S0273117705006733},
author = {Norbert Bartel and Michael F. Bietenholz}
}

@article{Bruni_2025,
   title={Discovery of a persistent radio source associated with FRB 20240114A},
   volume={695},
   ISSN={1432-0746},
   url={http://dx.doi.org/10.1051/0004-6361/202453233},
   DOI={10.1051/0004-6361/202453233},
   journal={Astronomy \&; Astrophysics},
   publisher={EDP Sciences},
   author={Bruni, G. and Piro, L. and Yang, Y.-P. and Palazzi, E. and Nicastro, L. and Rossi, A. and Savaglio, S. and Maiorano, E. and Zhang, B.},
   year={2025},
   month=mar, pages={L12} }

@article{Perley_2017,
   title={An Accurate Flux Density Scale from 50 MHz to 50 GHz},
   volume={230},
   ISSN={1538-4365},
   url={http://dx.doi.org/10.3847/1538-4365/aa6df9},
   DOI={10.3847/1538-4365/aa6df9},
   number={1},
   journal={The Astrophysical Journal Supplement Series},
   publisher={American Astronomical Society},
   author={Perley, R. A. and Butler, B. J.},
   year={2017},
   month=may, pages={7} }

@ARTICLE{1968AJ.....73..535T,
       author = {{Trimble}, Virginia},
        title = "{Motions and Structure of the Filamentary Envelope of the Crab Nebula}",
      journal = {\aj},
         year = 1968,
        month = sep,
       volume = {73},
        pages = {535},
          doi = {10.1086/110658},
       adsurl = {https://ui.adsabs.harvard.edu/abs/1968AJ.....73..535T}
}

@ARTICLE{1984ApJ...278..630R,
       author = {{Reynolds}, S.~P. and {Chevalier}, R.~A.},
        title = "{Evolution of pulsar-driven supernova remnants.}",
      journal = {\apj},
         year = 1984,
        month = mar,
       volume = {278},
        pages = {630-648},
          doi = {10.1086/161831},
       adsurl = {https://ui.adsabs.harvard.edu/abs/1984ApJ...278..630R}
}

@ARTICLE{2000MNRAS.318...58G,
       author = {{Gaensler}, B.~M. and {Stappers}, B.~W. and {Frail}, D.~A. and {Moffett}, D.~A. and {Johnston}, S. and {Chatterjee}, S.},
        title = "{Limits on radio emission from pulsar wind nebulae}",
      journal = {\mnras},
         year = 2000,
        month = oct,
       volume = {318},
       number = {1},
        pages = {58-66},
          doi = {10.1046/j.1365-8711.2000.03626.x},
archivePrefix = {arXiv},
       eprint = {astro-ph/0004273},
 primaryClass = {astro-ph},
       adsurl = {https://ui.adsabs.harvard.edu/abs/2000MNRAS.318...58G}
}

@ARTICLE{2009ApJ...703.2051G,
       author = {{Gelfand}, Joseph D. and {Slane}, Patrick O. and {Zhang}, Weiqun},
        title = "{A Dynamical Model for the Evolution of a Pulsar Wind Nebula Inside a Nonradiative Supernova Remnant}",
      journal = {\apj},
         year = 2009,
        month = oct,
       volume = {703},
       number = {2},
        pages = {2051-2067},
          doi = {10.1088/0004-637X/703/2/2051},
archivePrefix = {arXiv},
       eprint = {0904.4053},
 primaryClass = {astro-ph.HE},
       adsurl = {https://ui.adsabs.harvard.edu/abs/2009ApJ...703.2051G}
}

@ARTICLE{2005ApJ...619..839C,
       author = {{Chevalier}, Roger A.},
        title = "{Young Core-Collapse Supernova Remnants and Their Supernovae}",
      journal = {\apj},
         year = 2005,
        month = feb,
       volume = {619},
       number = {2},
        pages = {839-855},
          doi = {10.1086/426584},
archivePrefix = {arXiv},
       eprint = {astro-ph/0409013},
 primaryClass = {astro-ph},
       adsurl = {https://ui.adsabs.harvard.edu/abs/2005ApJ...619..839C}
}

@ARTICLE{vanstraten_2011_pasa,
       author = {{van Straten}, W. and {Bailes}, M.},
        title = "{DSPSR: Digital Signal Processing Software for Pulsar Astronomy}",
      journal = {\pasa},
         year = 2011,
        month = jan,
       volume = {28},
       number = {1},
        pages = {1-14},
          doi = {10.1071/AS10021},
archivePrefix = {arXiv},
       eprint = {1008.3973},
 primaryClass = {astro-ph.IM},
       adsurl = {https://ui.adsabs.harvard.edu/abs/2011PASA...28....1V}
}

@ARTICLE{agarwal_2020_mnras,
       author = {{Agarwal}, Devansh and {Aggarwal}, Kshitij and {Burke-Spolaor}, Sarah and {Lorimer}, Duncan R. and {Garver-Daniels}, Nathaniel},
        title = "{FETCH: A deep-learning based classifier for fast transient classification}",
      journal = {\mnras},
         year = 2020,
        month = sep,
       volume = {497},
       number = {2},
        pages = {1661-1674},
          doi = {10.1093/mnras/staa1856},
archivePrefix = {arXiv},
       eprint = {1902.06343},
 primaryClass = {astro-ph.IM},
       adsurl = {https://ui.adsabs.harvard.edu/abs/2020MNRAS.497.1661A}
}
\bibliographystyle{aasjournal}

\end{document}